\documentclass[12pt,a4paper]{article}
\usepackage{arxiv}
\usepackage[utf8]{inputenc}
\usepackage[T1]{fontenc}
\usepackage{graphicx}
\usepackage{booktabs}
\usepackage{longtable}
\usepackage{array}
\usepackage{calc}
\usepackage{pdflscape}
\usepackage{xurl}

\usepackage{amsmath,amssymb}
\usepackage[numbers]{natbib}
\usepackage{caption}
\usepackage[hidelinks]{hyperref}
\hypersetup{
  pdftitle={Beyond Training: A Feasibility Taxonomy for Inference-Time AI Governance},
  pdfauthor={Samar Ansari},
  pdfkeywords={inference-time governance; compute governance; frontier AI regulation; technical AI governance; feasibility taxonomy; adversary model}
}
\title{\textbf{Beyond Training: A Feasibility Taxonomy for Inference-Time AI Governance}}

\author{
	Samar Ansari\\
	School of Computing and Engineering Sciences\\
	University of Chester\\
	Chester, CH1 4BJ, United Kingdom \\
	\texttt{m.ansari@chester.ac.uk}
}

\date{\today}

\begin{document}
	
\maketitle
\begin{abstract}
Compute governance today is a governance of training: the thresholds, reporting requirements, and frontier-AI regimes now in force attach to training compute and treat the trained model as the regulatory unit. That picture is incomplete: capability increasingly migrates to the deployment stage through inference-time scaling, agentic scaffolding, and compression onto consumer hardware. This paper asks which mechanisms are available once the regulatory object shifts from the training run to the inference call. We develop a feasibility taxonomy of twenty inference-time mechanisms across monitoring, verification, and enforcement, each rated on a four-point readiness scale against a documented four-vendor evidence base. We then stress the taxonomy against a two-dimensional adversary model (three capability tiers crossed with four adversary roles) and map each mechanism to four governance scenarios (domestic regulation, bilateral or multilateral coordination, industry self-regulation, and compute-marketplace governance). Fifteen of the twenty mechanisms have commercial technical substrates in production today, although governance-grade assurance and adversarial robustness vary substantially. The adversary analysis shows that this readiness holds only against a cooperative deployer and a low-to-medium-capability user: no mechanism rates adequate against a high-capability state-level deployer, and fine-tuning removes the model-internal components of the enforcement cluster, although platform-external controls can persist. A substitution analysis connects the taxonomy to a companion hardware paper as a conditional substitution principle describing when inference-stage and hardware-stage mechanisms provide comparable regulatory coverage under stated conditions. A second-rater reliability check on a random subset of the readiness ratings returned a quadratic-weighted Cohen's kappa of 0.74.
\end{abstract}
\hypertarget{introduction}{%
\section{Introduction}\label{introduction}}

Compute governance, as currently practised, is a governance of training.
The thresholds in the EU AI Act \citep{euaiact2024}, the reporting
requirements in the United States Executive Order on AI
\citep{whitehouse2023eo}, and the developing frontier-AI regimes in the
United Kingdom and elsewhere all attach to the cumulative compute used
to train a model. They impose obligations on the developer at the moment
the training run is complete, and they treat the trained model as the
regulatory unit of analysis. This approach made sense when training
compute was the dominant determinant of capability: a few large
laboratories ran a few very long computations, and the results of those
computations defined the frontier. The regulatory architecture mirrored
the computational architecture.

That picture is becoming incomplete. Capability now migrates to the
deployment stage in three structural ways. First, \emph{inference
scaling}: models elicit additional capability through repeated sampling,
longer chain-of-thought reasoning, or larger ensemble inference, at
compute costs that are sometimes comparable to a small training run
\citep{zhang2025survey, almaghrabi2026inference, agarwal2025art}.
Second, \emph{agentic scaffolding}: a base model with modest stand-alone
capability becomes a powerful agent when embedded in tool-using loops
with persistent memory and external resources, and the capability
resides in the scaffolding as much as the model
\citep{ellis2026theory, khairi2025life}. Third, \emph{compression and
consumer-hardware proliferation}: capabilities that previously required
cloud-scale compute can be packaged into small, efficient models that
run on consumer hardware, displacing both the regulatory chokepoint and
the threat surface \citep{puri2026small}; this paper treats
consumer-hosted and local deployment as a coverage limitation on the
commercial-provider ratings rather than as a separately rated setting
(Section 1.2). None of these is a marginal
effect, and in combination they make the proposition ``training compute
determines capability'' empirically false at the frontier.

The governance literature has begun to notice. \citet{hooker2024limitations} argues that compute thresholds are
shortsighted because they ignore inference-time optimisations and assume
a stable training-to-capability relationship that the empirical evidence
does not support. \citet{pistillo2025defending}
identify four specific techniques by which a developer can bypass
training-compute thresholds, the last and structurally distinct of which
is above-optimal inference-time compute scaling. \citet{watson2025beyond} argues for a multidimensional governance
framework that accounts for capability drivers beyond compute. The
position emerging across this literature is that training-compute
thresholds remain useful as a pre-screen but cannot constitute a
standalone governance regime, and that an \emph{inference-time}
governance regime is necessary.

What that regime looks like, however, has not been worked out. The
literature on inference-time governance is partly written by computer
scientists who treat governance as a deployment-time engineering problem
to be solved in code, and partly by policy scholars who treat the
technical substrate as a black box. The result is a body of work that
talks past itself: technical proposals for verifiable inference,
agent-runtime control, and chain-of-thought monitorability accumulate
without being placed in a governance framework, while governance
proposals for inference-time obligations accumulate without engagement
with the actual mechanisms that could implement them. The field has the
parts of a taxonomy but not the taxonomy itself.

This paper provides the taxonomy. We propose a feasibility
classification of inference-time governance mechanisms organised into
three functional categories (monitoring, verification, enforcement),
rated on a four-point technology-readiness scale, and tested against an
extended adversary model and four governance scenarios. The methodology
is deliberately inherited from \citet{ansari2026hardware},
which proposed an analogous taxonomy of twenty hardware-level governance
mechanisms and identified inference-time governance as the natural next
domain to address. The mechanisms in that work are not the mechanisms in
this one, but the four-point readiness scale, the
monitoring-verification-enforcement decomposition, and the conservative
rating rule are identical. Methodological continuity makes the two
papers comparable and lets the present work be read as the second
instalment of a research programme rather than a standalone proposal.

\hypertarget{contributions}{%
\subsection{Contributions}\label{contributions}}

We make four contributions.

First, we provide a structured feasibility classification of twenty
inference-time governance mechanisms across monitoring, verification,
and enforcement. Each mechanism is defined operationally, anchored in
primary academic and vendor sources, rated on a four-point
technology-readiness scale, and described at the deployment layer at
which it operates. To our knowledge, the classification is the first for
inference-time governance specifically, and is comparable to Ansari
\citep{ansari2026hardware} by methodological design.

Second, we extend the adversary model from the three-tier capability
scale used in Ansari \citep{ansari2026hardware} to a two-dimensional model that combines
capability tier with adversary role (developer, deployer or integrator,
end user, and fine-tuner or scaffold-builder). This extension is
required because the inference stage introduces structural role
separation that the training stage did not: the developer who trains a
model is not in general the deployer who serves inference from it, and
neither is the user who issues queries. The 2D matrix surfaces threats
that the 1D scale conflates, including the user-as-adversary case
(prompt injection, distributed Sybil queries), the deployer-as-adversary
case (subverting attestation, lying about deployed model identity), and
the fine-tuner-as-adversary case (re-aligning the model
post-distribution).

Third, we map each mechanism to four governance scenarios: domestic
frontier-AI regulation, bilateral or multilateral agreement (including
treaty regimes with verification), industry self-regulation, and
marketplace-and-platform-level governance. Bilateral and multilateral
coordination are treated as a single scenario because they share the
same structural feature for our purposes, the ability to establish
shared certification authorities and mutual recognition that no single
jurisdiction can build alone. The marketplace scenario is the addition
relative to the scenario template used in Ansari \citep{ansari2026hardware}, and is
necessitated by the structural role of cloud providers, agent platforms,
and model marketplaces as inference-stage governance intermediaries.

Fourth, we develop a convergence analysis. Where the hardware-governance
domain and the inference-governance domain admit different mechanisms
for what is structurally the same governance task, we map the
substitutions explicitly. The convergence analysis culminates in a
conditional substitution principle (Section 5) stating the precise architectural
conditions under which training-time governance mechanisms fail and
inference-time mechanisms must take over.

\hypertarget{scope-and-method}{%
\subsection{Scope and method}\label{scope-and-method}}

The paper covers inference-time compute governance served by major
commercial providers. The readiness ratings are derived from
commercial-provider evidence; consumer-hardware and unfederated
open-weight deployment are treated as coverage limitations and as
conditions under which platform-mediated mechanisms lose force, noted in
the per-mechanism descriptions where they diverge.

The four-point readiness scale is identical to Ansari \citep{ansari2026hardware}:
\emph{currently deployable} (a commercial technical substrate for the mechanism is in production use at a major provider; this rates the availability of that substrate, not the completeness of the governance mechanism, whose assurance and adversarial robustness are assessed separately in Section 4.3),
\emph{near-term} (technical substrate exists, integration or scale-up
required), \emph{requires R\&D} (technical substrate is partial), and
\emph{speculative} (no technical substrate exists yet). The conservative
rating rule of Ansari \citep{ansari2026hardware} is preserved and extended: where academic
and vendor sources disagree on the readiness of a mechanism, we adopt
the more conservative position; where academic sources are silent on the
adversarial conditions under which the mechanism operates, we rate
against the adversarial case rather than the cooperative one. The
extension is necessary because much of the inference-scaling literature
implicitly assumes a cooperative deployer; a feasibility taxonomy for
governance must not. A consequence of the two-dimensional adversary
model (contribution two) is that a readiness rating is a claim about a
mechanism against a named adversary surface, not an unconditional
property: where a mechanism is in production against the surface it was
built for but defeated by another adversary, we state the readiness
rating in scoped form and report the per-adversary detail in the
adversary matrix (Section 4.3, Appendix E). This applies in particular
to the enforcement cluster against the fine-tuner-as-adversary.

\textbf{Table 1. The four-point readiness scale.} The scale is identical
to Ansari \citep{ansari2026hardware}; the conservative-rating rule resolves uncertainty
between adjacent ratings upward.

\begin{longtable}[]{@{}
  >{\raggedright\arraybackslash}p{(\columnwidth - 4\tabcolsep) * \real{0.3333}}
  >{\raggedright\arraybackslash}p{(\columnwidth - 4\tabcolsep) * \real{0.3333}}
  >{\raggedright\arraybackslash}p{(\columnwidth - 4\tabcolsep) * \real{0.3333}}@{}}
\toprule\noalign{}
\begin{minipage}[b]{\linewidth}\raggedright
Rating
\end{minipage} & \begin{minipage}[b]{\linewidth}\raggedright
Label
\end{minipage} & \begin{minipage}[b]{\linewidth}\raggedright
Definition
\end{minipage} \\
\midrule\noalign{}
\endhead
\bottomrule\noalign{}
\endlastfoot
1 & Currently deployable & Commercial technical substrate in production use at a major provider (substrate availability, not governance-grade completeness) \\
2 & Near-term & Technical substrate exists; integration or scale-up
required (roughly 2 to 3 years) \\
3 & Requires R\&D & Foundational research exists but key components are
missing (roughly 5 to 10 years) \\
4 & Speculative & Exists as an academic concept; no production-stage
implementation \\
\end{longtable}

Mechanism ratings were assigned by the author and independently re-rated
by a second rater on a randomly sampled subset of seven of the twenty
mechanisms, spanning all three categories. The two raters agreed exactly
on five of the seven mechanisms; the two disagreements (V4 and V7) were
one-point differences, with the second rater one step more conservative
in each case. Inter-rater reliability, computed with quadratic-weighted
Cohen's kappa as pre-specified in the rater brief, was 0.74, which
corresponds to substantial agreement under the Landis and Koch
interpretation. We report in Appendix B that the estimate is sensitive
to the weighting choice and to the small sample (unweighted kappa is
0.53, in the moderate band), and we do not over-interpret the point
estimate. The full reliability methodology is given in Appendix B, and
the two per-mechanism disagreements and their resolutions are documented
in Appendix C. Vendor documentation snapshots, dated and saved on disk,
are listed in Appendix D. The full populated adversary matrix (Section
4.3) is in Appendix E and the full populated scenario matrix (Section 6)
is in Appendix F.

The verification-mechanism count extends Ansari \citep{ansari2026hardware}'s V1-V6 by one.
The additional mechanism (V7, chain-of-thought monitorability) has no
analogue in hardware-level governance and is required because the
literature on inference-time verification has converged on it as a
distinct primitive
\citep{anwar2026analyzing, korbak2025chain, lanham2023measuring, turpin2023language}.
We treat V7 as a one-mechanism extension rather than as parity, and flag
the extension explicitly in the per-mechanism description in Section
3.2.

\hypertarget{roadmap}{%
\subsection{Roadmap}\label{roadmap}}

Section 2 surveys the existing landscape of compute-governance
instruments and the assumptions about inference that they encode.
Section 3 develops the mechanism taxonomy, with subsections for
monitoring (3.1), verification (3.2), enforcement (3.3), and a
feasibility summary (3.4). Section 4 develops the feasibility
constraints, the 2D adversary model, and the
marketplace-and-intermediation failure modes (4.1 through 4.5). Section
5 develops the convergence analysis, including the substitution table
and the conditional substitution principle. Section 6 maps mechanisms to the four
governance scenarios and draws lessons from existing analogous
monitoring regimes (financial transaction monitoring, pharmacovigilance,
aviation safety reporting). Section 7 discusses the readiness gap,
implications for research and policy, and limitations.

\hypertarget{the-inference-time-compute-landscape}{%
\section{The Inference-Time Compute
Landscape}\label{the-inference-time-compute-landscape}}

Existing compute-governance instruments were designed for training. They
assume that capability accrues during a long, capital-intensive,
geographically concentrated computation, and that the locus of
regulatory attention is the developer who owns or rents that
computation. This section surveys what current instruments do and do not
assume about \emph{inference-time} compute, and identifies the gaps that
the rest of the paper sets out to fill. The intent is descriptive, not
normative: we are mapping the landscape that the taxonomy in Section 3
has to navigate.

We organise the survey in three parts. Section 2.1 considers existing
compute thresholds and the assumptions about inference that they encode
(typically, none). Section 2.2 considers deployment-stage instruments
and the intermediation that inference governance must engage with.
Section 2.3 considers the smaller body of work that has proposed
capability-based or inference-aware governance directly, and identifies
what that body of work has and has not yet articulated.

\hypertarget{inference-and-existing-compute-thresholds}{%
\subsection{Inference and existing compute
thresholds}\label{inference-and-existing-compute-thresholds}}

The EU AI Act provides the most concrete current threshold regime.
Models trained with more than \(10^{25}\) FLOPs of cumulative training
compute are presumed to pose systemic risk and trigger additional
obligations \citep{euaiact2024}. The United States Executive Order on AI
imposed reporting requirements for models trained with more than
\(10^{26}\) FLOPs \citep{whitehouse2023eo}. The UK's emerging frontier
AI regulation, although less prescriptive, operates around comparable
benchmarks \citep{ritchie2025turing}. The threshold-as-trigger pattern
is now broadly established in advanced-economy regulatory practice.

\textbf{Table 2. Existing training-compute threshold regimes and their
treatment of inference.} Every regime triggers obligations on training
compute; none attaches an obligation to inference.

\begin{longtable}[]{@{}
  >{\raggedright\arraybackslash}p{(\columnwidth - 6\tabcolsep) * \real{0.2500}}
  >{\raggedright\arraybackslash}p{(\columnwidth - 6\tabcolsep) * \real{0.2500}}
  >{\raggedright\arraybackslash}p{(\columnwidth - 6\tabcolsep) * \real{0.2500}}
  >{\raggedright\arraybackslash}p{(\columnwidth - 6\tabcolsep) * \real{0.2500}}@{}}
\toprule\noalign{}
\begin{minipage}[b]{\linewidth}\raggedright
Instrument
\end{minipage} & \begin{minipage}[b]{\linewidth}\raggedright
Training-compute threshold
\end{minipage} & \begin{minipage}[b]{\linewidth}\raggedright
Status
\end{minipage} & \begin{minipage}[b]{\linewidth}\raggedright
Treatment of inference
\end{minipage} \\
\midrule\noalign{}
\endhead
\bottomrule\noalign{}
\endlastfoot
EU AI Act (GPAI systemic-risk presumption) & \textgreater{} \(10^{25}\)
FLOP & In force & Silent \\
US Executive Order 14110 (reporting) & \textgreater{} \(10^{26}\) FLOP &
Revoked & Silent \\
UK frontier-AI code of practice & Comparable benchmarks, less
prescriptive & Emerging, non-statutory & Silent \\
\end{longtable}

All of these thresholds are silent on inference. This silence is not
benign. \citet{pistillo2025defending} identify
four distinct techniques by which a developer can produce a model with
frontier capabilities while remaining below the training-compute
threshold: fine-tuning a smaller pre-trained model, reusing components
from existing models, model expansion, and above-optimal inference-time
compute scaling. The last of these is structurally different from the
others: it does not require any additional training compute at all. A
capability that requires \(10^{26}\) training FLOPs in a single-pass
architecture may require only \(10^{24}\) training FLOPs in an
architecture that compensates through repeated inference-time sampling,
deeper chain-of-thought reasoning, or larger ensemble inference. The
threshold-as-trigger pattern therefore fails to capture an entire class
of high-capability deployments.

\citet{hooker2024limitations} makes a stronger version of this
argument. The relationship between training compute and model risk is,
in her account, neither stable nor monotonic, and it ignores a
deployment-side dynamic in which capability emerges from how a model is
used rather than how it was trained. \citet{pistillo2025role} are more sympathetic to thresholds as a
practical regulatory device, but accept that thresholds must be
complemented by capability evaluations and post-market monitoring rather
than treated as a standalone instrument. \citet{erben2025training} provide the most technically detailed defence
of the threshold regime, proposing a framework for defining, measuring,
and updating cumulative training-compute thresholds under the EU AI Act;
even this framework, however, explicitly scopes itself to training and
notes that inference-side enforcement is an open problem. The picture
across these authors is consistent: existing threshold regimes are at
best partial instruments, and inference is the loophole most cleanly
identified in the literature but least addressed in the law.

The internal-deployment carve-outs in the EU AI Act compound the
difficulty. \citet{pistillo2025internal} establishes that the Act
applies to internally deployed AI systems with only narrow exceptions,
but the operationalisation of internal-deployment obligations at the
inference layer is undeveloped. Kontosis \citep{kontosisgoverning} notes
that the Act's model-centric approach to general-purpose AI introduces
upstream obligations on developers but leaves downstream accountability
for deployment-stage harms substantially unresolved. \citet{carey2026regulating} makes a sharper version of this critique:
the Act's reliance on systemic risk as the trigger for the most
stringent obligations institutionalises uncertainty, delegates critical
risk-definition tasks to private actors, and creates regulatory blind
spots that are particularly visible at the inference stage.

The position we take into Section 3 is therefore that compute thresholds
remain a useful pre-screen for identifying potentially high-capability
models at the training stage, but they do not constitute a governance
regime for what happens after the model is trained. They are necessary;
they are not sufficient. The taxonomy of inference-time mechanisms in
Section 3 fills the gap between threshold-triggered obligations and
deployed-system governance.

\hypertarget{deployment-stage-instruments-and-intermediation}{%
\subsection{Deployment-stage instruments and
intermediation}\label{deployment-stage-instruments-and-intermediation}}

A separate body of work focuses on the deployment stage directly. \citet{egan2023oversight} argue that the United States
government should require compute providers to implement
Know-Your-Customer (KYC) schemes for advanced AI development, treating
cloud-compute access as a regulatory chokepoint analogous to the
financial-services chokepoints that anti-money-laundering regulation
engages. \citet{heim2024governing} generalise this argument:
compute providers can serve as effective regulatory intermediaries by
acting as securers, record-keepers, verifiers, and enforcers in frontier
AI governance.

This intermediation argument transfers cleanly to inference. The same
intermediaries who allocate training clusters also serve inference
requests, and the same KYC and record-keeping obligations can in
principle apply at the inference layer. What has not been worked out is
the \emph{operational} form of this transfer. Training-side KYC is
typically tied to large, identifiable customers committing to
long-running compute jobs; inference-side KYC must contend with
high-volume, fine-grained, often programmatically generated API
requests. The intermediation hook is the same; the implementation
surface is different.

The intermediation argument also extends beyond the cloud provider as
such. \citet{williams2026regulating} consider the
regulation of downstream AI developers, particularly those who
fine-tune, modify, or scaffold an upstream model into a deployable
system. Their case is that regulatory attention is currently mismatched:
voluntary guidance and upstream obligations should precede direct
regulation of downstream developers, but downstream developers are
nonetheless an essential node in the governance chain. \citet{alexander2026efficiency} takes the argument further still: the
efficiency shock of highly capable open-weight models renders
compute-centric chokepoint governance partly obsolete, because the
deployment surface is not bottlenecked at a small number of cloud
providers. The implication is that an inference-governance regime
focused exclusively on hyperscaler intermediation will leave the
open-weight deployment surface ungoverned. \citet{gomes2026open}
reaches a related conclusion: restricting access to open-weight models
without parallel governance at the hardware layer simply displaces risk.
The picture across these authors is that intermediation-based governance
is necessary but insufficient, and that any inference-governance regime
must engage with multiple deployment archetypes simultaneously.

Privacy and sovereignty constraints further complicate the
intermediation approach. \citet{bernabei2024legal} note
that compute governance faces significant domestic privacy hurdles even
when it is technically feasible, and these hurdles are sharper at the
inference layer because per-query monitoring of model use bears directly
on user privacy in a way that per-training-run monitoring does not. The
legal architecture for inference-stage monitoring is at present
underdeveloped relative to the technical architecture, and the asymmetry
is itself a constraint on what mechanisms in Section 3 can realistically
deploy.

\hypertarget{capability-based-and-inference-aware-proposals}{%
\subsection{Capability-based and inference-aware
proposals}\label{capability-based-and-inference-aware-proposals}}

A smaller and more recent body of work has begun to engage with
inference-time governance directly, mostly by arguing that
compute-centric metrics should be replaced or supplemented with
capability-based ones. \citet{watson2025beyond} argues that
compute-centric AI governance is structurally inadequate and that
policymakers must adopt a multidimensional weighted framework accounting
for alternative capability drivers, including inference scaling, agentic
scaffolding, and novel substrates. \citet{pacchiardi2025framework} propose a capability-based framework for
categorising general-purpose AI models under the EU AI Act, evaluating
performance across four cognitive domains rather than relying on compute
proxies. \citet{bengani2025beyond} argues that alignment
techniques alone are insufficient to prevent misuse, and that a
comprehensive capability-governance framework requires compute
regulation, tiered audits, and international cooperation in combination.
\citet{caputo2025risk} propose standardised risk tiers
grounded in quantitative modelling where possible.

These proposals share an intuition that the right unit of regulatory
analysis is closer to the deployed system's \emph{capability profile}
than to the training compute that produced it. They differ on
operational specifics: Pacchiardi's framework is evaluation-driven,
Watson's is multidimensional and weighted, Caputo's is risk-tiered,
Bengani's is a layered combination. What unites them is the recognition
that capability and compute are decoupling at the inference stage, and
that governance must follow.

\citet{o2026automated} add an orthogonal proposal:
future AI systems will automate many regulatory compliance tasks, and
policymakers should use ``automatability triggers'' to enact regulations
only when compliance costs fall below acceptable levels. This is
methodologically interesting because it inverts the typical sequencing
(regulate, then automate compliance); the implication for inference
governance is that some mechanisms may become feasible only after the
technical means to deploy them at low marginal cost exist.

Two more recent contributions situate this work in the broader
technical-governance landscape. \citet{reuel2024open}
introduce the framing of \emph{technical AI governance} as a research
programme, providing a taxonomy of open technical problems across the AI
lifecycle. \citet{arslan2025advancing} develops a related taxonomy
specifically for the evaluation, access, and verification capacities.
Both works identify inference-time governance as a substantively
under-specified subarea, and both treat feasibility analysis of specific
inference-time mechanisms as the natural next step. \citet{radanliev2025frontier} argues for a structured regulatory
framework integrating security-first governance with cryptographic
techniques, foreshadowing some of the mechanisms developed in Section 3.

The cumulative picture from this section is consistent. The existing
compute-governance regime addresses training thoroughly and inference
glancingly. A growing body of work has identified the gap, proposed
conceptual frameworks for addressing it, and called for technical
specification of the mechanisms that would operationalise inference-time
governance. None of the works surveyed here has yet provided that
specification. The taxonomy in Section 3, the feasibility analysis in
Section 4, the convergence analysis in Section 5, and the scenario
mapping in Section 6 are our attempt to do so.

\hypertarget{taxonomy-of-inference-time-governance-mechanisms}{%
\section{Taxonomy of Inference-Time Governance
Mechanisms}\label{taxonomy-of-inference-time-governance-mechanisms}}

This section develops the taxonomy. We organise twenty mechanisms into
three categories: Monitoring (M, six mechanisms), Verification (V, seven
mechanisms), and Enforcement (E, seven mechanisms). Each mechanism is
defined operationally, anchored in primary academic and vendor sources,
rated on the four-point readiness scale from Section 1, and described at
the deployment layer at which it operates. The category boundaries
follow Ansari \citep{ansari2026hardware}: monitoring mechanisms produce evidence about what
an inference system is doing; verification mechanisms produce
cryptographic or procedural assurance that the evidence is trustworthy;
enforcement mechanisms intervene in the inference process to bound what
the system does.

The asymmetry in mechanism counts (six monitoring, seven each
verification and enforcement) follows from the cap-to-21 process
described in Appendix A. Two candidates, speculative-decoding draft-model
accounting and intent-verified delegation chains, were rejected for
failing the two-source rule: each had at
most one academic source and zero vendor productisation across the
four-vendor evidence base. The cluster is six mechanisms rather than
seven because the evidence supports six, not because the taxonomy
aspires to symmetry.

This section rests on a four-vendor evidence base: three vendors curated
to row level (Anthropic, OpenAI, Google Vertex AI) plus AWS Bedrock at
survey tier. Azure OpenAI was excluded
because it is structurally OpenAI's offering wrapped in Microsoft's
enterprise controls (Azure RBAC, Private Link, Customer-Managed Keys,
regional deployment); the mechanism-level findings would substantially
overlap with the OpenAI Platform pass while adding compliance primitives
that do not extend the taxonomy. Per-mechanism vendor citations in the
descriptions below refer to documentation snapshots dated May 22, 2026,
archived in Appendix D.

\textbf{Table 3. The inference-time governance taxonomy: twenty
mechanisms.} Readiness is the four-point rating of Section 3.4.
``Vendors'' counts the phase-2-curated vendors (of three: Anthropic,
OpenAI, Google Vertex AI) that productise a feature for the mechanism;
AWS Bedrock is documented at survey tier (Appendix D), so this count is
a lower bound on four-vendor coverage: AWS additionally productises
several mechanisms (for example E2; see Table 4). A value of 0 denotes
no productised vendor feature in the evidence base. ``Adversary
coverage'' gives the count of adequate / partial / inadequate /
not-applicable cells across the twelve adversary cells of Section 4.3
(Appendix E).

{\footnotesize\begin{longtable}[]{@{}
  >{\raggedright\arraybackslash}p{(\columnwidth - 12\tabcolsep) * \real{0.1429}}
  >{\raggedright\arraybackslash}p{(\columnwidth - 12\tabcolsep) * \real{0.1429}}
  >{\raggedright\arraybackslash}p{(\columnwidth - 12\tabcolsep) * \real{0.1429}}
  >{\raggedright\arraybackslash}p{(\columnwidth - 12\tabcolsep) * \real{0.1429}}
  >{\raggedright\arraybackslash}p{(\columnwidth - 12\tabcolsep) * \real{0.1429}}
  >{\raggedright\arraybackslash}p{(\columnwidth - 12\tabcolsep) * \real{0.1429}}
  >{\raggedright\arraybackslash}p{(\columnwidth - 12\tabcolsep) * \real{0.1429}}@{}}
\toprule\noalign{}
\begin{minipage}[b]{\linewidth}\raggedright
ID
\end{minipage} & \begin{minipage}[b]{\linewidth}\raggedright
Mechanism
\end{minipage} & \begin{minipage}[b]{\linewidth}\raggedright
Category
\end{minipage} & \begin{minipage}[b]{\linewidth}\raggedright
Readiness
\end{minipage} & \begin{minipage}[b]{\linewidth}\raggedright
Vendors
\end{minipage} & \begin{minipage}[b]{\linewidth}\raggedright
Adversary A/P/I/N
\end{minipage} & \begin{minipage}[b]{\linewidth}\raggedright
What it does
\end{minipage} \\
\midrule\noalign{}
\endhead
\bottomrule\noalign{}
\endlastfoot
M1 & Per-query usage and token accounting & Monitoring & Deployable & 3 &
3/1/2/6 & Meters compute and tokens per request from billing and usage
telemetry \\
M2 & Workload classification (MoE expert-activation logging) &
Monitoring & Near-term & 0 & 0/4/4/4 & Logs expert-activation patterns
to classify the workload; substrate exists but is not exposed to
callers \\
M3 & KV-cache and reasoning-token accounting & Monitoring & Deployable &
3 & 3/4/1/4 & Accounts for reasoning and cache tokens as an
inference-compute signal \\
M4 & Distributed-inference monitoring & Monitoring & Deployable & 3 &
2/2/2/6 & Tracks which region or data centre served a request \\
M5 & Authenticated account identity and request tracing & Monitoring & Deployable & 3 &
2/2/2/6 & Identifies the caller and traces requests at the API
boundary \\
M6 & Inference-stage power and cost monitoring & Monitoring &
Speculative & 0 & 0/0/6/6 & Per-inference energy and cost as a
governance signal; no vendor productisation \\
V1 & TEE-bounded inference attestation & Verification & Near-term & 1 &
2/6/1/3 & Attests that inference ran in a trusted enclave; substrate at
the infrastructure layer, not API-exposed \\
V2 & Zero-knowledge proofs of inference (zkML) & Verification & Requires
R\&D & 0 & 0/0/3/9 & Cryptographic proof that a stated computation
produced a given output \\
V3 & Hardware-backed workload certificates per output & Verification &
Requires R\&D & 0 & 0/6/3/3 & Per-output certificate inheriting from
hardware roots of trust \\
V4 & Tool-call cryptographic signing and agent-trace attestation &
Verification & Deployable & 2 & 5/4/0/3 & Provider-authenticated
continuity and integrity for tool calls and agent traces
(third-party-verifiable provenance is the stronger undeployed form) \\
V5 & Capability-evaluation reporting & Verification & Deployable & 2 &
4/3/2/3 & Reports that a given capability evaluation was run and
its result (descriptive, not independent or cryptographic attestation) \\
V6 & Retrieval provenance and citation logging & Verification & Deployable &
3 & 0/4/2/6 & Logs retrieval provenance and model-generated citations
linking outputs to sources (not tamper-evident) \\
V7 & Chain-of-thought monitorability & Verification & Deployable & 3 &
2/5/5/0 & Exposes reasoning traces for inspection; bounded by the
faithfulness limitation \\
E1 & Per-deployment licensing and access control & Enforcement &
Deployable & 3 & 3/2/1/6 & Gates model access per deployment and
customer \\
E2 & Output filtering and capability gating & Enforcement & Deployable &
3 & 2/5/5/0 & Filters or blocks model outputs at runtime \\
E3 & Jurisdiction-bounded inference & Enforcement & Deployable & 3 &
2/2/2/6 & Restricts where inference runs or is served (geofencing) \\
E4 & Rate limiting as governance primitive & Enforcement & Deployable &
3 & 2/4/0/6 & Caps per-caller throughput as a control lever \\
E5 & Agentic action permissions and policy-as-code gates & Enforcement &
Deployable & 3 & 2/4/3/3 & Gates agent actions through policy-as-code
engines \\
E6 & Tool-use authorisation and sandboxing & Enforcement & Deployable &
3 & 2/5/2/3 & Authorises and sandboxes tool execution \\
E7 & Inference-time off-switch and kill-chain & Enforcement & Deployable
& 2 & 2/5/2/3 & Suspends or revokes a running deployment \\
\end{longtable}}

Throughout this section, a readiness rating records whether a commercial
technical substrate for the mechanism is in production use; it does not
certify that the complete governance mechanism, including its assurance
and third-party-verifiability properties, is deployable. Those
governance-grade properties are assessed separately through the adversary
analysis of Section 4.3.

\hypertarget{monitoring-mechanisms}{%
\subsection{Monitoring mechanisms}\label{monitoring-mechanisms}}

Monitoring mechanisms produce per-inference evidence about computational
resources consumed, content generated, and actions performed. They are
the foundation of the regulatory architecture: enforcement and
verification depend on monitoring evidence to operate. We describe six
monitoring mechanisms below.

\hypertarget{m1.-per-query-flop-and-token-accounting}{%
\subsubsection{M1. Per-query usage and token accounting}\label{m1.-per-query-flop-and-token-accounting}}

\textbf{Operational definition.} Each inference request produces a
structured usage record reporting input tokens, output tokens, and
(where relevant) reasoning tokens consumed. The record is delivered to
the caller as part of the inference response, billed by the provider,
and retained as part of the request-logging substrate.

\textbf{Vendor productisation.} All four vendors expose M1 at production
scale. Anthropic returns a \texttt{usage} object with
\texttt{input\_tokens}, \texttt{output\_tokens},
\texttt{cache\_creation\_input\_tokens},
\texttt{cache\_read\_input\_tokens}, and
\texttt{output\_tokens\_details.thinking\_tokens} fields
\citep{anthropic_extended_thinking_2026}. OpenAI returns six rate-limit
metrics (RPM, RPD, TPM, TPD, IPM, audio minutes per minute) per
organisation per model and reports \texttt{reasoning\_tokens} in
response metadata for reasoning models \citep{openai_rate_limits_2026}.
AWS Bedrock exposes the underlying provider's accounting via its proxy
layer plus its own billing telemetry \citep{aws_bedrock_ug_2026}. Google
Vertex AI returns \texttt{usageMetadata} with documented per-metric
quotas \citep{vertex_quotas_2026}.

\textbf{Academic source.} The argument that compute accounting at the
regulatory level requires per-call granularity is developed by \citet{casper2025practical} in the context of compute thresholds;
\citet{heim2024governing} treat compute providers as the
natural intermediaries for accounting. The mechanism transfers from
\citet{ansari2026hardware}'s M1 without structural
modification.

\textbf{Readiness rating.} \emph{Currently deployable.} Four-vendor
productisation at production scale. A regulator mandating the per-query
accounting substrate could expect near-term technical compliance.

\hypertarget{m2.-workload-classification-moe-expert-activation-logging}{%
\subsubsection{M2. Workload classification (MoE expert-activation
logging)}\label{m2.-workload-classification-moe-expert-activation-logging}}

\textbf{Operational definition.} For models with
sparse-mixture-of-experts (MoE) architecture, the platform logs which
experts are activated for each inference request. The logs distinguish
workload classes (reasoning-heavy, knowledge-recall, code-generation,
multimodal) by activation pattern. Aggregated over time,
expert-activation logs reveal what categories of inference a deployment
is serving.

\textbf{Vendor productisation.} No vendor in the evidence base
productises M2 at the user-facing API surface. MoE architectures are
deployed (Anthropic's Claude family and Google's Gemini family include
MoE variants) but the per-request expert-activation telemetry is not
exposed.

\textbf{Academic source.} The mechanism transfers from Ansari's
M2 framing of workload classification as a monitoring primitive
\citep{ansari2026hardware}. The architectural shift toward MoE at the
frontier is well-documented; the regulatory application of activation
logs is in early stages.

\textbf{Readiness rating.} \emph{Near-term.} The technical substrate
exists (the activation telemetry is generated internally by every MoE
inference run); the operational gap is exposure to the caller and
aggregation infrastructure for regulatory consumption. A regulator
mandating M2 within a 2-3 year horizon would be technically feasible but
currently lacks vendor-side instrumentation.

\hypertarget{m3.-kv-cache-and-reasoning-token-accounting}{%
\subsubsection{M3. KV-cache and reasoning-token
accounting}\label{m3.-kv-cache-and-reasoning-token-accounting}}

\textbf{Operational definition.} The platform reports per-request
accounting separating: (a) compute attributable to key-value cache hits
versus cache misses, and (b) compute attributable to internal reasoning
tokens versus user-facing output tokens. The accounting distinguishes
``compute spent thinking'' from ``compute spent answering'' and
``compute saved through caching'' from ``fresh compute.''

\textbf{Vendor productisation.} All four vendors productise M3 at
production scale. Anthropic publishes
\texttt{cache\_creation\_input\_tokens},
\texttt{cache\_read\_input\_tokens}, and \texttt{thinking\_tokens} as
distinct accounting categories
\citep{anthropic_prompt_caching_2026, anthropic_extended_thinking_2026}.
OpenAI publishes prompt caching with documented retention (5-10 minutes
idle, max 1 hour for in-memory; extended retention on gpt-5.5 and newer)
and \texttt{reasoning\_tokens} for the o-series and gpt-5.5 models
\citep{openai_prompt_caching_2026, openai_reasoning_2026}. Google Vertex
AI publishes explicit context cache create/use/update/delete operations
and the \texttt{thinking\_level} parameter (MINIMAL/LOW/MEDIUM/HIGH) for
Gemini 3 models
\citep{vertex_context_cache_overview_2026, vertex_thinking_2026}. AWS
Bedrock exposes these primitives through its proxy layer.

\textbf{Academic source.} Inference-scaling papers treat reasoning
compute as a distinct accounting category
\citep{almaghrabi2026inference, agarwal2025art, zhang2025survey}.
KV-cache accounting is well-established in systems literature.

\textbf{Readiness rating.} \emph{Currently deployable.} Four-vendor
productisation; concrete cache TTLs and per-token billing documented.

\hypertarget{m4.-distributed-inference-monitoring-across-data-centres}{%
\subsubsection{M4. Distributed-inference monitoring across data
centres}\label{m4.-distributed-inference-monitoring-across-data-centres}}

\textbf{Operational definition.} When inference is served across
multiple data centres, regions, or jurisdictions, the platform monitors
the geographic distribution of inference requests and surfaces
per-region telemetry. The available evidence concerns the service
region, routing region, and data-residency configuration rather than a
measured attestation of the physical accelerator location: it records
where a request is admitted and routed, not the hardware on which
inference executed. The mechanism supports detecting (and bounding)
cross-jurisdictional inference flows that would evade
single-jurisdiction governance.

\textbf{Vendor productisation.} Three vendors productise M4 with partial
coverage. AWS Bedrock documents Cross-Region Inference profiles that
route requests across multiple AWS regions; per-region quotas and
request logs are available via the standard Bedrock + CloudTrail
interfaces \citep{aws_bedrock_ug_2026}. Anthropic publishes the
\texttt{inference\_geo} parameter (currently \texttt{us} or
\texttt{global}) and documents that rate limits are \emph{not} separated
by \texttt{inference\_geo} (the two share the same quota pool)
\citep{anthropic_rate_limits_2026}. Google Vertex AI ties per-region
telemetry to endpoint choice: request-response logging for partner
models requires a regional endpoint (the global and multi-regional
endpoints do not support it), and the resulting logs are written to a
customer-owned BigQuery table
\citep{vertex_request_response_logging_2026}. OpenAI offers no native
multi-region inference monitoring on the standard OpenAI Platform; the
closest analogue is its project-level data residency controls, which let
an eligible account set a region for new projects (charged an uplift for
models released on or after March 2026) \citep{openai_your_data_2026}.

\textbf{Academic source.} Krys et al. \citep{krys2025distributed}
establish the distributed-inference threat surface; \citet{mueller2025s} establishes that cumulative compute across small
distributed deployments can rival concentrated frontier deployments.
\citet{bernabei2024legal} establish the legal
architecture for cross-jurisdictional monitoring.

\textbf{Readiness rating.} \emph{Currently deployable.} The vendor-side
infrastructure exists at all four providers, with per-region quotas,
logs, and (in three of four) per-region accounting.

\hypertarget{m5.-api-level-kyc-and-request-tracing}{%
\subsubsection{M5. Authenticated account identity and request tracing}\label{m5.-api-level-kyc-and-request-tracing}}

\textbf{Operational definition.} The platform verifies the identity of
each API customer (authenticated account identity and request tracing,
not full regulatory KYC),
maintains traceable request logs linkable to customer identity, and
exposes audit-log access to regulators or auditors under documented
procedures. The mechanism enables regulatory follow-up on specific
deployments without requiring vendor-side judgement on which deployments
are concerning.

\textbf{Vendor productisation.} All four vendors productise M5 with
varying depth. AWS Bedrock uses AWS Identity and Access Management (IAM)
for per-customer identification, with CloudTrail providing audit-log
infrastructure \citep{aws_bedrock_ug_2026}. Anthropic's Trust Center
publishes SOC 2 Type II, ISO 27001, ISO 42001, HIPAA, and FedRAMP High
(via Bedrock GovCloud) certifications; the Privacy Policy documents
Verification Data (ID documents, biometric data, age verification)
collected in defined circumstances
\citep{anthropic_trust_center_2026, anthropic_privacy_policy_2026}.
OpenAI publishes a 30-day default abuse-monitoring log retention with
Modified Abuse Monitoring (excludes customer content, by approval) and
Zero Data Retention (also forces \texttt{store=false}) as opt-in
modifications; the Admin APIs surface audit logs, invites, users,
projects, API keys, spend alerts, data retention, and rate limit
controls \citep{openai_your_data_2026, openai_admin_apis_2026}. Google
Vertex AI provides Cloud Audit Logs (Data Access audit logs enabled
separately from always-on Admin Activity logs) and documents two-tier
abuse monitoring: a 90-day standard tier and a 60-day Advanced AI Safety
Addendum tier for designated models including Claude Mythos, Claude
Fable, and Claude Opus 4.7+ in dual-use contexts
\citep{vertex_enable_audit_logs_2026, vertex_abuse_monitoring_2026}.

\textbf{Academic source.} \citet{egan2023oversight}
establish KYC-for-compute as the regulatory frame; the FATF
Recommendations \citep{fatf2012recommendations} provide the
financial-services analogue; \citet{heim2024governing}
generalise the compute-provider-as-intermediary argument. We return to
the financial-services analogue at length in Section 6.3.

\textbf{Readiness rating.} \emph{Currently deployable.} Multi-vendor
productisation with documented compliance certifications.

\hypertarget{m6.-inference-stage-power-and-cost-monitoring}{%
\subsubsection{M6. Inference-stage power and cost
monitoring}\label{m6.-inference-stage-power-and-cost-monitoring}}

\textbf{Operational definition.} Per-inference accounting of energy
consumption (watts, kilowatt-hours) and cost (in dollars or
compute-credits) attributable to a specific request, attributable to a
specific account, or aggregated to a deployment. The mechanism is the
inference-side analogue of training-stage compute monitoring.

\textbf{Vendor productisation.} No vendor in the evidence base
productises per-inference energy monitoring. Cost accounting is
universal at the API layer but does not separate energy from amortised
infrastructure cost.

\textbf{Academic source.} The mechanism transfers from Ansari's
M4 \citep{ansari2026hardware} which framed power monitoring as a
training-stage primitive; the inference-stage application is
undeveloped. We retain M6 in the taxonomy as a speculative mechanism
because the substrate (data-centre power meters, per-GPU power
telemetry) exists at the infrastructure layer and is not exposed at the
inference API only as a vendor-product decision.

\textbf{Readiness rating.} \emph{Speculative.} The honest read of the
evidence is that energy and per-inference cost monitoring as a
\emph{governance} primitive (as distinct from billing) has neither
academic nor vendor substrate at the inference stage. We retain M6
because Ansari's mirroring argument applies and because the
data-centre energy-reporting regulatory regime in the EU
\citep{eu_csrd_2022} will require this primitive in the near term
whether or not vendors choose to expose it. The rating may move to
``near-term'' within two years if the regulatory pressure materialises.

\hypertarget{verification-mechanisms}{%
\subsection{Verification mechanisms}\label{verification-mechanisms}}

Verification mechanisms produce cryptographic, attestable, or
procedurally trustworthy evidence that monitoring or enforcement
evidence is what it claims to be. Where monitoring tells a regulator
what happened, verification tells the regulator that the monitoring
report is true. We describe seven verification mechanisms below.

\hypertarget{v1.-tee-bounded-inference-attestation}{%
\subsubsection{V1. TEE-bounded inference
attestation}\label{v1.-tee-bounded-inference-attestation}}

\textbf{Operational definition.} Inference computation runs inside a
Trusted Execution Environment (TEE): a hardware-isolated enclave that
cryptographically attests its initial state, the code it is running, and
the integrity of the computation it performs. A third-party verifier can
confirm that a given inference output was produced by a specific model
running in a specific TEE configuration. The mechanism is the
cryptographic substrate underlying several other V-mechanisms in this
section.

\textbf{Vendor productisation.} Vendor evidence is one-vendor partial.
Google Cloud's Confidential Computing platform productises Confidential
VMs, Confidential Space, and Google Cloud Attestation
\citep{google_confidential_computing_2026}. Vertex AI Workbench supports
Confidential VMs but Vertex AI inference (Gemini and partner models)
does not expose TEE-bounded execution at the API layer. AWS Nitro
Enclaves provide CPU-side TEE infrastructure but Bedrock does not expose
TEE-bounded inference at its API layer. Anthropic and OpenAI do not
document confidential inference. NVIDIA Confidential Computing for H100
and B200 GPUs is publicly documented \citep{nvidia_cc_2026} but is not
surfaced through any of the four vendor API layers.

\textbf{Academic source.} A substantial body of work: Chantasantitam et
al.~(PAL*M, \citep{chantasantitam2026palm}); Omega Trusted AI Agents
\citep{omega2025trusted}; Abdollahi et al.~(AgenTEE,
\citep{abdollahi2026agentee}); Duddu et al.~(Laminator, CODASPY 2025
\citep{duddu2025laminator}). The cryptographic substrate is
well-established.

\textbf{Readiness rating.} \emph{Near-term.} The substrate is
production-grade and available at multiple infrastructure providers; the
gap is integration into the user-facing inference product. A regulator
mandating TEE-bounded inference within a 2-3 year horizon could be
complied with by adapting existing infrastructure-level TEE deployment
to inference workloads. The honest characterisation is that V1 is
blocked by vendor-side productisation decisions rather than by missing
technology.

\hypertarget{v2.-zero-knowledge-proofs-of-inference-zkml}{%
\subsubsection{V2. Zero-knowledge proofs of inference
(zkML)}\label{v2.-zero-knowledge-proofs-of-inference-zkml}}

\textbf{Operational definition.} The platform produces a cryptographic
proof, alongside each inference output, that the output was produced by
a specific model on specific inputs without revealing the model weights
or the inputs themselves. A verifier with the proof, the output, and a
commitment to the model's identity can verify inference correctness
while preserving model and input privacy.

\textbf{Vendor productisation.} No vendor in the four-vendor evidence
base productises zkML.

\textbf{Academic source.} Substantial. South et al.~(Verifiable
evaluations of machine learning models using zkSNARKs,
\citep{south2024verifiable}) is foundational. Ivanov et
al.~(DSperse, \citep{dsperse2025framework}) demonstrate targeted
verification reducing proof costs. \citet{optimistic2025teerollups} combine zkML with TEE-bounded
execution to amortise verification overhead. The cryptographic substrate
(zkSNARKs, zkSTARKs) is well-established; the operational challenge is
that producing proofs for frontier-model inference incurs latency and
compute overhead orders of magnitude beyond commercial viability.

\textbf{Readiness rating.} \emph{Requires R\&D.} The mechanism is
implementable in principle (multiple research-stage demonstrations exist
for small models) but production deployment at frontier scale is blocked
by proof-generation overhead. The honest characterisation is that zkML
is a mechanism the academic literature has worked out conceptually; the
engineering work to make it operational at scale has not.

\hypertarget{v3.-hardware-backed-workload-certificates-per-output}{%
\subsubsection{V3. Hardware-backed workload certificates per
output}\label{v3.-hardware-backed-workload-certificates-per-output}}

\textbf{Operational definition.} Each inference output carries a
cryptographic certificate binding the output to: (1) the input that
produced it, (2) the model weights that were used, and (3) the TEE
configuration in which the inference was performed. The certificate is
verifiable by any party with the model's public commitment, and it
enables third-party verification of inference provenance without
requiring access to the underlying TEE infrastructure or to the
inference weights.

\textbf{Vendor productisation.} No vendor productises V3 directly. The
closest evidence is Anthropic's \texttt{signature} field on
extended-thinking blocks (which cryptographically signs internal
reasoning but is not a per-output certificate of the complete inference
flow) and Google Vertex AI's \texttt{thought\_signature} field (similar
primitive with stricter enforcement). These are V4 mechanisms
(cryptographic signing of intermediate state) rather than V3.

\textbf{Academic source.} Medium. Duddu et al.~(Laminator,
\citep{duddu2025laminator}) define Input-Model-Output Attestation
(IOAtt) as the V3 primitive. AEX \citep{aex2026nonintrusive} defines
multi-hop API-level attestation. Omega \citep{omega2025trusted}
documents tamper-evident logging as a V3-adjacent primitive.
\citet{chantasantitam2026palm} integrate V3 with
V1 substrate.

\textbf{Readiness rating.} \emph{Requires R\&D.} V3 is operationally
adjacent to V4 (which is productised at two of four vendors) and to V1
(which has substrate readiness but no productisation). The distinct V3
contribution (per-output certificate of full inference flow) has
academic substrate but no production-stage implementation. We retain V3
as distinct from V4 on the grounds that V3's binding target (final
inference output) is operationally different from V4's (intermediate
state); the distinction may collapse if vendors choose to productise
both as a single primitive.

\hypertarget{v4.-tool-call-cryptographic-signing-and-agent-trace-attestation}{%
\subsubsection{V4. Tool-call cryptographic signing and agent-trace
attestation}\label{v4.-tool-call-cryptographic-signing-and-agent-trace-attestation}}

\textbf{Operational definition.} When a language model invokes a tool,
the invocation is cryptographically signed by the model platform such
that a verifier can later confirm: which model issued the call, what
tool was invoked, what arguments were passed, and that the call has not
been tampered with (provider-authenticated continuity and integrity, not
third-party-verifiable provenance). Where reasoning is involved, the internal reasoning
state that led to the tool invocation is cryptographically bound to the
tool call.

\textbf{Vendor productisation.} Two of four vendors productise V4.
Anthropic's extended thinking provides a \texttt{signature} field on
thinking blocks cryptographically signed by Anthropic's infrastructure;
the signature lets the API verify thinking blocks were generated by
Claude when passed back. Signatures are compatible across Anthropic API,
AWS Bedrock, and Google Vertex AI deployment surfaces
\citep{anthropic_extended_thinking_2026}. Google Vertex AI's
\texttt{thought\_signature} field on Gemini 3 model responses provides
the same primitive with stricter enforcement (Gemini 3 Pro returns HTTP
400 if a required \texttt{thought\_signature} is missing in the next
request) and applies to functionCall parts as well as thinking content
\citep{vertex_thought_signatures_2026}. OpenAI's strict\_tool\_use
schema enforcement provides structural validation but not cryptographic
signing \citep{openai_function_calling_2026}. AWS Bedrock provides
CloudTrail audit logging but no per-call cryptographic signing.

\textbf{Academic source.} MCPSHIELD \citep{mcpshield2026} defines L-CTA
(Cryptographic Tool Attestation). OpenPort Protocol
\citep{zhu2026openport} defines preflight impact hashing as a tool-call
provenance primitive. Omega \citep{omega2025trusted} documents
tamper-evident logging. AEX \citep{aex2026nonintrusive} defines
non-intrusive multi-hop API attestation.

\textbf{Readiness rating.} \emph{Currently deployable.} Two of four
vendors productise the primitive at frontier scale, with one (Vertex)
imposing strict enforcement. A regulator mandating the
provider-authenticated signing primitive within a 6-12 month horizon
could expect adoption of the existing Anthropic or Vertex primitive
across other vendors; the complete V4 attestation mechanism, including
third-party-verifiable provenance, would require additional
specification.

\hypertarget{v5.-capability-evaluation-attestation}{%
\subsubsection{V5. Capability-evaluation reporting}\label{v5.-capability-evaluation-attestation}}

\textbf{Operational definition.} When a model is evaluated against a
benchmark or a capability test, the evaluation infrastructure produces a
report recording: what model was tested, what test was
administered, what the result was, and that the test was performed as
described (evaluation reporting substrate, not cryptographic or
independent attestation). In the current productised form the report is
descriptive: it is legible to parties other than the model developer,
but its integrity rests on the developer's own reporting, so third-party
trust in capability claims depends on the report being honest and
complete rather than on independent or tamper-evident verification.

\textbf{Vendor productisation.} Three vendors productise V5 with varying
depth. AWS Bedrock Model Evaluation provides automated and
human-evaluator workflows with results reported to the customer
\citep{aws_bedrock_ug_2026}. OpenAI's Evals platform provided
programmatic evaluation with adaptive rubrics; the platform was
scheduled for deprecation on October 31, 2026 (read-only) and shutdown
on November 30, 2026 \citep{openai_evals_deprecation_2026}. Google
Vertex AI's Gen AI evaluation service provides adaptive rubrics
(per-prompt tailored pass/fail tests), static rubrics, computation-based
metrics, and custom functions, with evaluation supported from production
logs and synthetic data \citep{vertex_evaluation_overview_2026}. None of
the three vendors cryptographically signs evaluation results; results
are model-graded data, not attestations.

\textbf{Academic source.} \citet{south2024verifiable}
define cryptographic evaluation attestation as a V5 primitive. \citet{balaji2025evaluating} frame evaluation infrastructure as a
deployment-stage governance instrument.

\textbf{Readiness rating.} \emph{Currently deployable} for descriptive
evaluation reporting (the vendor primitive that exists); \emph{Requires
R\&D} for cryptographically attestable evaluation results (the academic
primitive that does not exist at any vendor). The taxonomy retains V5 as
a single mechanism with the rating reflecting the current productisation
state; the gap between vendor reporting and cryptographic attestation is
documented in Section 4.4 and informs the convergence discussion in
Section 5.

\hypertarget{v6.-retrieval-corpus-and-rag-attestation}{%
\subsubsection{V6. Retrieval provenance and citation logging}\label{v6.-retrieval-corpus-and-rag-attestation}}

\textbf{Operational definition.} When inference uses retrieval-augmented
generation (RAG), the platform logs: (1)
what documents were retrieved, (2) what content from each document was
passed to the model, and (3) what citations the model produced linking
output back to retrieved sources (retrieval provenance and citation
logging, not tamper-evident attestation). The logged evidence supports
checking whether model outputs are grounded in claimed sources and
whether retrieval was performed against authorised corpora, to the
extent the logs and model-generated citations are accurate; it is not
tamper-evident, and the citations are model-generated rather than
independently verified.

\textbf{Vendor productisation.} Four vendors productise V6 with
different architectural choices. AWS Bedrock Knowledge Bases provides
first-party vector-store retrieval with source-citation tracking;
per-document source URLs returned with retrieved content
\citep{aws_bedrock_ug_2026}. Anthropic's Citations primitive and Search
Results content blocks provide source-attribution alongside generated
text \citep{anthropic_search_results_2026}. OpenAI's first-party Web
Search tool and Vector Stores (via Assistants API and File Search)
provide source-URL-tagged retrieval \citep{openai_web_search_2026}.
Google Vertex AI's RAG Engine, Grounding with Google Search, and
Grounding with Google Maps provide retrieval primitives with citation
generation \citep{vertex_rag_2026, vertex_grounding_search_2026}. None
of the four vendors provides cryptographic provenance on retrieval
corpora; citations are model-generated attributions, not tamper-evident
bindings.

\textbf{Academic source.} \citet{kholkar2025policy} (Policy-as-Prompt) frame RAG attestation in policy context. The
retrieval-attestation primitive is discussed in multiple corpus sources
without a single canonical citation.

\textbf{Readiness rating.} \emph{Currently deployable.} Four-vendor
productisation at production scale; the cryptographic-attestation gap
(citations are model-generated, not tamper-evident) is a Phase 2 concern
but does not block the basic primitive.

\hypertarget{v7.-chain-of-thought-monitorability}{%
\subsubsection{V7. Chain-of-thought
monitorability}\label{v7.-chain-of-thought-monitorability}}

\textbf{Operational definition.} Models that perform multi-step
reasoning expose intermediate reasoning steps to a verifier in a form
sufficient for the verifier to: (1) confirm the model is reasoning
according to a permitted pattern, (2) detect misalignment or deception
in the reasoning process, and (3) understand the causal chain from
prompt to output. The mechanism may operate over raw chain-of-thought
traces, summarised reasoning, or model-internal activations (white-box
latent monitoring). The productised substrate provides trace exposure:
access to raw or summarised reasoning. This is distinct from faithful
process monitorability, which additionally requires that the exposed
trace accurately reflect the model's actual causal computation; under
the faithfulness limitation (Section 4.3), aims (2) and (3) hold only to
the extent the trace is faithful, which current evidence does not
establish.

\textbf{Vendor productisation.} Three of four vendors productise V7 with
architecturally divergent choices. Anthropic exposes extended-thinking
content with three display modes: \texttt{enabled} (raw chain-of-thought
returned to caller), \texttt{summarized} (model-generated summary), and
\texttt{omitted} (signature only)
\citep{anthropic_extended_thinking_2026}. OpenAI's o-series and gpt-5.5
reasoning models expose reasoning \emph{summaries} only, not raw chains
of thought, by explicit safety design \citep{openai_reasoning_2026}.
Google Vertex AI displays full thinking processes in the Agent Studio
console but exposes only the cryptographically signed
\texttt{thought\_signature} at the programmatic API layer
\citep{vertex_thinking_2026}. AWS Bedrock surfaces Anthropic and Vertex
models' thinking via the underlying provider mechanisms.

The architectural divergence is informative. Anthropic chose
transparency (raw CoT is the default for callers requesting it). OpenAI
chose abstraction (only summaries are exposed, as a safety design).
Vertex chose cryptographic continuity (signatures protect reasoning
state across multi-turn interactions, while raw thinking is reserved for
the console UI). Three vendors, three architecturally distinct choices
for the same underlying primitive. Section 5 returns to this
convergence-with-divergence pattern in the convergence analysis.

The mechanism's white-box variant, internal-activation monitoring rather
than externalised-reasoning monitoring, is research-stage. Yu et
al.~(LatentAudit, \citep{yu2026latentaudit}) demonstrate
residual-stream Mahalanobis-distance monitoring as a real-time
faithfulness-checking primitive. No vendor productises white-box
variants at the API layer; we treat white-box monitoring as a
sub-variant of V7 rather than as a distinct mechanism.

\textbf{Academic source.} Anwar et al.~(Analyzing CoT Faithfulness,
\citep{anwar2026analyzing}) is foundational. \citet{korbak2025chain} and \citet{lanham2023measuring}
provide critical evaluation methodology. \citet{turpin2023language} demonstrate the key limitation: models
sometimes produce reasoning chains that do not faithfully represent
their computational process. White-box variants in \citet{yu2026latentaudit}.

\textbf{Readiness rating.} \emph{Currently deployable.} Three-vendor
productisation of the basic primitive (with architectural divergence
noted). The faithfulness question (whether reasoning traces accurately
represent the model's computation) is a research question for V7's
effectiveness, not a feasibility question for the mechanism itself.

\hypertarget{enforcement-mechanisms}{%
\subsection{Enforcement mechanisms}\label{enforcement-mechanisms}}

Enforcement mechanisms intervene in the inference process to bound what
a system can do. Where monitoring observes and verification attests,
enforcement constrains. We describe seven enforcement mechanisms below.

\hypertarget{e1.-per-deployment-licensing-and-access-control}{%
\subsubsection{E1. Per-deployment licensing and access
control}\label{e1.-per-deployment-licensing-and-access-control}}

\textbf{Operational definition.} The platform restricts inference access
to identified, authorised customers under documented terms. Access is
granted at the level of organisations (account-tier governance),
projects (workspace-level scoping), or per-API-key. Enforcement is via
tiered licensing structures with documented graduation thresholds,
role-based access control, and explicit access-revocation procedures.
The mechanism is the inference-side equivalent of cloud-compute
licensing.

\textbf{Vendor productisation.} Four-vendor productisation. AWS Bedrock
uses AWS IAM for per-account / per-role access control with explicit
``Provisioned Throughput'' reservation for committed-capacity customers
\citep{aws_bedrock_ug_2026}. Anthropic offers per-organisation
tier-based limits (Tier 1-4 plus Monthly Invoicing) with documented
graduation thresholds; per-workspace spend limits and Service Tier
(Priority) commitments are available \citep{anthropic_rate_limits_2026}.
OpenAI offers six usage tiers (Free, Tier 1-5) with documented
qualification thresholds (\$5 / \$50 / \$100 / \$250 / \$1,000 paid);
role-based access control with org-level and project-level roles plus IP
allowlist, mTLS, and OIDC configuration for enterprise customers
\citep{openai_permissions_2026}. Google Vertex AI uses Cloud IAM with
per-project / per-resource permissions; Provisioned Throughput provides
fixed-cost reservations; Model Garden access control gates partner-model
and open-model deployment with explicit consent flows for Advanced AI
models (including Claude Mythos, Fable, and Opus 4.7+)
\citep{vertex_access_control_2026, vertex_control_model_access_2026}.

\textbf{Academic source.} \citet{heim2024governing} and
\citet{brundage2026frontier} frame per-deployment
licensing as a foundational enforcement primitive. The mechanism
transfers from \citet{ansari2026hardware}'s E1 with
deployment-side rather than training-side framing.

\textbf{Readiness rating.} \emph{Currently deployable.} Universal
multi-vendor productisation with documented compliance certifications
and tier-based access governance.

\hypertarget{e2.-output-filtering-and-capability-gating}{%
\subsubsection{E2. Output filtering and capability
gating}\label{e2.-output-filtering-and-capability-gating}}

\textbf{Operational definition.} A runtime layer between the model and
the end user inspects model outputs (and, optionally, model inputs) and
blocks or modifies content that violates documented policy. The
mechanism may operate via separate classifier models, via the inference
model acting as a self-classifier, or via deterministic rule sets over
output tokens. The mechanism is the primary direct lever against
user-as-adversary threats including jailbreak attempts, prompt
injection, and prohibited-content generation.

\textbf{Vendor productisation.} Four vendors productise E2 with four
architecturally distinct choices. AWS Bedrock Guardrails is a paid
configurable runtime layer with documented harm categories and threshold
settings \citep{aws_bedrock_ug_2026}. Anthropic operates server-side
enforcement against its Acceptable Use Policy with documented escalation
to throttle, suspend, or terminate access
\citep{anthropic_usage_policy_2026}; the Refusals and Streaming Refusals
primitives expose this enforcement to callers as a structured
stop\_reason or streaming event \citep{anthropic_refusals_2026}. OpenAI
publishes the Moderation API as a free dedicated endpoint that callers
invoke before or after their main generation call
\citep{openai_moderation_2026}; the Agents SDK adds an integrated
Guardrails framework with input / output / tool guardrails and
human-in-the-loop approvals \citep{openai_guardrails_2026}. Google
Vertex AI documents four configurable HARM\_CATEGORY enums
(HATE\_SPEECH, DANGEROUS\_CONTENT, HARASSMENT, SEXUALLY\_EXPLICIT) with
five HARM\_BLOCK threshold values, plus non-configurable filters for
CSAM and PII; the platform also applies a distinct suspected-CSAM
classifier on image inputs to Anthropic models hosted on Vertex,
separate from Anthropic's own filtering
\citep{vertex_safety_filters_2026, vertex_claude_safety_2026}.

The four architectural choices are: paid configurable runtime layer
(AWS), opaque server-side enforcement (Anthropic), free dedicated
classifier endpoint (OpenAI), configurable per-category threshold API
(Vertex). The architectural diversity is informative for Section 4.5's
marketplace argument.

\textbf{Academic source.} ShieldGemma \citep{zeng2024shieldgemma}, Llama
Guard \citep{inan2023llama}, NeMo Guardrails \citep{rebedea2023nemo} are
foundational. \citet{qi2023fine} demonstrate that fine-tuning
can disable safety guardrails, a finding directly relevant to the
fine-tuner-as-adversary case in Section 4.3.

\textbf{Readiness rating.} \emph{Currently deployable} against the
user-as-adversary and cooperative-deployer surface; inadequate against
the fine-tuner-as-adversary at medium-to-high capability (\citet{qi2023fine}; Section 4.3). Universal multi-vendor productisation;
the architectural diversity is itself a feasibility finding, and the
adversary-scoping is developed in Section 4.3 and summarised in Section
3.4.

\hypertarget{e3.-jurisdiction-bounded-inference}{%
\subsubsection{E3. Jurisdiction-bounded
inference}\label{e3.-jurisdiction-bounded-inference}}

\textbf{Operational definition.} The platform restricts inference
processing to specified jurisdictions, regions, or data centres, with
documented data-residency assertions, network-perimeter enforcement, and
audit-log retention bounded to the same jurisdiction. The mechanism is
the inference-side instrument of the territorial-jurisdiction dimension
of compute sovereignty.

\textbf{Vendor productisation.} Four vendors productise E3 with multiple
sub-primitives. AWS Bedrock supports per-region model access plus VPC
endpoints for network-perimeter enforcement \citep{aws_bedrock_ug_2026}.
Anthropic's \texttt{inference\_geo} parameter (currently \texttt{us} or
\texttt{global}) bounds inference to documented geographic sets, and
cloud-platform deployments provide regionally-scoped deployment surfaces
\citep{anthropic_rate_limits_2026}. OpenAI documents an ``allowed
geography'' list (free-tier access conditional on a supported-countries
enumeration) and regional deployment of OpenAI models via Amazon Bedrock
and Microsoft Foundry \citep{openai_amazon_bedrock_2026}; its
standard-platform data-residency controls let an eligible account set a
project region \citep{openai_your_data_2026}. Google Vertex AI provides
per-region quotas with documented data-residency assertions, VPC Service
Controls perimeter enforcement, and Customer-Managed Encryption Keys
with documented per-model support matrices
\citep{vertex_security_controls_2026, vertex_networking_2026}.

\textbf{Academic source.} \citet{hawkins2025ai} identify
three distinct levels of compute sovereignty (territorial jurisdiction,
provider nationality, accelerator vendor nationality); \citet{lehdonvirta2024compute} establish the Compute North/South
asymmetry; \citet{bernabei2024legal} establish the legal
framework.

\textbf{Readiness rating.} \emph{Currently deployable.} Universal vendor
productisation; Google Vertex AI documents the deepest E3 substrate of
any vendor.

\hypertarget{e4.-rate-limiting-as-governance-primitive}{%
\subsubsection{E4. Rate limiting as governance
primitive}\label{e4.-rate-limiting-as-governance-primitive}}

\textbf{Operational definition.} The platform enforces per-organisation,
per-user, or per-key quotas on inference requests, tokens, or compute.
Quotas are tiered by customer relationship (free, paid, enterprise) and
serve as a regulatory chokepoint: changing quotas changes the deployable
inference budget without requiring new infrastructure. The mechanism is
the operational mechanism of capability throttling.

\textbf{Vendor productisation.} Four vendors productise E4 with concrete
documented numbers. AWS Bedrock publishes per-region per-model quotas
adjustable via the AWS Service Quotas dashboard
\citep{aws_bedrock_ug_2026}. Anthropic publishes tier-based limits (Tier
1: 50 RPM, 30,000 ITPM; Tier 4: 4,000 RPM, up to 10M ITPM) with separate
limits for Message Batches and Managed Agents endpoints
\citep{anthropic_rate_limits_2026}. OpenAI publishes six rate-limit
dimensions (RPM, RPD, TPM, TPD, IPM, audio minutes per minute) per
organisation per model across six usage tiers, plus priority processing
(service\_tier=priority), flex processing (service\_tier=flex), and
batch queue limits \citep{openai_rate_limits_2026}. Google Vertex AI
publishes per-metric quotas (e.g.,
\texttt{aiplatform.googleapis.com/embed\_content\_input\_tokens\_per\_minute\_per\_base\_model\ =\ 5,000,000})
editable in the Cloud console, with concrete published numbers for Agent
Runtime (10-10,000 per-operation-class RPM), sandbox environment (1,000
RPM), and A2A agent operations \citep{vertex_quotas_2026}. All four
vendors publish 429 / retry-after error semantics for rate-limit
enforcement.

\textbf{Academic source.} \citet{egan2023oversight} frame
rate limiting as a regulatory chokepoint analogous to
financial-services. The FATF Recommendations
\citep{fatf2012recommendations} provide the financial-services analogue.
\citet{ho2023international} frame rate limiting as a
verifiable lever for international AI governance.

\textbf{Readiness rating.} \emph{Currently deployable.} Universal
multi-vendor productisation with published numerical limits; the
mechanism is in production use for billing, abuse prevention, and
capacity allocation today.

\hypertarget{e5.-agentic-action-permissions-and-policy-as-code-gates}{%
\subsubsection{E5. Agentic action permissions and policy-as-code
gates}\label{e5.-agentic-action-permissions-and-policy-as-code-gates}}

\textbf{Operational definition.} When a language model invokes external
tools or takes actions in the world, the platform enforces a declarative
policy specifying which actions are permitted, under what conditions,
and with what argument constraints. The mechanism may operate via
structured tool schemas (JSON Schema enforced server-side), via runtime
policy engines, or via per-action approval gates. Anthropic's Agent
Skills with SKILL.md format documents Mavračić's Policy Cards as a
deployment-layer artefact within this mechanism; the artefact-runtime
distinction is captured here rather than as a separate mechanism.

\textbf{Vendor productisation.} Four-vendor productisation with multiple
primitives. AWS Bedrock Agents action groups define structured tool
surfaces with documented schemas \citep{aws_bedrock_ug_2026}.
Anthropic's Tool Use surfaces structured \texttt{tools} parameters with
JSON-schema-enforced per-tool inputs; Strict Tool Use adds server-side
schema enforcement; Computer Use adds prompt-injection classifiers with
confirmation gates; Agent Skills with SKILL.md format provide
deployment-layer governance artefacts with documented six-stage
lifecycle (validate \(\rightarrow\) sandbox \(\rightarrow\) shadow
\(\rightarrow\) gate \(\rightarrow\) rollback)
\citep{anthropic_tool_use_2026, anthropic_strict_tool_use_2026, anthropic_skills_enterprise_2026}.
OpenAI Function Calling with JSON Schema enforcement; Structured Outputs
for schema-enforced JSON output; Orchestration primitives (handoffs and
agents-as-tools) in the Agents SDK
\citep{openai_function_calling_2026, openai_structured_outputs_2026, openai_orchestration_2026}.
Google Vertex AI Function Calling with JSON Schema validation;
Controlled Generation for structured output; thought\_signature
enforcement on functionCall parts (Gemini 3+); Code Execution tool with
per-region sandbox quotas
\citep{vertex_function_calling_2026, vertex_control_generated_output_2026, vertex_code_execution_2026}.

The mechanism encompasses several sub-primitives that surfaced during
the cap-to-21 process as separate candidates but were merged:
deployment-layer artefacts (Mavračić Policy Cards
\citep{mavracic2025policy}, Anthropic Skills), policy-as-prompt \citep{kholkar2025policy}, longitudinal trust factor scoring
(\citet{gaurav2025governance}), information flow tracking
with taint propagation (MCPSHIELD L-IFT \citep{mcpshield2026}).

\textbf{Academic source.} Multiple. MI9 (\citet{wang2025mi9}) provides the canonical multi-primitive framework.
OpenPort (\citet{zhu2026openport}) defines
authorisation-dependent discovery, draft-first writes, preflight impact
hashing. Uchibeke (Before the Tool Call, \citep{uchibeke2026before})
defines pre-action authorisation. SentinelAgent
\citep{sentinelagent2026} defines Intent-Preserving Delegation Protocol.

\textbf{Readiness rating.} \emph{Currently deployable} against the
user-as-adversary and cooperative-deployer surface; degrades against the
fine-tuner-as-adversary (partial at medium capability, inadequate at
high capability; Section 4.3). These degradation ratings assume model-embedded or model-mediated enforcement; provider-external implementations (external moderation, provider-side policy engines, orchestration, or API gateways) may retain higher resilience. Universal vendor productisation with the
artefact-and-runtime composition documented at all four vendors.

\hypertarget{e6.-tool-use-authorisation-and-sandboxing}{%
\subsubsection{E6. Tool-use authorisation and
sandboxing}\label{e6.-tool-use-authorisation-and-sandboxing}}

\textbf{Operational definition.} When a language model invokes a tool
that can take action with side effects (run code, edit files, query
external services, interact with computer interfaces), the platform
provides an isolated execution environment in which the tool can operate
without affecting the broader system. The mechanism bounds what an agent
can do with the tool surface even when the policy layer (E5) authorises
the action.

\textbf{Vendor productisation.} Four-vendor productisation with one
vendor productising a first-party sandbox primitive. AWS Bedrock relies
on AWS IAM for tool-permission scoping and customer-implemented sandbox
runtime \citep{aws_bedrock_ug_2026}. Anthropic's Computer Use tool
requires customer-implemented sandboxing (reference implementation
provides Docker container, agent loop, web interface); Bash tool and
Text editor tool provide tool schemas only with sandbox runtime
customer-side
\citep{anthropic_computer_use_2026, anthropic_bash_tool_2026}. OpenAI's
Sandbox Agents primitive in the TypeScript and Python Agents SDKs
provides a first-party isolated execution environment with filesystem,
shell, installed packages, mounted data, exposed ports, and snapshots;
Computer Use offers three integration paths (built-in loop, custom
harness, code-execution harness)
\citep{openai_sandbox_agents_2026, openai_computer_use_2026}. Google
Vertex AI provides VPC Service Controls perimeter enforcement plus
per-region sandbox environment quotas (Code Execution: 1,000 RPM
execute, 1,000 entities, 500 RPM write)
\citep{vertex_networking_2026, vertex_code_execution_2026}.

OpenAI's Sandbox Agents is the cleanest first-party sandbox primitive
across the four vendors. The other three vendors document sandboxing as
guidance for customer implementation rather than as a first-party
platform service.

\textbf{Academic source.} Multiple. MCPSHIELD \citep{mcpshield2026}
defines L-CAC (Capability-Based Access Control) and L-CTA (Cryptographic
Tool Attestation). OpenPort \citep{zhu2026openport} defines draft-first
writes, state witness preconditions. SentinelAgent
\citep{sentinelagent2026} defines Scope Enforcement Proxy.

\textbf{Readiness rating.} \emph{Currently deployable} against the
user-as-adversary and cooperative-deployer surface; degrades against the
fine-tuner-as-adversary (partial at medium capability, inadequate at
high capability; Section 4.3). These degradation ratings assume model-embedded or model-mediated enforcement; provider-external implementations (external moderation, provider-side policy engines, orchestration, or API gateways) may retain higher resilience. The sandboxing substrate is universally
implementable and documented across the four vendors, but as first-party
productisation it is demonstrated at one vendor: OpenAI's first-party
Sandbox Agents is the strongest single-vendor evidence, while the other
three vendors document sandboxing as guidance for customer
implementation rather than as a first-party primitive.

\hypertarget{e7.-inference-time-off-switch-and-kill-chain}{%
\subsubsection{E7. Inference-time off-switch and
kill-chain}\label{e7.-inference-time-off-switch-and-kill-chain}}

\textbf{Operational definition.} The platform provides documented
procedures for: (1) rate-limiting access in response to detected misuse
or capability concerns, (2) suspending access to a specific customer or
workspace under documented conditions, (3) deprecating a specific model
version on a documented timeline with notice to deployed customers, and
(4) terminating the platform itself under specified catastrophic
conditions. The mechanism is the inference-side
enforcement-of-last-resort and supports the
regulatory-enforcement-action primitive in Section 6's scenarios.

\textbf{Vendor productisation.} Four vendors productise E7 partially.
AWS Bedrock documents standard AWS service-deprecation procedures with
customer notification and migration support. Anthropic's Usage Policy
documents explicit escalation procedures from throttle to suspend to
terminate, applied by the Safeguards Team under documented criteria
\citep{anthropic_usage_policy_2026}. OpenAI's Deprecations page
documents advance notice (email plus page entry) before model retirement
with documented transition windows \citep{openai_deprecations_2026}.
Google Vertex AI documents standard GCP service-deprecation procedures;
the Advanced AI Safety Addendum provides specific consent-and-revocation
procedures for designated models including Claude Mythos, Fable, and
Opus 4.7+ \citep{vertex_abuse_monitoring_2026}. The capability-update
lifecycle (validate \(\rightarrow\) sandbox \(\rightarrow\) shadow
\(\rightarrow\) gate \(\rightarrow\) rollback) surfaced during the
cap-to-21 process from \citet{qin2026governed} is
incorporated here as the rollback portion of the kill-chain.

\textbf{Academic source.} Wang et al.~(MI9 Graduated Containment,
\citep{wang2025mi9}) provides the foundational framework. Qin et
al.~(Governed Capability Evolution, \citep{qin2026governed}) provides
the rollback-at-deployment-time argument.

\textbf{Readiness rating.} \emph{Currently deployable} against the
user-as-adversary and cooperative-deployer surface; degrades against the
fine-tuner-as-adversary (partial at medium capability, inadequate at
high capability; Section 4.3). These degradation ratings assume model-embedded or model-mediated enforcement; provider-external implementations (external moderation, provider-side policy engines, orchestration, or API gateways) may retain higher resilience. Multi-vendor productisation of
throttle/suspend/terminate with documented procedures; the rollback
portion (Qin et al.) is implementable as a composition of existing
deployment-pipeline primitives.

\hypertarget{feasibility-summary}{%
\subsection{Feasibility summary}\label{feasibility-summary}}

Of the twenty mechanisms in the taxonomy, the readiness ratings
distribute as follows.

\textbf{Currently deployable substrate (fifteen mechanisms): M1, M3, M4, M5, V4,
V5, V6, V7, E1, E2, E3, E4, E5, E6, E7.} The majority of the taxonomy
has been productised at production scale by multiple major commercial
vendors. A regulator mandating the corresponding productised substrates
could expect near-term technical compliance, although governance-grade
assurance would require additional specification. Two qualifications attach to this category. First,
V5 carries a split rating: currently deployable for descriptive
evaluation reporting (which all four vendors provide) and requires R\&D
for cryptographically attestable evaluation results (which no vendor
provides); we count it in the deployable category on the strength of the
productised primitive and flag the attestation gap in Section 4.4 and
Section 5. Second, the readiness ratings in this category are stated
against the adversary surface each mechanism was built for, not as
unconditional claims. This matters most for the enforcement cluster (see
below).

\textbf{Near-term (two mechanisms): M2, V1.} Workload classification
(M2) has substrate (MoE inference is happening today) but exposure to
callers is undeveloped. TEE-bounded inference attestation (V1) has
substrate (Confidential VMs, Confidential GPUs, Nitro Enclaves at
infrastructure level) but is not exposed at the inference API surface.

\textbf{Requires R\&D (two mechanisms): V2, V3.} Zero-knowledge proofs
of inference have well-established cryptographic substrate but
production deployment at frontier scale is blocked by proof-generation
overhead. Hardware-backed workload certificates per output have academic
substrate but no production-stage implementation.

\textbf{Speculative (one mechanism): M6.} Per-inference energy and cost
monitoring as a governance primitive has neither vendor productisation
nor strong academic substrate at the inference stage; we retain it on
Ansari-mirroring grounds and on anticipated near-term regulatory
pressure.

These four categories sum to the full twenty (fifteen + two + two +
one).

\textbf{A note on adversary-scoped readiness for the enforcement
cluster.} The ``currently deployable'' rating for the enforcement
mechanisms E2, E5, E6, and E7 is scoped, not unconditional. Each is
currently deployable against the user-as-adversary and
cooperative-deployer surface it was designed for, and each is defeated
by the fine-tuner-as-adversary (E2 from medium capability upward; E5,
E6, E7 at high capability) who retrains the deployed model to evade the
runtime defence, per \citet{qi2023fine}. We state these
ratings in scoped form because the two-dimensional adversary model in
Section 4.3 makes a single scalar readiness rating insufficient: a
mechanism can be in production and still inadequate against an adversary
it was not built for. The scoping is not a downgrade of the readiness
rating; it is the rating format the 2D model implies, and Section 4.3
reports the per-adversary detail. The same logic underlies the V5 split
rating (deployable as descriptive reporting, R\&D as cryptographic
attestation): readiness is a claim about a mechanism against a named
purpose or adversary, not a property of the mechanism in the abstract.

The distribution is the headline finding of the paper: \textbf{fifteen
mechanisms have a commercial substrate deployable at production scale,
though the adversary analysis (Section 4.3) shows their governance-grade
coverage is uneven} (one of them,
V5, only in its descriptive form), and the gap between
deployment-readiness and the academic literature's full taxonomy is in
cryptographic verification (V1, V2, V3) and in late-stage productisation
decisions (M2, M6). The substrate readiness for inference-time
governance is substantially more mature than the existing regulatory
architecture would suggest. The qualification that Section 4.3 develops
is that deployment-readiness and adversary-coverage are distinct: the
fifteen deployable mechanisms are deployable against the surface they
were built for, and the coverage of the high-capability deployer and
fine-tuner surfaces is much thinner.

The next section develops the feasibility constraints under which this
readiness distribution holds, beginning with the substitutability
between training and inference compute that erodes any rating resting on
a compute signature.

\hypertarget{feasibility-constraints-and-adversarial-considerations}{%
\section{Feasibility Constraints and Adversarial
Considerations}\label{feasibility-constraints-and-adversarial-considerations}}

The readiness ratings in Section 3 hold under a set of feasibility
constraints that this section makes explicit. We treat five in turn: the
substitutability between training-time and inference-time compute that
erodes compute-signature-based detection (4.1); distributed and
decentralised inference, which undercuts the assumption of a detectable
concentrated workload (4.2); the two-dimensional adversary model that
replaces the cooperative-deployer assumption implicit in much of the
vendor productisation (4.3); privacy, sovereignty, and political
feasibility at the per-query level (4.4); and the
marketplace-and-intermediation failure modes that emerge where a
concentrated supply side meets a fragmented demand side (4.5).

\hypertarget{substitutability-between-training-time-and-inference-time-compute}{%
\subsection{Substitutability between training-time and inference-time
compute}\label{substitutability-between-training-time-and-inference-time-compute}}

Ansari \citep{ansari2026hardware} Section 4.1 establishes a structural feasibility
constraint on hardware-level governance: algorithmic efficiency
improvements continuously erode the capability threshold at which a
given training-compute budget produces a frontier-capable model. The
training compute required to reach any fixed capability target falls by
roughly a factor of three per year on the most demanding benchmarks, and
the threshold below which the most-capable open-weight model sits is now
several orders of magnitude lower than the EU AI Act's \(10^{25}\) FLOP
trigger \citep{hooker2024limitations, jones2025overcoming}. The
implication is that compute thresholds, treated as static numerical
triggers, lose their regulatory power as algorithmic efficiency
advances.

The argument transfers to inference-time governance with one important
sharpening: \emph{inference is itself the channel through which the
substitution operates}. Where the algorithmic-efficiency argument is
about training-side improvements that reduce the compute cost of
reaching a capability target, the inference-substitution argument
identifies a category of techniques that achieve frontier capabilities
without spending additional \emph{training} compute at all.
Above-optimal inference-time scaling, in the framing of \citet{pistillo2025defending}, is the fourth and structurally
distinct loophole through which training-compute thresholds are
bypassed. Models trained below the \(10^{25}\) FLOP threshold can be
made to demonstrate frontier-level capability at inference time by
spending compute on repeated sampling, chain-of-thought reasoning,
ensemble inference, or agentic scaffolding.

Three quantitative observations make the substitution argument concrete.
First, the per-query inference compute used by frontier reasoning models
on the most demanding benchmarks now sometimes approaches the per-query
compute used during training \citep{almaghrabi2026inference}. Second,
the cost-per-capability curve for inference is improving even faster
than for training, and the open-weight portion of the deployment surface
is reaching capability levels that previously required cloud-scale
infrastructure \citep{puri2026small, alexander2026efficiency}. Third,
\citet{casper2025practical} establish the principle that
consistent compute accounting at the regulatory level requires explicit
accounting \emph{across} the training-inference boundary, not in
isolation on either side. The substitution channel is not a
hypothetical: it is the dominant capability-production mode at the
frontier today.

The feasibility implication for governance is direct. Any mechanism in
the inference-time taxonomy that depends on identifying high-capability
inference \emph{by its compute signature alone} will face the same
erosion that training-side thresholds face: as algorithmic efficiency
improves, the compute signature of high-capability inference falls, and
a mechanism calibrated to today's signature will be circumventable
tomorrow. Mechanisms that rate as currently deployable on the strength
of their compute-signature detectability (a class that includes parts of
M1 and M4, and the near-term M2, in Section 3.1) should therefore be
read as deployable \emph{at present capability levels}, not as durably
deployable. The convergence analysis in Section 5 takes this point
further by examining the conditions under which the training-inference
boundary collapses entirely.

\hypertarget{distributed-and-decentralised-inference}{%
\subsection{Distributed and decentralised
inference}\label{distributed-and-decentralised-inference}}

Ansari \citep{ansari2026hardware} Section 4.2 documents the parallel challenge for training:
distributed and decentralised training architectures that spread the
computation across multiple data centres, jurisdictions, or networked
devices undercut the assumption that frontier training is detectable as
a concentrated, high-power signature in a single location.
Low-communication algorithms in particular \citep{krys2025distributed}
enable training runs in which the per-node compute and bandwidth profile
is indistinguishable from non-AI workloads, defeating mechanisms that
rely on detecting concentrated training compute.

The structural problem transfers directly to inference but with a
different signal profile. Inference at frontier scale, particularly for
the largest mixture-of-experts and reasoning-heavy models, has
historically required tightly coupled GPU clusters with high-bandwidth
interconnects, which produces a detectable signature. But three trends
are changing this. First, mixture-of-experts architectures with sparse
activation reduce the per-query bandwidth profile of frontier-scale
inference; the cluster signature converges toward that of routine
inference \citep{krys2025distributed}. Second, decentralised inference
networks (Petals, Bittensor, Together AI, Gensyn-class platforms)
operationalise distributed inference across heterogeneous, untrusted
nodes, splitting a single inference request across geographies. Third,
the cumulative compute spent across many small, geographically
distributed inference deployments may rival a single concentrated
frontier deployment without any single deployment crossing a
detectability threshold \citep{mueller2025s}.

The implication for the mechanism taxonomy is that the
distributed-and-decentralised case must be treated as a first-class
scenario, not as an exception to be handled in a footnote. Several
mechanisms in Section 3 that rate well against cluster-served frontier
inference (V1 TEE-bounded attestation, V3 hardware-backed workload
certificates, M4 distributed-inference monitoring) face structural
challenges when the workload is itself distributed across untrusted or
untrusted-but-attested nodes. The challenge is sharpest when the
distributed inference is \emph{across jurisdictions}, because mechanisms
whose enforcement substrate is jurisdiction-specific (E3
jurisdiction-bounded inference, E1 per-deployment licensing) lose
effectiveness when each node holds only a fragment of the inference. We
return to this in Section 4.5 (marketplace and intermediation) and again
in Section 6.4 (industry self-regulation), where decentralised inference
is the scenario in which industry-led governance is most stressed.

Open-weight deployment compounds the distributed-inference challenge in
a specific way. Where decentralised inference networks at least have a
network identity that can be attested or monitored, fully unfederated
open-weight inference (a model running on a consumer machine with no
network supervisor) has no shared monitoring substrate at all
\citep{gomes2026open}. The fallback for these cases is not technical
attestation but hardware-layer governance of the accelerators the
inference runs on. This is the boundary at which the inference-time
governance framework hands back to the hardware-level framework
developed in Ansari \citep{ansari2026hardware}, and we return to this handoff in Section 5.

\hypertarget{two-dimensional-adversary-analysis}{%
\subsection{Two-Dimensional Adversary
Analysis}\label{two-dimensional-adversary-analysis}}

The feasibility ratings in Section 3 assess each mechanism against the
question ``is this mechanism deployable today, with the documented
productisation?'' They do not assess against the question ``deployable
against whom?'' Most of the inference-governance literature collapses
the latter question into the former by implicitly assuming a cooperative
deployer: the operator of the inference platform follows documented
procedures, returns honest telemetry, applies declared filters, and
accepts regulatory oversight. Under that assumption the mechanism
ratings in Section 3.4 hold and the headline finding (fifteen mechanisms
with commercially available production substrates) describes the regulatory
toolbox accurately.

The cooperative-deployer assumption is, however, not the only adversary
model relevant to inference-time governance. We follow \citet{ansari2026hardware} in developing a two-dimensional adversary
analysis: capability tier \(\times\) adversary role. The first dimension
distinguishes adversaries by resource scale (low, medium, high); the
second distinguishes by position in the deployment chain (developer,
deployer, user, fine-tuner). The cross product produces twelve cells.
For each of the twenty mechanisms, we ask whether the mechanism is
adequate, partial, inadequate, or not applicable against each cell. The
resulting twenty-by-twelve matrix is reported in full in Appendix E and
synthesised in this section (Figure 1).

\hypertarget{methodology}{%
\subsubsection{Methodology}\label{methodology}}

The matrix is populated against the same four-point conservative-rating
rule that governs Section 3's readiness ratings: where uncertain between
two ratings, we choose the more conservative. Three additional rating
conventions are specific to this section.

First, we distinguish \emph{inadequate} (the mechanism is ineffective
against this adversary) from \emph{not applicable} (the adversary role
does not exercise the threat surface the mechanism targets). A
fine-tuner-as-adversary is not applicable to retrieval-corpus
attestation (V6) because the fine-tuner operates pre-inference on
weights, not at inference-time on retrieval. Conflating not-applicable
with inadequate would systematically overstate the gaps in mechanism
coverage.

Second, we rate against the productised state of each mechanism (per
Section 3's vendor evidence base) rather than against a hypothetical
fully-deployed state. For mechanisms with partial productisation (V1,
V3, M2, M6) we rate against the near-term productisation hypothesis
articulated in Section 3.4, with the rating explicitly marked as
contingent on productisation. This is the compromise stated in the
matrix specification (Appendix E); a stricter
rating-against-current-productisation would understate the eventual
coverage of mechanisms whose substrate exists but whose user-facing API
does not.

Third, where vendor implementations diverge architecturally (E2, V7) we
rate against the most defensive available architecture. The
architectural-diversity finding itself is treated in Section 4.5; here
we accept that an attacker can choose the weakest vendor while a
regulator can mandate the strongest, and we rate against the
regulator-favoured case. This is the convention that supports the
asymmetry: the matrix tells us what the regulatory toolbox \emph{could}
defend if used; Section 4.5 tells us what it would defend if applied
uniformly. One deliberate exception applies. Against the
fine-tuner-as-adversary (R4), the enforcement cells (E2, E5, E6, E7) are
scored on the conservative whole-mechanism reading set out below rather
than the most-defensive-architecture rule, because a weight-modifying
adversary defeats the model-embedded component of these mechanisms
regardless of which vendor's architecture is chosen. The
platform-external components that survive are noted but not credited in
those cells; a reader who credits them would rate some R4 enforcement
cells higher. Where these two conventions would otherwise disagree, the
conservative R4 reading takes precedence for the enforcement cluster.

\begin{figure}
\centering
\includegraphics[width=\linewidth]{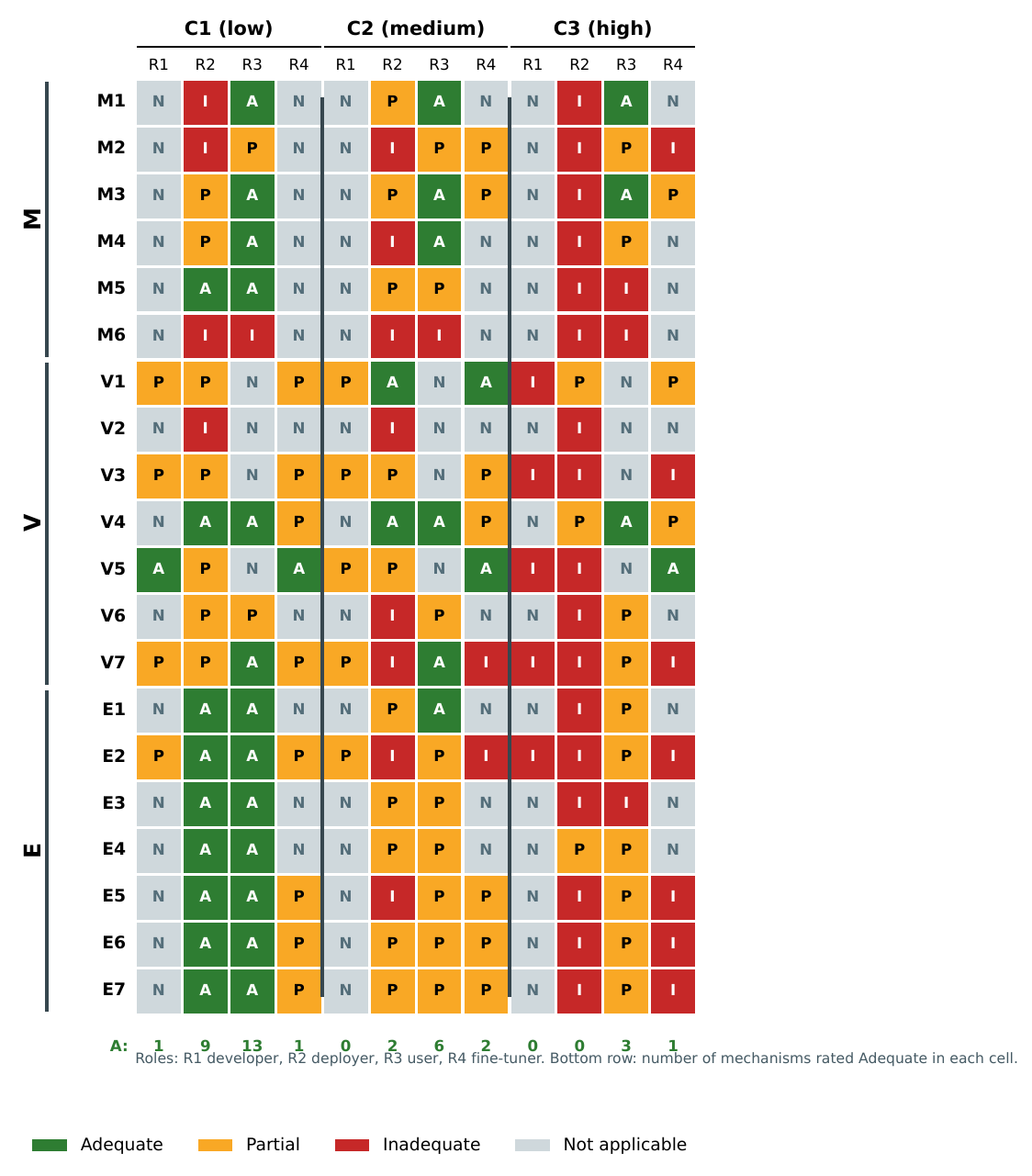}
\caption{Adversary-coverage heatmap. The twenty mechanisms (rows, grouped by category) against the twelve adversary cells (columns: capability tier C1 to C3 by role R1 to R4). Green adequate, amber partial, red inadequate, grey not applicable. The bottom row gives the number of mechanisms rated adequate in each cell: coverage peaks at the low-capability user (C1.R3, thirteen of twenty) and falls to zero at the high-capability deployer (C3.R2). The full populated matrix is in Appendix E.}
\end{figure}

\hypertarget{headline-findings}{%
\subsubsection{Headline findings}\label{headline-findings}}

Three findings emerge from the populated matrix, each load-bearing for
subsequent sections.

\hypertarget{finding-1-the-user-as-adversary-r3-column-is-the-strongest-covered}{%
\subsubsection{Finding 1: The user-as-adversary (R3) column is the
strongest-covered}\label{finding-1-the-user-as-adversary-r3-column-is-the-strongest-covered}}

The user-as-adversary column has thirteen of twenty mechanisms rated
adequate at low capability (C1.R3), and six of twenty at medium
capability (C2.R3). The drop at C3.R3 (three of twenty) reflects the
well-documented finding that jailbreak research and prompt-injection
sophistication scale with adversary resources. The adequate ratings at
C1.R3 reflect the fact that several mechanisms in the taxonomy were
\emph{designed} against this adversary: E2 (output filtering and
capability gating), E4 (rate limiting as a governance primitive), parts
of E6 (sandboxing), and parts of V7 (chain-of-thought monitorability as
a jailbreak detector). The user-as-adversary case is what the deployed
mechanisms were optimised for.

The implication for governance is mixed. The strong coverage of R3 at
low capability tiers means a regulator mandating user-facing safeguards
(E2, E4, E6) can expect compliance to produce intended effects against
the population of casual or moderately-sophisticated misuse attempts.
The degradation at C3.R3, however, means the same regulatory regime
offers weaker assurance against state-level user adversaries
(intelligence services, sophisticated criminal networks). The taxonomy
as currently productised does not solve the high-capability-user
problem; it bounds the moderate-capability-user problem.

\hypertarget{finding-2-the-state-level-deployer-c3.r2-cell-is-the-weakest-covered}{%
\subsubsection{Finding 2: The state-level-deployer (C3.R2) cell is the
weakest-covered}\label{finding-2-the-state-level-deployer-c3.r2-cell-is-the-weakest-covered}}

Seventeen of twenty mechanisms are rated inadequate against C3.R2. The
remaining three are not inadequate but only partial: V1 (TEE-bounded
attestation), V4 (tool-call cryptographic signing), and E4 (rate
limiting). No mechanism in the taxonomy rates adequate against the
state-level deployer.

The structural pattern is that almost every mechanism in the taxonomy
depends on cooperative platform behaviour, and a state-level deployer is
precisely the actor for which cooperative behaviour cannot be assumed.
M1 (per-query accounting), M3 (KV-cache/reasoning accounting), M5 (KYC
and tracing), M6 (power monitoring), E1 (per-deployment licensing), E2
(output filtering), E3 (geofencing), E5 (policy-as-code), E6
(sandboxing), and E7 (kill-chain) all rate inadequate at C3.R2 because
the state-level deployer self-hosts inference infrastructure outside the
platform whose telemetry, filters, and quotas the mechanism instruments.
The three mechanisms that retain partial coverage do so for different
reasons. V1 (TEE attestation) and V4 (cryptographic signing) rely on
cryptographic primitives the deployer cannot trivially forge, but both
fall to partial rather than adequate because the state-level R2
adversary can in principle compromise vendor key or attestation
infrastructure (and, for V1, because the primitive is not yet
productised at the inference API). E4 (rate limiting) retains partial
coverage because a quota is enforced at the platform boundary the
deployer must still cross to reach commercial scale, even when
self-hosting displaces the other platform-dependent controls. The
remaining verification mechanisms do not help here: V2 and V3 are rated
inadequate at C3.R2 because their cryptographic substrate is not
productised, and V5 because evaluation reporting is descriptive rather
than tamper-evident at every vendor.

V4 is the most robust of the three at the state-level-deployer cell. Of
the twelve cells in V4's profile, five are adequate, four are partial,
and zero are inadequate; the remaining three are not-applicable. V4's
strength against R2 specifically reflects the design property that
cryptographic signatures are unforgeable by definition once the signing
infrastructure is operational; a state-level R2 adversary cannot defeat
the signature primitive directly, only by compromising the signing key
infrastructure (which is itself a high-capability operation but is
distinct from the everyday deployer-as-adversary threat).

The implication is sharp: the existing taxonomy is structurally unable
to defend adequately against a state-level deployer adversary. No
mechanism rates adequate at C3.R2; only three (V1, V4, E4) retain even
partial coverage, and the strongest of these is cryptographic signing of
intermediate reasoning and tool-call state (V4). Section 5's substitution
principle returns to this finding; it is the empirical foundation for the
corollary that layered-substitution mechanisms become structurally
necessary when the substitutability channel is open.

\hypertarget{finding-3-the-fine-tuner-as-adversary-r4-defeats-most-enforcement-mechanisms-at-c2-and-c3}{%
\subsubsection{Finding 3: The fine-tuner-as-adversary (R4) defeats the
model-internal enforcement mechanisms at C2 and C3, while
platform-external controls can persist}\label{finding-3-the-fine-tuner-as-adversary-r4-defeats-most-enforcement-mechanisms-at-c2-and-c3}}

\citet{qi2023fine} demonstrated that fine-tuning can
disable safety alignment in deployed models. The populated matrix shows
this attack across the R4 column: the enforcement mechanisms degrade as
fine-tuner capability rises. E2 (output filtering) is already inadequate
at C2.R4 and remains inadequate at C3.R4; E5 (policy-as-code), E6
(sandboxing), and E7 (kill-chain) hold partial coverage at C2.R4 but
fall to inadequate at C3.R4, against the high-capability fine-tuner who
modifies the deployed base model to evade them. E1 (licensing) and E3
(geofencing) are not-applicable to R4 because the fine-tuner operates
pre-deployment.

Enforcement is not homogeneous under this attack, and three cases should
be separated. \emph{Model-internal enforcement}, meaning behaviour
aligned into the weights and the output tendencies that depend on them,
is directly vulnerable to weight modification: a fine-tuner who controls
the weights can remove it. \emph{Runtime-orchestration enforcement},
meaning policy-as-code gates, sandboxing, and tool-use authorisation
enforced by the serving harness, is potentially independent of the
weights and survives weight modification unless the fine-tuner also
controls the orchestration layer. \emph{Provider- or infrastructure-level
enforcement}, meaning a separate moderation classifier, per-account rate
limits, external sandboxing, provider access control, and platform
suspension, sits outside the fine-tuner's reach entirely unless the
adversary also controls the deployment platform. The R4 enforcement
failures the matrix records are failures of the first case, and of the
second where the serving harness re-runs the modified weights; the
platform-external controls of the third case can persist. The matrix
rates these R4 enforcement cells on a conservative, whole-mechanism
reading, scoring each mechanism where it is embedded in or bypassable
through the model rather than crediting the platform-external controls
that may survive, so a reader who credits those surviving controls would
rate some of these cells higher.

The pattern is that the affected components are the model-internal and
model-mediated parts of these enforcement mechanisms. A fine-tuner produces a
model that, when run on the same platform, no longer triggers the
model-mediated runtime defences (it has been retrained not to); the
platform-external components identified above are not evaded in this
way. E2 makes the point
sharply. Section 3 rates E2 ``currently deployable,'' and the matrix
shows it inadequate against the fine-tuner at C2.R4 and C3.R4. These are
not in tension once the readiness rating is read as adversary-scoped,
which is the discipline the two-dimensional model exists to impose. A
single scalar readiness rating is exactly what the 2D model replaces: E2
is currently deployable against the user-as-adversary and
cooperative-deployer surface it was built for, and inadequate against
the fine-tuner-as-adversary at medium-to-high capability. We therefore
state the E2 readiness rating in scoped form in Section 3.4 rather than
leaving it unqualified and contradicted here. The same scoping applies
across the enforcement cluster: E5, E6, and E7 degrade against R4 as
capability rises (partial at C2, inadequate at C3) to the extent their
enforcement is model-mediated rather than platform-external. Their
external components (external policy engines, sandboxes, and provider
suspension) can persist, which is why the whole-mechanism reading above
scores these cells conservatively rather than crediting the surviving
controls; they are scoped accordingly in Section 3.4. The
conservative-rating rule is not violated by this move; the rule guards
against overstating readiness against a \emph{named} adversary, and a
scoped rating names the adversary the rating holds against.

The verification mechanisms defeat R4 in a way the enforcement
mechanisms do not. V5 (evaluation reporting) is rated adequate against the
fine-tuner at all three capability tiers on the reasoning that it detects
capability divergence from baseline. This rating reflects the mechanism's
best case and is flagged for reconsideration (Section 7.4): because the
current productised substrate is descriptive reporting that is neither
independent nor tamper-evident, a high-capability adversary could decline
to run the evaluation, substitute the evaluated model, alter the reporting
pipeline, selectively disclose results, or misrepresent model identity, so
adequate detection is not guaranteed at the higher tiers. A future split of
V5 into descriptive reporting and independent or cryptographic evaluation
attestation, rated separately, would resolve the tension; we retain the
single mechanism here and flag the R4 ratings rather than silently
re-coding them. V1 (TEE attestation) is adequate at medium
capability and partial elsewhere; V3 and V4 provide partial coverage.
The common property is that these mechanisms bind or test \emph{model
identity and capability} rather than constraining runtime behaviour: a
fine-tuner cannot deploy a modified model in the same attested
environment without the attestation or the evaluation reflecting the
change. This is the structural reason verification matters: the
fine-tuner is one of the few adversaries for which monitoring alone
(telling us what the model is doing) and enforcement alone (filtering
what the model produces) both fail, while verification (attesting what
model is actually being run) succeeds.

\hypertarget{per-mechanism-profiles-worth-detailed-discussion}{%
\subsubsection{Per-mechanism profiles worth detailed
discussion}\label{per-mechanism-profiles-worth-detailed-discussion}}

Three mechanisms have profiles that are particularly informative for the
paper's overall argument.

\textbf{V4 (tool-call cryptographic signing).} Five adequate, four
partial, zero inadequate, three not-applicable. The cleanest mechanism
profile in the matrix. V4 defends against deployers at C1-C2 (the
signature is unforgeable at these tiers) and against users at C1-C3 (the
user cannot forge a signature regardless of capability). The partial
ratings at C3.R2 (state-level deployer who might compromise vendor
signing-key infrastructure) and the not-applicable ratings for R1
(developer-as-adversary, who operates pre-inference) are honest
acknowledgements. V4 is the most structurally robust mechanism in the
taxonomy across the adversary surface.

The profile reinforces Section 3.2's emphasis on V4 as currently
deployable at two of four vendors (Anthropic, Google Vertex) and
supports the convergence argument in Section 5: V4 is the kind of
mechanism whose effectiveness derives from cryptography rather than from
platform cooperation, and is therefore robust to the kinds of evasions
that defeat cooperative-deployer-dependent mechanisms.

\textbf{V7 (chain-of-thought monitorability).} Two adequate, five
partial, five inadequate, zero not-applicable. V7's two adequate cells
are both in the user-as-adversary column at low and medium capability
(C1.R3 and C2.R3), where a jailbreak attempt often reveals itself in the
reasoning trace. It degrades to partial or inadequate everywhere else,
and is inadequate against the medium-capability deployer and fine-tuner
(C2.R2, C2.R4) and across most of the high-capability column, because a
deployer can disable thinking exposure and a fine-tuner can train the
model away from legible reasoning. V7 has no not-applicable cells: it
engages every adversary role, but defends few of them well. The
faithfulness limitation (\citet{turpin2023language})
bounds even the adequate cells: a model can produce reasoning traces
that read as legitimate but do not faithfully reflect the computation
that produced the output, which is why V7's adequate coverage does not
extend past the user-as-adversary case.

The profile means V7's ``currently deployable'' rating in Section 3 is a
statement about the mechanism's productisation (three vendors expose
reasoning traces), not about its adversarial coverage. As a defence it
reaches adequate only against the low- and medium-capability user
(C1.R3, C2.R3) and is partial or inadequate against every other
adversary, including every deployer and fine-tuner cell above low
capability. Against the deployer and fine-tuner cases, V7 requires
substantial further work (white-box latent monitoring, formal
faithfulness guarantees) that is research-stage. The honest summary is
that V7 is deployable as a primitive and weak as a defence outside the
user-as-adversary case: a regulator can mandate reasoning-trace
exposure, but should not treat the resulting traces as a reliable
control against a capable deployer or fine-tuner.

\textbf{E2 (output filtering and capability gating).} Two adequate, five
partial, five inadequate, zero not-applicable. E2's two adequate cells
are where it was designed to work: against the low-capability deployer
(C1.R2) and the low-capability user (C1.R3). It degrades to partial or
inadequate as capability rises, and is inadequate against the
medium-capability deployer, the medium-capability fine-tuner (C2.R2,
C2.R4), and most of the high-capability column, because the self-hosting
deployer disables filters and the fine-tuner trains the model not to
trigger them (the Qi et al.~attack). Like V7, E2 has no not-applicable
cells: it engages every adversary role and defends few well past low
capability. E2 is also the most architecturally heterogeneous mechanism
in the taxonomy: a regulator mandating it must specify which vendor's
architecture is meant, since AWS Bedrock Guardrails, Anthropic's
server-side enforcement, OpenAI's Moderation API, and Vertex's
HARM\_CATEGORY enums all provide different operational guarantees
(Section 4.5).

\hypertarget{cross-cutting-findings}{%
\subsubsection{Cross-cutting findings}\label{cross-cutting-findings}}

Beyond the per-cell and per-mechanism findings, three cross-cutting
patterns emerge.

\textbf{Pattern 1: Adequate coverage of the deployer-as-adversary is
thin and concentrates at low capability.} The R2 column has only eleven
adequate ratings across all sixty cells (three capability tiers
\(\times\) twenty mechanisms); the rest are twenty partial and
twenty-nine inadequate. Of those eleven adequate ratings, seven are
enforcement mechanisms (E1-E7) and all seven sit at the lowest
capability tier (C1.R2), one is a monitoring mechanism (M5 at C1.R2),
and only three are verification mechanisms (V1 at C2.R2, V4 at C1.R2 and
C2.R2). The pattern is the opposite of what a cooperative-deployer
reading would predict: against the low-capability deployer the runtime
enforcement cluster works, but adequacy collapses as deployer capability
rises, and no mechanism of any category rates adequate against the
deployer at C3. Verification mechanisms do not dominate R2 adequacy;
they provide the only adequate coverage that survives past the lowest
capability tier (V1 and V4 at C2.R2), which is a narrower and more
accurate claim. This still supports the convergence argument in Section
5: where deployer-adversary coverage survives rising capability at all,
it survives because of cryptography (V1, V4), not because of runtime
enforcement, which is defeated as soon as the deployer can self-host.

\textbf{Pattern 2: At C3, mechanisms are inadequate against R2 and R4
and only weakly constraining against R3.} A state-level adversary cannot
be defeated as deployer or fine-tuner by individual mechanisms in the
existing taxonomy, and even as a user it is only partially constrained
at best: at C3.R3 just three of twenty mechanisms rate adequate and
several rate inadequate. This is the asymmetric coverage pattern that
makes inference-time governance a complement to rather than substitute
for hardware-level governance: the C3.R2 and C3.R4 gaps, and most of the
C3.R3 surface, must be filled by other regulatory layers (hardware
substrate per Ansari \citep{ansari2026hardware}; international agreement; export controls).
The honest characterisation of the matrix is that the inference-time
taxonomy provides substantial coverage of the low-capability surface
across most roles and of the moderate-capability user, and limited
coverage of the high-capability deployer and fine-tuner surfaces. We
avoid the stronger word ``complete'' for the low-capability surface,
because even at C1-C2 many cells are partial rather than adequate;
``substantial'' is what the matrix supports and ``complete'' is not.

\textbf{Pattern 3: The not-applicable ratings concentrate on R1
(developer-as-adversary) and R4 (fine-tuner-as-adversary).} Of the
eighty-six not-applicable cells, seventy-four are in the R1 and R4
columns (forty-five in R1, twenty-nine in R4). This reflects the
structural property that inference-time mechanisms operate at deployment
and inference time, not at training and modification time. The R1 and R4
adversary cases are largely out of scope for inference-time governance
by construction; they are within scope for the hardware-level
training-stage governance of Ansari \citep{ansari2026hardware} and for the export-controls
regime that constrains who can modify weights.

\hypertarget{implications-for-the-readiness-summary}{%
\subsubsection{Implications for the readiness
summary}\label{implications-for-the-readiness-summary}}

Section 3.4 reports fifteen mechanisms as currently deployable. The
two-dimensional adversary analysis refines this: the ``currently
deployable'' mechanisms are deployable \emph{against} the
cooperative-deployer plus low-to-medium-capability-user adversary
surface. Against the state-level-deployer surface (C3.R2), no mechanism
rates adequate and only three (V1, V4, E4) retain partial coverage.
Against the fine-tuner-as-adversary surface, the verification cluster is
the exception that proves the rule: V5 (capability-evaluation
attestation) is adequate at all three capability tiers because it
detects capability divergence from baseline, V1 (TEE attestation) is
adequate at medium capability, and V3 and V4 provide partial coverage,
while the enforcement cluster degrades against the fine-tuner as
capability rises: E2 is inadequate at both C2.R4 and C3.R4, and E5, E6,
and E7 are partial at C2.R4 and inadequate at C3.R4. The structural
reason is that the fine-tuner changes the model, and only mechanisms
that attest \emph{which model is actually running} (the verification
cluster) can detect the change; mechanisms that filter or constrain a
model's behaviour at runtime (the enforcement cluster) are retrained
around.

This is not a refutation of the Section 3.4 ratings. It is a refinement:
the ratings hold against the adversary they were assigned to, and
Section 4.3 surfaces the structural gaps in adversary coverage that the
ratings alone do not show. The discussion in Section 7 returns to the
gap between deployment-readiness and full-coverage-readiness as a
research and policy implication.

\hypertarget{forward-connections}{%
\subsubsection{Forward connections}\label{forward-connections}}

The architectural-diversity finding in Section 4.5 amplifies the C3.R2
weakness identified here: even within the cooperative-deployer model,
vendor heterogeneity fragments E2 coverage in ways that make uniform
regulatory application difficult. Section 5's substitution principle builds on
the C3.R2 finding to argue that layered-substitution mechanisms (V1, V3,
M4, E3) become structurally necessary when training-inference
substitution is open, because hardware-level governance is the only
layer below which the C3.R2 adversary cannot evade.

The full populated matrix is in Appendix E.

\hypertarget{privacy-sovereignty-and-political-feasibility-for-inference}{%
\subsection{Privacy, sovereignty, and political feasibility for
inference}\label{privacy-sovereignty-and-political-feasibility-for-inference}}

Ansari \citep{ansari2026hardware} Section 4.4 notes a recurring tension between governance
mechanisms that maximise verifiable transparency and those that preserve
user, customer, or sovereign privacy. Perfect mutual verification of an
AI system's behaviour by a regulator and perfect privacy of the system's
inputs are, in general, in tension; tradeoffs are unavoidable. The
inference-time domain sharpens this tension because the unit of analysis
shifts from ``the developer's training computation'' to ``the user's
individual query.'' Privacy at the query level is a substantively
different and more contested category than privacy at the training-run
level.

The shift has three concrete implications.

First, the legal architecture for inference monitoring is underdeveloped
relative to the technical architecture. \citet{bernabei2024legal} establish that compute governance faces
significant privacy hurdles even when technically feasible, and these
hurdles are sharper at the per-query level. A regulatory regime that
monitors API queries for compute and content properties is closer in
legal architecture to communications surveillance than to financial
reporting. Whether existing privacy regimes (GDPR in the EU, sectoral US
privacy law, jurisdiction-specific provisions elsewhere) permit
per-query monitoring at the granularity that effective enforcement
requires is unresolved and varies across jurisdictions. Several of the
verification mechanisms in Section 3 (V1 TEE-bounded attestation, V3
hardware-backed workload certificates) can in principle preserve query
privacy through enclave-bounded computation, but the regulatory
framework for compelling such attestation in privacy-protective form
does not yet exist.

Second, compute sovereignty arguments transfer to inference with
structural changes. \citet{hawkins2025ai} identify three
distinct levels of compute sovereignty: territorial jurisdiction,
provider nationality, and accelerator vendor nationality. At the
training stage, these three levels mostly coincide for a given training
run. At the inference stage, they routinely separate: an inference query
may be issued from one jurisdiction, served by a provider headquartered
in a second, on accelerators manufactured by a vendor in a third, in a
data centre physically located in a fourth. Each split creates a
sovereignty seam at which inference governance must either align across
jurisdictions or accept jurisdictional fragmentation. \citet{lehdonvirta2024compute} further establish that the global
distribution of cloud compute creates a structural asymmetry between a
Compute North that can govern training and a Compute South whose
effective leverage is largely limited to governing deployment.
Inference-time governance is therefore the \emph{only} layer at which
much of the world has effective regulatory leverage at all; this is not
a marginal consideration.

Third, the political feasibility of inference monitoring is in some
respects easier and in some respects harder than for training. Easier,
because the existing regulatory infrastructure for API-level KYC,
transaction monitoring, and content moderation has analogues in finance,
telecommunications, and pharmaceuticals that inference governance can
borrow from (we develop this argument in Section 6.3). Harder, because
inference is closer to ordinary user activity than training is, and the
political coalitions that support training-side governance do not
automatically support per-query monitoring of user activity. The
mechanisms in Section 3 that score well technically may face
political-feasibility constraints that are structurally separate from
their technical readiness rating. We address this gap explicitly by
noting, for each mechanism in the scenario mapping in Section 6, where
technical feasibility and political feasibility diverge.

The position taken by \citet{sastry2024computing} and \citet{heim2024govern}: compute governance can be an effective
lever, but only with careful guardrails against privacy erosion and
concentration of power. The position is well-established for training;
it requires re-derivation, not re-statement, for inference. The
mechanism taxonomy in Section 3 is necessary but not sufficient to make
the inference-time application of this position concrete. The
convergence analysis in Section 5 and the scenario mapping in Section 6
together provide what we hope is the start of the re-derivation.

\hypertarget{marketplace-and-intermediation-failure-modes}{%
\subsection{Marketplace and intermediation failure
modes}\label{marketplace-and-intermediation-failure-modes}}

The four-vendor concentration on the supply side of frontier inference
(Section 3) meets a fragmented demand side: thousands of downstream
developers, integrators, and end deployers who consume inference through
cloud platforms, agent platforms, and model marketplaces. The
intermediation structure has consequences for governance that the
per-mechanism feasibility analysis in Section 3 does not surface. We
develop three.

\textbf{First, the architectural-diversity finding from Section 3
produces a marketplace-fragmentation finding here.} The same mechanism
is productised at multiple major vendors with architecturally distinct
surfaces. The E2 (output filtering) cluster is the cleanest example: AWS
Bedrock Guardrails is a paid configurable runtime layer; Anthropic
enforces server-side against an Acceptable Use Policy with documented
Safeguards Team escalation; OpenAI publishes the Moderation API as a
free dedicated endpoint that callers invoke separately; Google Vertex AI
documents four configurable HARM\_CATEGORY enums with five HARM\_BLOCK
threshold values. Each architecture has different implications for the
deployer-as-adversary case, for the user-as-adversary case, for
jurisdictional applicability, and for regulatory compliance reporting.
The V7 (chain-of-thought monitorability) cluster shows a similar
pattern: Anthropic chose transparency with three display modes, OpenAI
chose abstraction with summaries-only output, Vertex chose cryptographic
continuity with thought\_signature enforcement. A downstream developer
integrating across two or more vendors faces three governance surfaces
for what the taxonomy calls one mechanism. The mechanism is deployable
in any vendor in isolation; the cross-vendor portfolio is not coherent.
Regulatory rules written against ``the'' E2 mechanism cannot be
uniformly applied; they must be re-derived per architecture.

\textbf{Table 4. Architectural divergence in two productised mechanisms
across the four-vendor evidence base.} The same mechanism is offered
with operationally distinct surfaces, so a rule written against ``the''
mechanism cannot be applied uniformly (Section 4.5).

{\footnotesize\begin{longtable}[]{@{}
  >{\raggedright\arraybackslash}p{(\columnwidth - 4\tabcolsep) * \real{0.3333}}
  >{\raggedright\arraybackslash}p{(\columnwidth - 4\tabcolsep) * \real{0.3333}}
  >{\raggedright\arraybackslash}p{(\columnwidth - 4\tabcolsep) * \real{0.3333}}@{}}
\toprule\noalign{}
\begin{minipage}[b]{\linewidth}\raggedright
Mechanism
\end{minipage} & \begin{minipage}[b]{\linewidth}\raggedright
Vendor
\end{minipage} & \begin{minipage}[b]{\linewidth}\raggedright
Productised surface and operational guarantee
\end{minipage} \\
\midrule\noalign{}
\endhead
\bottomrule\noalign{}
\endlastfoot
E2 Output filtering & AWS Bedrock & Guardrails: a paid, configurable
runtime filtering layer \\
E2 Output filtering & Anthropic & Server-side enforcement against an
Acceptable Use Policy, with documented Safeguards Team escalation \\
E2 Output filtering & OpenAI & Moderation API: a free, dedicated
endpoint the caller invokes separately \\
E2 Output filtering & Google Vertex AI & Four configurable
HARM\_CATEGORY enums with five HARM\_BLOCK threshold values \\
V7 Chain-of-thought monitorability & Anthropic & Transparency: three
reasoning-trace display modes \\
V7 Chain-of-thought monitorability & OpenAI & Abstraction:
summaries-only output \\
V7 Chain-of-thought monitorability & Google Vertex AI & Cryptographic
continuity: thought\_signature enforcement \\
\end{longtable}}

\textbf{Second, the deployment-platform-as-governance-overlay pattern is
structural, not incidental.} Google Vertex AI applies a Vertex-specific
suspected-CSAM classifier on image inputs to Anthropic models hosted on
Vertex, separate from Anthropic's own Trust and Safety filters
\citep{vertex_claude_safety_2026}. The classifier exists because the
deployment platform has its own governance obligations that the
partner-model provider's filtering does not satisfy. A similar pattern
appears in the Advanced AI Safety Addendum at Vertex, which imposes
60-day prompt-and-response logging for designated models (including
Claude Mythos, Fable, and Opus 4.7+) regardless of the partner-model
provider's data-retention practices
\citep{vertex_abuse_monitoring_2026}. The implication is that an
inference passed through a deployment platform is governed by the
\emph{intersection} of the partner-model's governance and the deployment
platform's governance, not by the partner-model's governance alone.
Customers cannot rely on the partner-model's published primitives as a
complete description of what governs their deployment.

The intersection has both protective and adversarial directions. From
the regulator's perspective, the deployment-platform overlay strengthens
enforcement: it adds a second layer of monitoring that the partner-model
provider cannot disable. From the deployer's perspective, the
deployment-platform overlay introduces governance behaviour that the
deployer cannot configure or appeal. From the user's perspective, the
same inference produces different outputs and different refusal patterns
depending on which deployment surface served the request. The governance
landscape is not the cooperative-deployer model that much of the
academic literature implicitly assumes.

\textbf{Third, the open-weight deployment surface escapes the
marketplace architecture entirely.} Where the four-vendor concentration
provides intermediation hooks for governance, fully unfederated
open-weight inference (a model running on a customer machine with no
network supervisor, accessing no commercial inference endpoint) has no
shared monitoring substrate at all
\citep{gomes2026open, alexander2026efficiency}. The mechanisms that
depend on platform intermediation (E1 per-deployment licensing, M5 KYC,
V4 cryptographic signing) lose force when the inference platform is the
deployer's own laptop. The fallback for these cases is not
platform-level governance but hardware-level governance of the
underlying accelerators, which is the boundary at which the
inference-time governance framework hands back to the hardware-level
framework developed in \citet{ansari2026hardware}. The
capability-evolution arguments of \citet{alexander2026efficiency} and \citet{puri2026small}
establish that the open-weight surface is converging toward frontier
capability; the marketplace-intermediation framework therefore loses
coverage of a growing share of the deployed inference base.

The combination of these three failure modes produces a structural
finding: \textbf{the marketplace architecture is necessary but not
sufficient for inference-time governance.} A regulatory regime that
relies entirely on the four-vendor intermediation will leave the
open-weight surface ungoverned and the cross-architecture portfolio
unregulated; a regime that aspires to comprehensive coverage must
combine platform-level mechanisms (Section 3's taxonomy) with
hardware-level mechanisms (Ansari's taxonomy \citep{ansari2026hardware}) and with the
regulatory infrastructure necessary to apply mechanisms uniformly across
architectural diversity (which does not yet exist). We return to the
architectural-diversity finding in Section 5's convergence analysis and
to the open-weight handoff in Section 6.4 (industry self-regulation) and
Section 6.5 (marketplace-and-platform-level governance).

\hypertarget{convergence-between-hardware-level-and-inference-time-governance}{%
\section{Convergence between Hardware-Level and Inference-Time
Governance}\label{convergence-between-hardware-level-and-inference-time-governance}}

This section develops the convergence analysis between Ansari's
hardware-level governance taxonomy and the inference-time taxonomy
developed in this paper \citep{ansari2026hardware}. The two taxonomies cover overlapping but
distinct regulatory domains. The convergence is the structured mapping
of which mechanisms in one domain substitute for which mechanisms in the
other, when, and at what cost. The analysis culminates in the substitution principle,
which states the structural conditions under which the two domains
converge.

We develop the analysis in three steps. Section 5.1 establishes the
structural mapping between the two taxonomies and presents the
substitution table. Section 5.2 identifies three distinct convergence
patterns surfaced by the substitution analysis: parallel substitution,
layered substitution, and architectural divergence. Section 5.3 states
the substitution principle.

\hypertarget{structural-mapping-between-the-two-taxonomies}{%
\subsection{Structural mapping between the two
taxonomies}\label{structural-mapping-between-the-two-taxonomies}}

Ansari \citep{ansari2026hardware} defined twenty hardware-level governance mechanisms
organised across the same three categories (Monitoring, Verification,
Enforcement) at the same readiness scale used in this paper
\citep{ansari2026hardware}. The structural mirroring is methodologically
deliberate: it permits direct cross-domain comparison without
compatibility translation.

The two taxonomies do not have identical scopes. Hardware-level
governance operates on the accelerator (the chip), the data centre (the
rack), and the infrastructure provider (the cloud platform).
Inference-time governance operates on the inference request (the API
call), the deployed system (the model + tools + scaffolding), and the
marketplace (the cloud + agent + skill providers). The two scopes
overlap at the cloud-platform layer, where a single regulatory actor (a
hyperscaler) sits simultaneously in both taxonomies. The overlap is not
coincidental; it reflects the operational reality that a cloud provider
is the natural point of intervention for both training-side and
inference-side mechanisms.

Twelve of the twenty inference-time mechanisms have a direct or
near-direct hardware-level analogue. Eight do not. We tabulate the
mapping in Table 5.

\textbf{Table 5. Substitution mapping between hardware-level (Ansari
\citep{ansari2026hardware} and inference-time governance mechanisms.}

\begin{longtable}[]{@{}
  >{\raggedright\arraybackslash}p{(\columnwidth - 8\tabcolsep) * \real{0.2000}}
  >{\raggedright\arraybackslash}p{(\columnwidth - 8\tabcolsep) * \real{0.2000}}
  >{\raggedright\arraybackslash}p{(\columnwidth - 8\tabcolsep) * \real{0.2000}}
  >{\raggedright\arraybackslash}p{(\columnwidth - 8\tabcolsep) * \real{0.2000}}
  >{\raggedright\arraybackslash}p{(\columnwidth - 8\tabcolsep) * \real{0.2000}}@{}}
\toprule\noalign{}
\begin{minipage}[b]{\linewidth}\raggedright
Inference mech
\end{minipage} & \begin{minipage}[b]{\linewidth}\raggedright
Name
\end{minipage} & \begin{minipage}[b]{\linewidth}\raggedright
Hardware-level analogue
\end{minipage} & \begin{minipage}[b]{\linewidth}\raggedright
Substitution type
\end{minipage} & \begin{minipage}[b]{\linewidth}\raggedright
Notes
\end{minipage} \\
\midrule\noalign{}
\endhead
\bottomrule\noalign{}
\endlastfoot
M1 & Per-query token accounting & A-M1 Cloud metadata and billing
records & Parallel & Operationally identical at different scales \\
M2 & MoE workload classification & A-M2 Workload classification &
Parallel & Hardware-level monitors GPU activation patterns; inference
monitors expert activations \\
M3 & KV-cache and reasoning-token accounting & (no direct analogue) &
None & Reasoning compute is an inference-stage concept \\
M4 & Distributed-inference monitoring & A-M6 Chip location tracking &
Layered & Inference monitoring depends on hardware-level location
attestation \\
M5 & Authenticated account identity and request tracing & A-M3 KYC for compute access &
Parallel & Inference KYC is per-request; hardware KYC is per-account \\
M6 & Per-inference energy monitoring & A-M4 Power consumption monitoring
& Parallel & Same underlying telemetry; different aggregation level \\
V1 & TEE-bounded inference attestation & A-V1 TEEs for workload
attestation & Parallel & Same substrate (Confidential VM, Nitro
Enclaves, NVIDIA CC); deployment-stage application \\
V2 & zkML proofs of inference & A-V2 Proof of Training & Architectural
divergence & Stochastic gradient descent admits proof-of-training;
inference is closer to deterministic \\
V3 & Hardware-backed workload certificates & A-V3 FlexHEG-based
verifiable claims & Layered & Inference certificates inherit from the
hardware FlexHEG certificate substrate \\
V4 & Tool-call signing and agent-trace attestation & (no direct
analogue) & None & Tool use is an inference-stage concept \\
V5 & Capability-evaluation reporting & A-V1 TEEs for workload
attestation & Parallel & Prior V1 attests that a specific evaluation was
run and produced a given result; A-V1 thus underwrites both V1 and V5 \\
V6 & RAG attestation & (no direct analogue) & None & Retrieval is an
inference-stage concept \\
V7 & Chain-of-thought monitorability & (no direct analogue) & None & The
V7 extension; introduced in this paper \\
E1 & Per-deployment licensing & A-E1 Cloud provider access control &
Parallel & Both gate compute access at the provider layer; different
consumption units \\
E2 & Output filtering and capability gating & (no direct analogue) &
None & Output is an inference-stage concept \\
E3 & Jurisdiction-bounded inference & A-M6 Chip location tracking &
Layered & Inference geofencing depends on hardware-level location
attestation \\
E4 & Rate limiting as governance & A-E4 Remote performance degradation
and disablement & Parallel & Throughput throttling at different
scales \\
E5 & Agentic action permissions & (no direct analogue) & None & Agentic
action is an inference-stage concept \\
E6 & Tool-use authorisation and sandboxing & (no direct analogue) & None
& Tool use is an inference-stage concept \\
E7 & Inference-time kill-chain & A-E3 Hardware-embedded off-switches &
Parallel & Suspend/off-switch primitive at the platform layer and
hardware layer \\
\end{longtable}

Twelve mechanisms substitute (nine parallel plus three layered). Eight
do not: seven have no hardware-level analogue, and one (V2) is
architecturally divergent and requires qualitative discussion. The
substitution structure is consequential for the convergence argument:
the parallel-substitution column is where the same regulatory primitive
operates at two different lifecycle stages, and the layered-substitution
column is where inference governance is \emph{built on} hardware
governance.

\hypertarget{three-convergence-patterns}{%
\subsection{Three convergence
patterns}\label{three-convergence-patterns}}

The substitution table surfaces three distinct convergence patterns.
Each has different implications for what a regulator should do when
designing or applying mechanisms.

\textbf{Parallel substitution (nine mechanisms).} Where the same
primitive operates at training-side and inference-side with
operationally identical mechanics, a regulator can apply the mechanism
at either stage with comparable regulatory function under the stated assumptions. Per-query token
accounting (M1) and cloud metadata and billing records (A-M1) are the
cleanest example: the underlying telemetry is identical; the question is
only at what aggregation level the regulator chooses to consume the
evidence. The other eight parallel mechanisms follow the same logic at
their own substrates: MoE workload classification (M2) and
hardware-level workload classification (A-M2) monitor activation
patterns; per-request KYC (M5) and KYC for compute access (A-M3)
identify the customer; per-inference energy monitoring (M6) and
power-consumption monitoring (A-M4) read the same telemetry at different
aggregation levels; per-deployment licensing (E1) and cloud provider
access control (A-E1) gate access; rate limiting (E4) and remote
performance degradation (A-E4) cap throughput; inference-time kill-chain
(E7) and hardware-embedded off-switches (A-E3) are the
enforcement-of-last-resort; TEE-bounded inference (V1) and TEE workload
attestation (A-V1) operate on the same hardware substrate; and
capability-evaluation reporting (V5) and TEE-attested evaluation
results (A-V1) produce comparable evidence at different lifecycle stages,
drawing on the same A-V1 substrate that underwrites V1.

The regulatory implication of parallel substitution is that the choice
of \emph{when} to apply the mechanism is a policy decision rather than a
technical one. A regulator who can intervene at training has equivalent
leverage to a regulator who can intervene at inference, provided the
mechanism is in the parallel-substitution column.

\textbf{Layered substitution (three mechanisms).} Where the
inference-time mechanism depends on the hardware-level mechanism as a
substrate, the hardware-level mechanism is necessary but not sufficient
for the inference-time mechanism to operate. M4 (distributed-inference
monitoring) depends on A-M6 (chip location tracking)'s location
attestation: an inference-time monitor cannot establish where a query
was served if the hardware substrate does not assert it. V3
(hardware-backed workload certificates per output) inherits from A-V3
(FlexHEG-based verifiable claims) at the cryptographic level. E3
(jurisdiction-bounded inference) depends on the same hardware-level
location attestation that A-M6 (chip location tracking) implements. The
regulatory implication of layered substitution is that mechanisms in
this column cannot be deployed without the corresponding hardware-level
mechanism; intervention at inference is contingent on intervention at
hardware.

\textbf{Architectural divergence (one mechanism).} Where the same
regulatory intent applies to both training and inference but the
underlying mechanism is structurally different in the two domains, the
mechanisms cannot substitute and must be deployed independently if
regulatory coverage is desired at both stages. The single inference-side
case is V2 (zkML proofs of inference), whose hardware-side counterpart
is A-V2 (Proof of Training). Proof of Training exploits the stochastic
gradient descent dynamics specific to training: the proof binds a model
commitment to a deterministic-given-randomness training trace, which is
verifiable because the randomness is recorded. Inference does not
produce gradients and does not admit a structurally analogous proof. The
inference-side mechanism (zkML) operates on a structurally different
proof system (zkSNARKs over inference computation), with different
overhead characteristics, different scaling behaviour, and different
regulatory affordances. The two are not parallel; they are both
verification mechanisms operating at different lifecycle stages but with
different cryptographic substrates.

The regulatory implication of architectural divergence is that a
regulator who wishes to verify capability claims at both training and
inference must adopt two mechanisms; a single mechanism does not cover
both stages.

\hypertarget{proposition-1}{%
\subsection{A conditional substitution principle}\label{proposition-1}}

The above structure motivates the following principle.

\begin{quote}
\textbf{Conditional substitution principle.} Under the conditions below,
an inference-time mechanism in the parallel-substitution column and its
hardware-stage analogue provide comparable regulatory coverage: a
regulatory regime that applies the mechanism only at training produces
coverage comparable to a regime that applies it only at inference,
provided that (a) the regulated activity occurs predominantly at the
regulated lifecycle stage, and (b) substitution between training and
inference compute is bounded. When either (a) or (b) fails, the
parallel-substitution column
collapses: training-side application leaks regulated activity into
inference, and inference-side application leaks regulated activity
across the lifecycle boundary from inference back into a fine-tuning
(training-stage) step. Under the failure
conditions, mechanisms in the layered-substitution column gain regulatory
importance (rather than remaining merely \emph{available}), because the
hardware-level substrate becomes a less substitutable intervention
point, one at which regulatory coverage is harder to evade through stage
substitution. Mechanisms in the
architectural-divergence column require independent deployment at both
stages regardless of substitution dynamics.
\end{quote}

\begin{figure}
\centering
\includegraphics[width=\linewidth]{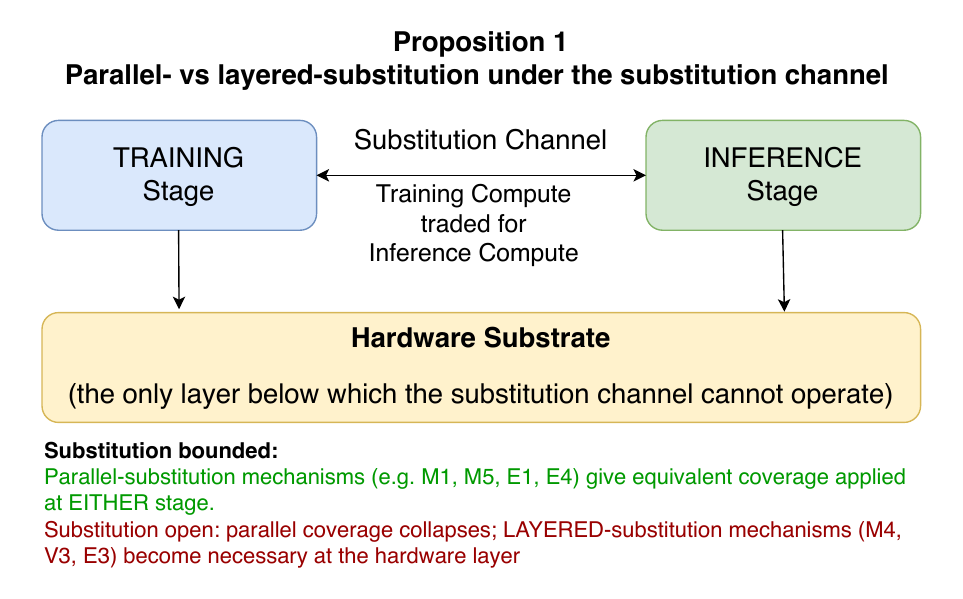}
\caption{The substitution principle in schematic form. While the substitution channel is bounded, parallel-substitution mechanisms give comparable regulatory coverage under the stated assumptions applied at either lifecycle stage; when the channel opens, that comparability collapses and the layered-substitution mechanisms (M4, V3, E3), which depend on hardware-derived evidence, gain regulatory importance rather than remaining merely available (they remain inference-governance mechanisms, not hardware-layer mechanisms).}
\end{figure}

The substitution principle makes precise what Section 4.1's substitutability argument
framed informally. Substitutability between training and inference
compute is the failure condition that turns parallel-substitution
mechanisms from ``either stage'' into ``neither stage adequate.''
Layered-substitution mechanisms become not just useful but structurally
necessary in the substitution regime, because hardware-level governance
is the only layer below which the substitution channel cannot operate
(the substitution operates \emph{between} training and inference, but it
cannot operate \emph{below} the hardware that runs both).

Three corollaries follow.

\textbf{Corollary 1.1.} As training-inference substitution becomes more
efficient (the central empirical finding of Section 4.1), the regulatory
weight on layered-substitution mechanisms increases. M4, V3, and E3
become more important relative to their parallel-substitution
counterparts. This corollary has direct implications for the readiness
ratings in Section 3: V3 (hardware-backed workload certificates) is
currently rated ``requires R\&D,'' but the regulatory pressure on V3 is
structurally rising as substitution efficiency increases. The
mechanism's near-term productisation may be more important than its
current rating suggests.

\textbf{Corollary 1.2.} Architecturally divergent mechanisms (V2 / A-V2)
are the most expensive to deploy under regulatory comprehensiveness
because they require two independent infrastructures. Regulators
choosing coverage at both stages must either fund both or accept that
one is uncovered. The choice has been made implicitly: the existing
regulatory architecture covers A-V2 (Proof of Training) thinly and V2
(zkML) not at all; both are research-stage. The implicit choice is
``neither stage covered for now.''

\textbf{Corollary 1.3.} For inference-only mechanisms (the seven with no
hardware-level analogue: M3, V4, V6, V7, E2, E5, E6), the principle is
silent. These mechanisms are required for inference-time governance
regardless of substitution dynamics. The architectural divergence
between Anthropic and OpenAI on V7 (Section 3.2, Section 4.5) is
internal to the inference-stage governance landscape and is not affected
by the substitution principle.

The substitution principle has a single substantive consequence. It establishes that
inference-time governance is \emph{not} a redundant alternative to
hardware-level governance; the two are coupled by substitution dynamics,
and the substitution dynamics make hardware-level mechanisms \emph{more}
important when inference-time mechanisms are available, not less. The
relationship is not ``either-or'' but ``necessary together when
substitution operates.'' The convergence is not a merger of the two
domains; it is a recognition that the substitution channel forces a
unified regulatory architecture.

\hypertarget{mapping-mechanisms-to-governance-scenarios}{%
\section{Mapping Mechanisms to Governance
Scenarios}\label{mapping-mechanisms-to-governance-scenarios}}

The preceding sections rated each mechanism on technical readiness
(Section 3), against an adversary surface (Section 4.3), and as a
substitute for or complement to hardware-level governance (Section 5).
This section asks a different question: within a given governance
\emph{scenario}, defined by which actor holds authority and what
compliance levers that actor controls, does each mechanism function as a
primary lever, a supporting element, an irrelevant addition, or a
structural impossibility? We rate the twenty mechanisms against four
scenarios: domestic frontier-AI regulation (6.1), bilateral or
multilateral agreement (6.2), industry self-regulation (6.4), and
marketplace-and-platform-level governance (6.5). Section 6.3 stands
between the second and third scenario subsections and draws lessons from
three existing monitoring regimes; it does not consume scenario-matrix
cells.

The rating is deliberately distinct from the adversary analysis. A
mechanism may defend an adversary cell weakly in Section 4.3 yet be a
primary lever in a scenario, because the scenario describes regulatory
\emph{fit}, not adversarial \emph{capability}. Distributed-inference
monitoring (M4) is the clearest case: it defends the
deployer-as-adversary surface only partially (Section 4.3), but it is
the single mechanism a multilateral agreement is uniquely positioned to
operate, because cross-jurisdictional inference flows are visible to
coordinating signatories in a way they are not visible to any one
domestic regulator. The two ratings answer two questions and should not
be conflated.

A further clarification concerns time horizon: the scenario matrix
assesses presently actionable levers given the May 2026 productisation
state. A mechanism rated Irrelevant partly because it is not yet
productised (for example M2, M6, V2, or V3) may become a lever in the
same scenario once standards, procurement, or R\&D mandates bring it
into production; the rating describes what an actor in that scenario can
operate now, not a permanent property of the mechanism.

We rate against a four-point scale: Lever (L), the mechanism is
structurally central to the scenario's regulatory toolkit; Supporting
(S), the mechanism contributes defence-in-depth but is not the central
lever; Irrelevant (I), the mechanism exists but the scenario's actors
cannot make it do regulatory work; and Blocks (B), the scenario's
structure conflicts with the mechanism's preconditions, so the mechanism
cannot be deployed within the scenario at all. The conservative-rating
rule applies: where uncertain between adjacent ratings, we choose the
more conservative (L \(\rightarrow\) S \(\rightarrow\) I). The full
populated scenario matrix is in Appendix F.

A note on B. Only three of eighty cells are rated Blocks, and all three
are in the industry-self-regulation column (V1, V2, V3). The rating is
reserved for the specific structural conflict in which a
cryptographic-verification mechanism requires a trusted third-party
verifier or certification authority that the scenario's structure cannot
constitute. We discuss the distinction between Blocks and Irrelevant
where it arises (6.4) and flag in Section 7.4 that a reviewer who
rejects the distinction can collapse B into I without disturbing the
rest of the analysis.

\textbf{Table 6. Mechanism-scenario mapping.} Each cell rates a
mechanism's role in a scenario: L lever, S supporting, I irrelevant, B
structurally blocked (Section 6; the populated matrix in Appendix
F). The totals row gives each scenario's L/S/I/B distribution.

{\footnotesize\begin{longtable}[]{@{}
  >{\raggedright\arraybackslash}p{(\columnwidth - 8\tabcolsep) * \real{0.2000}}
  >{\raggedright\arraybackslash}p{(\columnwidth - 8\tabcolsep) * \real{0.2000}}
  >{\raggedright\arraybackslash}p{(\columnwidth - 8\tabcolsep) * \real{0.2000}}
  >{\raggedright\arraybackslash}p{(\columnwidth - 8\tabcolsep) * \real{0.2000}}
  >{\raggedright\arraybackslash}p{(\columnwidth - 8\tabcolsep) * \real{0.2000}}@{}}
\toprule\noalign{}
\begin{minipage}[b]{\linewidth}\raggedright
Mechanism
\end{minipage} & \begin{minipage}[b]{\linewidth}\raggedright
Domestic (S1)
\end{minipage} & \begin{minipage}[b]{\linewidth}\raggedright
Bi/Multilateral (S2)
\end{minipage} & \begin{minipage}[b]{\linewidth}\raggedright
Self-regulation (S4)
\end{minipage} & \begin{minipage}[b]{\linewidth}\raggedright
Marketplace (S5)
\end{minipage} \\
\midrule\noalign{}
\endhead
\bottomrule\noalign{}
\endlastfoot
M1 & L & S & S & L \\
M2 & I & I & I & I \\
M3 & S & S & S & S \\
M4 & S & L & I & S \\
M5 & L & S & S & L \\
M6 & I & I & I & I \\
V1 & S & L & B & S \\
V2 & I & S & B & I \\
V3 & I & S & B & I \\
V4 & S & L & S & L \\
V5 & S & L & S & S \\
V6 & S & S & S & S \\
V7 & S & S & L & S \\
E1 & L & S & S & L \\
E2 & L & S & S & L \\
E3 & L & L & I & L \\
E4 & L & S & S & L \\
E5 & S & I & S & L \\
E6 & S & I & S & L \\
E7 & L & S & S & L \\
\textbf{Totals (L/S/I/B)} & \textbf{L 7 / S 9 / I 4 / B 0} & \textbf{L 5
/ S 11 / I 4 / B 0} & \textbf{L 1 / S 12 / I 4 / B 3} & \textbf{L 10 / S
6 / I 4 / B 0} \\
\end{longtable}}

\noindent\emph{Mechanism IDs.} M1 per-query usage and token accounting;
M2 workload classification; M3 KV-cache and reasoning-token accounting;
M4 distributed-inference monitoring; M5 authenticated account identity
and request tracing; M6 inference-stage power and cost monitoring;
V1 TEE-bounded inference attestation; V2 zero-knowledge proofs of
inference; V3 hardware-backed workload certification; V4 tool-call
cryptographic signing and agent-trace attestation; V5 capability-evaluation
reporting; V6 retrieval provenance and citation logging; V7
chain-of-thought monitorability; E1 per-deployment licensing and access
control; E2 output filtering and capability gating; E3
jurisdiction-bounded inference; E4 rate limiting as governance primitive;
E5 agentic action permissions and policy-as-code; E6 tool-use
authorisation and sandboxing; E7 inference-time off-switch and
kill-switch.

Six mechanisms are levers in no scenario (M2, M3, M6, V2, V3, V6);
jurisdiction-bounded inference (E3) is the only lever in three
scenarios, and no mechanism is a lever in all four.

\hypertarget{domestic-frontier-ai-regulation}{%
\subsection{Domestic frontier-AI
regulation}\label{domestic-frontier-ai-regulation}}

In the domestic scenario, a single jurisdiction governs inference at the
deployment stage through legislation, executive action, or regulatory
rule-making, enforced by domestic bodies and binding on foreign
providers only when they sell into or operate within the territory. The
reference cases are the EU AI Act's general-purpose provisions
\citep{euaiact2024, carey2026regulating}, the now-revoked US Executive
Order 14110 \citep{whitehouse2023eo}, the UK Code of Practice, and the
California SB 53 framework. The structural feature that distinguishes
the domestic scenario from the others is that the regulator holds
direct, legally backed authority over the deployers, users, and
licensees inside its territory, but no authority outside it.

The matrix shows the domestic scenario to be mechanism-broad: of the
twenty mechanisms, sixteen are actionable
(seven Levers, nine Supporting) and only four are Irrelevant, with no
mechanism structurally blocked. The seven Levers are the mechanisms a
regulator can mandate directly and that are already productised at the
four-vendor evidence base: per-query accounting (M1), API-level KYC and
request tracing (M5), per-deployment licensing (E1), output filtering
(E2), jurisdiction-bounded inference (E3), rate limiting (E4), and the
inference-time off-switch (E7). These are the operational core of a
domestic regime. Each is a control the regulator can write into law,
each is universally productised, and each binds the cooperative
in-territory deployer that the domestic scenario assumes. The KYC lever
in particular has the strongest external analogue: the financial
anti-money-laundering regime we discuss in 6.3 is built on exactly this
primitive operated through regulated intermediaries
\citep{egan2023oversight, fatf2012recommendations}.

The four Irrelevant ratings are informative. M2 (MoE workload
classification) and M6 (power monitoring) are Irrelevant in \emph{every}
scenario, for the reason established in Section 3: neither is
productised, so no actor can mandate consumption of telemetry that no
vendor exposes. Their irrelevance in the domestic column is therefore
not a property of the scenario but of the mechanism, and a regulator who
wished to mandate them would be mandating a substrate that does not yet
exist. The other two Irrelevant ratings, V2 (zkML) and V3 (workload
certificates), reflect the same productisation gap from the verification
cluster: the academic substrate is worked out but the production
substrate is not, so a domestic mandate would have nothing deployable to
require. The single most important consequence for domestic policy is
therefore that the readiness gap identified in Section 3.4 maps almost
exactly onto the Irrelevant column here: the mechanisms a domestic
regulator cannot use are the mechanisms that are not yet built, not the
mechanisms the scenario excludes.

The nine Supporting ratings are the defence-in-depth layer: the
monitoring and verification mechanisms (M3, M4, V1, V4, V5, V6, V7) and
the agentic-runtime controls (E5, E6) that a regulator can require but
that are more naturally configured at the platform than legislated in
detail. The pattern across the Levers and Supporting ratings is that the
domestic scenario makes best use of \emph{enforcement} mechanisms (five
of seven E-mechanisms are Levers) and treats verification mechanisms as
supporting. This is the inverse of the multilateral scenario, to which
we turn next, and the inversion is the central finding of the scenario
mapping: the same taxonomy reweights itself depending on which actor
holds authority. The domestic regulator's comparative advantage is
direct enforcement; the multilateral coordinator's comparative advantage
is verification.

The honest limitation of the domestic scenario is the one Section 4.3
established: every Lever here binds the cooperative in-territory
deployer and none reaches the state-level deployer who self-hosts
outside the platform (the C3.R2 cell) or the foreign provider who does
not sell into the jurisdiction. The domestic regime is broad but
territorially bounded, and the substitutability dynamics of Section 4.1
mean a determined actor can relocate inference rather than comply. This
is the structural reason the domestic scenario is necessary but not
sufficient, and the reason it pairs naturally with the marketplace
scenario (6.5) and the multilateral scenario (6.2), which reach where it
cannot.

\hypertarget{bilateral-or-multilateral-agreement}{%
\subsection{Bilateral or multilateral
agreement}\label{bilateral-or-multilateral-agreement}}

In the multilateral scenario, two or more jurisdictions coordinate
inference governance through a treaty, agreement, or harmonised
regulatory action, establishing shared rules, mutual recognition of
compliance, shared enforcement, or coordinated export controls. The
reference cases are the Hiroshima Process, the US-UK AI Safety Institute
memorandum, the Bletchley follow-on summits, and the proposed
International Network of AI Safety Institutes
\citep{ho2023international, scholefield2025international, aarne2025international}.
The structural feature that distinguishes this scenario is that
signatories can construct institutions no single jurisdiction can build
alone: shared certification authorities, mutual recognition of
cryptographic attestations, and coordinated visibility into cross-border
flows.

This is the scenario in which verification mechanisms come into their
own. The matrix records five Levers,
and three of them are verification mechanisms that rate only Supporting
in the domestic column: TEE-bounded inference attestation (V1),
tool-call cryptographic signing (V4), and capability-evaluation
attestation (V5). The reason is structural. A cryptographic attestation
is only as useful as the authority that verifies it, and a single
domestic regulator has little use for an attestation it must verify
alone; a coalition that agrees a shared verification authority and
mutually recognises each other's attestations converts the same
primitive into a cross-border trust instrument. V4 is the strongest
case: cryptographic signing is robust to the non-cooperative deployer
(it had zero Inadequate ratings against deployer adversaries in Section
4.3) and verifiable across borders once signatories recognise the
signer, key infrastructure and verification policy, which
is precisely the property a multilateral regime needs and a domestic
regime underuses. V5 is the verification primitive a capability-limit
treaty most requires, because third-party-verifiable capability claims
are what such a treaty would be built to check, and the shared
evaluation authority it needs is exactly what coordination can
establish.

The other two Levers are distributed-inference monitoring (M4) and
jurisdiction-bounded inference (E3). M4 is the single mechanism uniquely
suited to this scenario and to no other: it is the only mechanism whose
Lever rating appears in the multilateral column alone, because
cross-jurisdictional inference flows are visible to coordinating
signatories and invisible to any one regulator. E3 is a Lever in three
scenarios, but its multilateral form (mutual data-residency recognition,
coordinated regional restrictions) does work the domestic form cannot,
because jurisdiction-binding across borders requires the cooperation of
the bordering jurisdictions. The presence of M4 and E3 as Levers,
alongside the three verification mechanisms, gives the multilateral
scenario a distinctive profile: it is the scenario built for the
cross-border and the cryptographic, the two properties that the
substitutability and distributed-inference findings of Sections 4.1 and
4.2 establish as the hardest problems for any single jurisdiction.

The scenario is weaker on per-query runtime enforcement than the
domestic scenario, and the matrix records this honestly: rate limiting
(E4), output filtering (E2), licensing (E1), and the off-switch (E7) all
drop from Lever in the domestic column to Supporting here, because these
controls are implemented and enforced per provider under domestic law,
and an agreement harmonises them rather than operating them directly.
Two agentic-runtime mechanisms, E5 (policy-as-code) and E6 (sandboxing),
drop further to Irrelevant, because per-deployment technical
configuration is not something a treaty meaningfully actuates. The
division of labour the matrix implies is clean: the multilateral
scenario coordinates verification and cross-border structure; the
domestic and marketplace scenarios operate runtime enforcement. This is
the empirical content of the claim, made informally in Sections 4.1 and
5, that inference-time governance requires layered application across
scenarios rather than a single regulatory locus. The substitution principle's
layered-substitution mechanisms (M4, V3, E3) are the mechanisms whose
regulatory weight rises under substitution, and two of the three (M4,
E3) are Levers precisely in the scenario positioned to operate them.

\hypertarget{lessons-from-existing-inference-monitoring-regimes}{%
\subsection{Lessons from existing inference-monitoring
regimes}\label{lessons-from-existing-inference-monitoring-regimes}}

Ansari \citep{ansari2026hardware} Section 5.3 draws lessons from existing verification
regimes in arms control, specifically the bilateral and multilateral
treaty regimes governing nuclear, chemical, and biological weapons. The
arms-control analogue is appropriate to hardware-level governance
because the underlying substrate (a small number of large facilities
producing a small number of high-value items) has a structural
resemblance to the geopolitics of frontier-training compute. The same
analogue is \emph{less} appropriate for inference-time governance, where
the underlying substrate (large numbers of small transactions
distributed across many service providers and end users) has a different
structural shape. The natural analogues for inference-time governance
are different: financial transaction monitoring, pharmaceutical
post-market surveillance, and aviation safety reporting.

We work through each in turn.

\textbf{Table 7. Three existing monitoring regimes as analogues for
inference-time governance} (Section 6.3). None maps one-to-one; each
contributes structural patterns.

{\footnotesize\begin{longtable}[]{@{}
  >{\raggedright\arraybackslash}p{(\columnwidth - 8\tabcolsep) * \real{0.2000}}
  >{\raggedright\arraybackslash}p{(\columnwidth - 8\tabcolsep) * \real{0.2000}}
  >{\raggedright\arraybackslash}p{(\columnwidth - 8\tabcolsep) * \real{0.2000}}
  >{\raggedright\arraybackslash}p{(\columnwidth - 8\tabcolsep) * \real{0.2000}}
  >{\raggedright\arraybackslash}p{(\columnwidth - 8\tabcolsep) * \real{0.2000}}@{}}
\toprule\noalign{}
\begin{minipage}[b]{\linewidth}\raggedright
Analogue
\end{minipage} & \begin{minipage}[b]{\linewidth}\raggedright
Unit of analysis
\end{minipage} & \begin{minipage}[b]{\linewidth}\raggedright
Monitoring substrate
\end{minipage} & \begin{minipage}[b]{\linewidth}\raggedright
Lessons that transfer
\end{minipage} & \begin{minipage}[b]{\linewidth}\raggedright
Best fits
\end{minipage} \\
\midrule\noalign{}
\endhead
\bottomrule\noalign{}
\endlastfoot
Financial transaction monitoring (FATF AML/CFT) & Per transaction
(customer due diligence, suspicious-activity reports) & Regulated
intermediaries (banks, money-services businesses) & Per-transaction KYC
at scale is feasible; reporting thresholds are calibrated and updated,
not fixed in law; evasion is contained only by international
coordination infrastructure & Domestic (6.1), Marketplace (6.5) \\
Pharmaceutical post-market surveillance (FAERS, EudraVigilance) & The
deployed system as actually used & Regulator-administered adverse-event
reporting & Post-market surveillance is the right model for capability
monitoring; regulator-administered reporting addresses the
deployer-as-adversary; the architecture must extract rare severe signals
from a long tail & Multilateral capability monitoring (6.2) \\
Aviation safety reporting (ICAO, NTSB, EASA) & Per flight, per incident
& Independent investigation authorities & Reporting must be decoupled
from liability; operators, regulators, and investigators must be
institutionally separate; each failure is invested in as a learning
event & Incident-reporting infrastructure (all scenarios) \\
\end{longtable}}

\textbf{Financial transaction monitoring.} The international
anti-money-laundering and counter-terrorism financing regime, codified
in the Financial Action Task Force Recommendations
\citep{fatf2012recommendations}, is structurally the closest analogue to
inference governance among the three. It operates at the per-transaction
level (Customer Due Diligence, Suspicious Activity Reports), it relies
on regulated intermediaries (banks, money-services businesses) as the
primary monitoring substrate, and it has developed over thirty years of
incremental implementation across more than 200 jurisdictions. Three
specific lessons transfer to inference. First, per-transaction
Know-Your-Customer at scale is operationally feasible if the
intermediary substrate is well-defined; the existing scholarship has
already extended the analogue to AI compute providers
\citep{egan2023oversight}. Second, the reporting threshold
(suspicious-activity or above-amount triggers) is calibrated and updated
over time rather than fixed in legislation; inference-time thresholds
will need the same calibration discipline. Third, the regime survives
the substitutability problem (cross-border transfers, informal value
transfer systems, cryptocurrency) only because it has built
international coordination infrastructure (the FATF itself, mutual
evaluations, the Egmont Group of financial intelligence units). An
inference-monitoring regime without analogous coordination
infrastructure will face the same evasion dynamics with no analogous
response.

\textbf{Pharmaceutical post-market surveillance.} Adverse event
reporting systems (FAERS in the United States, EudraVigilance in the
European Union, comparable systems elsewhere) provide a model for
capability monitoring of deployed AI systems. Umaru and colleagues
\citep{umaru2026global} document the comparative pharmaceutical
regulatory landscape and the trend toward harmonisation across
jurisdictions despite persistent differences in mandates, timelines, and
data requirements. Three lessons transfer. First, post-market
surveillance is the right structural model for \emph{capability}
monitoring at the deployment stage; the regulatory unit is ``the
deployed system as actually used,'' not ``the system as approved at
release.'' Second, the reporting substrate can be regulator-administered
rather than provider-administered, which addresses the
deployer-as-adversary case that several mechanisms in Section 3 (notably
V3, V4, and V5) are sensitive to. Third, the pharmacovigilance regime is
calibrated to a long-tail risk profile in which most adverse events are
minor but a small number are severe and demand rapid response. Inference
governance has a similar profile: most deployments are routine, but a
small number warrant rapid intervention. The signal-extraction
architecture matters more than the average-case monitoring.

\textbf{Aviation safety reporting.} \citet{chatzipanagiotis2026incident} makes the most directly applicable
analogue argument: the EU AI Act's incident reporting framework can be
substantially improved by adopting aviation safety principles,
specifically the decoupling of reporting from liability and the
establishment of independent investigation authorities. Three lessons
transfer. First, the per-flight (per-deployment, per-incident)
attestation architecture in aviation is the closest analogue to
per-query or per-deployment attestation in inference; the technical and
procedural lessons translate with adjustments. Second, the institutional
separation between operators (airlines), regulators (the FAA, EASA), and
investigators (the NTSB, BEA) is a structural feature without which
incident reporting tends to be either suppressed by operators or
politicised by regulators. Inference governance faces the same risks if
the same actors play multiple roles. Third, the aviation regime accepts
a non-zero failure rate but invests heavily in learning from each
failure; inference governance has the option of the same posture but has
not yet committed to it.

The combined lessons are sobering. None of the three analogues maps onto
inference governance one-to-one; each provides structural patterns that
the inference-time framework can borrow. The financial-transaction
analogue is most operationally informative for the domestic and
marketplace scenarios (Sections 6.1 and 6.5), where per-transaction KYC
operated through regulated intermediaries is the closest structural fit.
The pharmacovigilance analogue is most informative for the
capability-monitoring component of the bilateral-or-multilateral
scenario (Section 6.2). The aviation analogue is most informative for
the incident-reporting infrastructure that any of these scenarios will
require. \citet{ho2023international} propose four
institutional models for international AI governance; we note in passing
that none of those four models corresponds cleanly to any of the three
analogues drawn here, and we read this as evidence that the
institutional design problem for inference governance is genuinely open.
The mechanisms in Section 3 and the scenarios in Section 6 give the
technical and political substrate; the institutional question is
unresolved.

The arms-control analogue retains some force for the
bilateral-or-multilateral scenario specifically (Section 6.2),
particularly for any regime that contemplates international verification
of compliance with capability limits
\citep{wasil2024verification, scholefield2025international, baker2025verifying}.
The technical-verification literature draws on arms-control precedent
and the lessons are not displaced by the financial, pharmaceutical, and
aviation analogues. The position we take is therefore additive rather
than substitutive: the arms-control analogue applies where Ansari \citep{ansari2026hardware}
Section 5.3 applies, the three analogues developed here apply where the
substrate of the regulated activity is distributed-and-transactional,
and the institutional question of \emph{which combination of analogues
fits which scenario} is what Section 6 (mechanism-scenario mapping) sets
out to make precise. \citet{vermeer2024historical} notes more
generally that historical governance models for nuclear, internet,
encryption, and genetic engineering each provide partial lessons for AI
but none provides a complete template; the same is true for our three
analogues in microcosm.

\hypertarget{industry-self-regulation}{%
\subsection{Industry self-regulation}\label{industry-self-regulation}}

In the self-regulation scenario, major commercial inference providers
voluntarily adopt governance practices through consortia, voluntary
frameworks, or coordinated commitments, with the participating actors
setting the standards and enforcement resting on reputation, contract,
and bilateral pressure rather than government action. The reference
cases are the Frontier Model Forum, the Responsible Scaling Policy
frameworks at individual companies, the 2023 voluntary White House
commitments, and the voluntary adoption of the NIST AI Risk Management
Framework. The structural feature that distinguishes this scenario, and
that drives its ratings, is the absence of legal authority to compel: a
self-regulatory body can bind its members but cannot reach a
non-participating provider, a self-hosting deployer, or a fine-tuner
operating on open weights. For the purposes of this matrix, the scenario
denotes provider self-regulation without an independent certification
institution, and the structurally Blocked ratings for V1-V3 hold under
that reading. Consortium arrangements that establish accredited
laboratories, independent auditors, or shared trust roots lie outside
the scenario as defined and could raise those ratings; we treat this
boundary as a definitional choice about what ``self-regulation'' denotes
here rather than a claim that industry coordination is technically or
institutionally incapable of building such an authority.

The matrix records the self-regulation scenario as the weakest of the
four, with a single Lever, twelve Supporting ratings, four Irrelevant,
and the only three Blocks ratings in the entire matrix. The structural
reason is that the mechanisms the scenario \emph{can} advance are those
already productised by the participating vendors, which a voluntary
commitment formalises rather than creates, hence the unusually high
count of Supporting ratings. Twelve mechanisms are Supporting because
the major vendors already operate them and could commit to a shared
baseline: per-query accounting, KYC, licensing, output filtering, rate
limiting, the off-switch, the agentic-runtime controls, and the
productised verification primitives (V4, V5, V6). None rises to Lever
because the defining limitation of voluntary governance is that
formalising existing practice among the willing does not produce a
primary regulatory lever; it produces a floor that binds only the
signatories.

The single Lever is chain-of-thought monitorability (V7), and the reason
it rises to Lever here while remaining Supporting in every other
scenario is worth stating, because it is the one place the
self-regulation scenario has a comparative advantage. Reasoning-trace
transparency is a commitment a voluntary frame can advance
\emph{without} needing legal compulsion: a provider can choose to expose
reasoning traces, as Anthropic does with its display modes, and the
research consensus on CoT monitorability has formed largely through
voluntary cross-lab collaboration rather than regulatory mandate
\citep{korbak2025chain, anwar2026analyzing}. The mechanism that is
hardest to legislate, because of the architectural divergence across
vendors documented in Sections 3.2 and 4.5, is the mechanism best suited
to voluntary coordination, because the same divergence is something a
consortium can converge by agreement where a regulator cannot converge
it by statute. This is the clearest case in the scenario mapping of a
mechanism finding its natural home in a single scenario.

The four Irrelevant and three Blocks ratings are where the scenario's
structural limit is sharpest, and they justify the Blocks rating that
the matrix specification flagged as reviewer-exposed. E3
(jurisdiction-bounded inference) is Irrelevant because jurisdictional
boundaries are a function of state law, not voluntary commitment: a
consortium cannot allocate or enforce them. The three Blocks ratings, V1
(TEE attestation), V2 (zkML), and V3 (workload certificates), are the
mechanisms whose preconditions the scenario cannot satisfy. Each
requires a trusted third-party verifier or certification authority to
give the attestation meaning, and pure industry self-regulation, in
which the regulated parties are also the only standard-setters,
structurally cannot constitute the independent authority that a credible
attestation regime needs. This is the distinction between Blocks and
Irrelevant: M2 and M6 are Irrelevant because they are unproductised (a
property of the mechanism), whereas V1, V2, and V3 are Blocked because
the scenario's structure conflicts with the mechanism's precondition (a
property of the scenario). We note that the Irrelevant coding here
partly tracks present productisation rather than a permanent scenario
mismatch: as the time-horizon note in the rating methodology records, M2
and M6 could become Supporting or Lever in this scenario once
productised, so the rating reflects the May 2026 state, not an
in-principle exclusion. A reviewer who finds the distinction
unconvincing can collapse the three Blocks into Irrelevant; the
substantive finding, that the verification cluster cannot be operated
through voluntary governance alone, survives either coding. The policy
reading is direct: self-regulation can formalise the productised floor
and can lead on reasoning transparency, but it cannot deliver the
cryptographic-verification layer that the multilateral and marketplace
scenarios can, and it is structurally incapable of reaching the
open-weight surface that Section 4.5 identifies as the growing coverage
gap \citep{williams2026regulating, gomes2026open}.

\hypertarget{marketplace-and-platform-level-governance}{%
\subsection{Marketplace-and-platform-level
governance}\label{marketplace-and-platform-level-governance}}

In the marketplace scenario, cloud platforms and inference marketplaces
act as governance intermediaries, imposing conditions on the partner
models and customer deployments routed through their infrastructure and
enforcing them through terms of service, technical controls, and
contract. Government action is indirect: regulators leverage the
marketplace as a chokepoint rather than acting on inference providers
directly \citep{heim2024governing}. The reference cases are AWS
Bedrock's per-model access controls, Vertex AI's Advanced AI Safety
Addendum and its CSAM classifier applied to partner-model image inputs,
and Anthropic's Acceptable Use Policy enforced across cloud partners.
The structural feature that distinguishes this scenario is direct
operational control over inference \emph{routing}: the platform sees and
can act on every request that passes through it, for any model,
regardless of the model provider's own governance.

The matrix records the marketplace scenario as the most Lever-dense of
the four, with ten Levers, the highest count in any column, and this is
the scenario mapping's most consequential finding. The ten Levers are
the per-query accounting and KYC monitoring pair (M1, M5), the full
enforcement cluster except the off-switch in its coordinated form (E1,
E2, E3, E4, E5, E6, plus E7), and the strongest productised verification
primitive (V4). The platform can mandate each of these on all routed
traffic because it owns the substrate each runs on: it meters every
request, operates per-customer identity through IAM, controls the quota
dashboards, runs its own filtering overlay, enforces per-region routing,
owns the agent-runtime and sandbox surfaces, and can revoke access
unilaterally. The marketplace scenario is, in effect, the operational
complement to the domestic scenario: six mechanisms (M1, M5, E1, E2, E4,
E7) are Levers in exactly these two scenarios and no other, which is the
matrix's quantitative signature of the claim that domestic regulation
and marketplace governance are the two scenarios that operate runtime
enforcement, one by legal mandate and one by operational control.

The marketplace scenario also exhibits a property the others do not: the
platform's governance is applied \emph{on top of} the model provider's,
producing the intersection pattern documented in Section 4.5. The Vertex
CSAM classifier on Claude image inputs is the concrete case: the
platform enforces a control the partner model's own filtering does not,
and the inference is governed by the intersection of both. This is why
E2 (output filtering) is a Lever here independent of whether the
underlying model provider operates filtering: the marketplace lever is
the platform's own overlay, not the model's. The same logic makes the
marketplace the strongest scenario for the runtime agentic controls (E5,
E6) that drop to Irrelevant in the multilateral scenario, because the
platform operates the execution environment that a treaty cannot reach.
The marketplace scenario is where the
deployment-platform-as-governance-overlay finding of Section 4.5 becomes
a regulatory asset rather than a complication.

The four Irrelevant ratings are the same unproductised and proof-stage
mechanisms that are Irrelevant elsewhere (M2, M6, V2, V3), confirming
again that the boundary of every scenario's reach is set by mechanism
readiness more than by scenario structure. The marketplace scenario's
distinctive limitation is the one Section 4.5 established and that no
marketplace lever can close: the open-weight surface escapes the
marketplace architecture entirely. A model running on a customer's own
hardware with no commercial inference endpoint routes through no
platform, so the ten Levers that make this the strongest enforcement
scenario all lose force at exactly the deployment surface that Section
4.5, drawing on \citet{alexander2026efficiency} and \citet{puri2026small}, identifies as converging toward frontier
capability. The marketplace scenario governs the routed inference base
comprehensively and the unrouted inference base not at all, and the
boundary between them is the handoff to the hardware-level governance of
Ansari \citep{ansari2026hardware} that Section 4.5 and the substitution principle's layered-substitution
corollary both identify as the layer below which the substitution
channel cannot operate.

\hypertarget{cross-cutting-findings}{%
\subsection{Cross-cutting findings}\label{cross-cutting-findings}}

Three patterns hold across the four scenario columns.

First, the scenarios reweight the taxonomy by actor authority rather
than by mechanism quality. The same twenty mechanisms produce a
different Lever set in each scenario: the domestic scenario leads with
enforcement (five of seven E-mechanisms are Levers), the multilateral
scenario leads with verification (three verification mechanisms become
Levers that are only Supporting domestically), the self-regulation
scenario leads with the one mechanism legal compulsion cannot improve
(V7), and the marketplace scenario leads with operational runtime
control (ten Levers, the most of any column). No mechanism is a Lever in
all four scenarios, and only one (E3, jurisdiction-bounded inference) is
a Lever in three. The practical consequence is that a comprehensive
regime is likely to require a combination of scenarios, not a choice among
them: the domestic and marketplace scenarios operate runtime
enforcement, the multilateral scenario operates cross-border
verification, and the self-regulation scenario operates the
reasoning-transparency layer, each covering what the others structurally
cannot.

Second, the boundary of every scenario's reach is set by mechanism
readiness, not by scenario structure. The four Irrelevant ratings are
nearly identical across all four columns: M2 and M6 (unproductised
monitoring) are Irrelevant in every scenario, and V2 and V3 (proof-stage
verification) are Irrelevant or Blocked in every scenario. The
mechanisms a scenario cannot use are overwhelmingly the mechanisms that
are not yet built (the readiness gap of Section 3.4), not the mechanisms
the scenario excludes by its structure. This means the single most
consequential intervention for \emph{every} scenario is the same:
closing the productisation gap on the verification cluster (V1, V2, V3),
which is currently the binding constraint on what any regulator,
coalition, consortium, or platform can do.

Third, six mechanisms are never Levers in any scenario: M2, M3, M6, V2,
V3, and V6. Four of the six are the unproductised or proof-stage
mechanisms already discussed. The other two, M3 (reasoning-token
accounting) and V6 (RAG attestation), are productised and useful but are
Supporting everywhere because they provide visibility and grounding
rather than a primary point of control; they are the defence-in-depth
instruments of the taxonomy. Their consistent Supporting rating is not a
weakness but a correct characterisation: not every mechanism needs to be
a lever, and a regime that tried to make accounting and provenance
primitives carry primary regulatory weight would be misusing them.

\hypertarget{discussion-and-implications}{%
\section{Discussion and
Implications}\label{discussion-and-implications}}

We draw together the readiness distribution (Section 3), the adversary
analysis (Section 4.3), the convergence argument (Section 5), and the
scenario mapping (Section 6) into four implications and a statement of
limitations. The throughline is the gap between two senses of readiness
that the paper has kept distinct: a mechanism can be deployable today
and still cover only a fraction of the adversary surface, and the
distance between those two facts is where the policy and research work
lies.

\hypertarget{the-readiness-gap}{%
\subsection{The readiness gap}\label{the-readiness-gap}}

The headline empirical finding is that inference-time governance has a
substantial deployable substrate. Fifteen of the twenty mechanisms have
commercially available production substrates, productised by multiple major
commercial vendors; only two require near-term integration (M2, V1), two
require research (V2, V3), and one is speculative (M6). A regulator who
assumed that inference-stage governance was a future research problem
would be wrong: the substrates for most of the toolbox exist and are in
production for billing, abuse prevention, and capacity allocation today.

The gap is not in deployment but in two places the deployment count
conceals. The first is the verification cluster. The mechanisms that are
not yet built are concentrated in cryptographic verification (V1
TEE-bounded attestation at the inference API, V2 zkML, V3 per-output
workload certificates), which is precisely the category that the
adversary analysis identifies as the only category whose coverage
survives a capable, non-cooperative deployer. The substrate that is
missing is the substrate that matters most against the hardest
adversary. The second is the distance between deployment-readiness and
adversary-coverage. The fifteen deployable mechanisms are deployable
against the surface they were built for, which is the cooperative
deployer and the low-to-moderate-capability user; against the
state-level deployer (C3.R2) no mechanism rates adequate, and against
the fine-tuner at high capability the model-internal enforcement cluster
fails.
The readiness gap, stated precisely, is that the inference-time toolbox
is mature where the threat is ordinary and thin where the threat is
sophisticated. The gap this paper documents is therefore not a shortage
of commercial substrate but a shortage of governance-grade assurance:
primitives are common, independently auditable and adversary-resistant
mechanisms are rare.

This reframing matters for how the finding is read. The optimistic
reading (most mechanisms are deployable) and the pessimistic reading
(coverage collapses against capable adversaries) are both true and
describe the same matrix. The paper's contribution is to make the
conditions under which each holds explicit, so that a regulator can tell
which claim applies to the threat they are actually trying to govern.

\hypertarget{implications-for-research}{%
\subsection{Implications for research}\label{implications-for-research}}

Three research priorities follow directly from the matrix, and they are
not the priorities the current literature emphasises.

The first is the productisation of the verification cluster, not the
invention of new mechanisms. V1 (TEE-bounded inference attestation) is
rated near-term rather than deployable for a single reason: the
substrate exists at the infrastructure layer (Confidential VMs,
Confidential GPUs, Nitro Enclaves, NVIDIA confidential computing) but is
not exposed at the inference API. This is a productisation decision, not
a research frontier, and it is the cheapest single intervention that
would change the matrix, because V1 is the substrate on which several
other mechanisms layer. V3 (per-output workload certificates) and V2
(zkML) are genuine research problems, the latter blocked by
proof-generation overhead at frontier scale, but the highest-value
near-term move is the engineering work to surface attestation that
already exists below the API.

The second is the structural rise of layered-substitution mechanisms.
The substitution principle and its first corollary establish that as
training-inference substitution becomes more efficient, the regulatory
weight on the layered-substitution mechanisms (M4, V3, E3) increases,
because the hardware substrate becomes a less substitutable
intervention point at which the substitution channel is harder to evade. This couples the inference-time
research agenda to the hardware-level one: V3's ``requires R\&D'' rating
understates its rising structural importance, and the research community
should treat per-output certification as more urgent than its current
readiness suggests.

The third is chain-of-thought monitorability as a defence rather than a
primitive. V7 is productised at three vendors but rates adequate only
against the low- and medium-capability user and partial or inadequate
against every more capable adversary, bounded by the faithfulness
limitation that a model's exposed reasoning traces may not faithfully
reflect the computation that produced its output
\citep{turpin2023language, korbak2025chain}. The research gap is not
exposing reasoning traces, which is done, but establishing whether
exposed traces can be trusted as evidence, which is open. White-box
latent monitoring \citep{yu2026latentaudit} and formal faithfulness
guarantees are the directions; until they mature, V7 should not be
relied on as a control against a capable deployer or fine-tuner, and
governance proposals that lean on CoT monitoring should say so.

\hypertarget{implications-for-policy}{%
\subsection{Implications for policy}\label{implications-for-policy}}

The scenario mapping yields a policy conclusion that the per-mechanism
analysis alone does not: a comprehensive inference-governance regime is
a combination of scenarios, not a choice among them. The same twenty
mechanisms reweight by which actor holds authority. The domestic
regulator's comparative advantage is direct enforcement (five of seven
enforcement mechanisms are primary levers domestically); the
multilateral coordinator's is verification (three verification
mechanisms that are merely supporting domestically become primary levers
under coordination, because a shared certification authority is what
gives a cross-border attestation its value); the marketplace's is
operational runtime control (the most lever-dense scenario, governing
all routed traffic); and industry self-regulation's narrow advantage is
the one mechanism legal compulsion cannot improve, reasoning-trace
transparency. No mechanism is a primary lever in all four scenarios, and
only one (jurisdiction-bounded inference) is a lever in three. The
practical instruction is to pair the scenarios that cover what the
others cannot: domestic and marketplace governance operate runtime
enforcement, multilateral coordination operates verification, and
self-regulation operates the transparency layer.

The second policy finding is that the binding constraint is the same
across every scenario, and it is the readiness gap. The mechanisms that
are non-actionable in one scenario are almost exactly the mechanisms
that are non-actionable in all of them: the unproductised monitoring
primitives (M2, M6) and the proof-stage verification primitives (V2, V3)
are irrelevant or blocked regardless of which actor holds authority. The
mechanisms a scenario cannot use are overwhelmingly the mechanisms that
are not yet built, not the mechanisms the scenario excludes by
structure. This means the single intervention that most expands what
every regulator, coalition, consortium, and platform can do is the same:
close the productisation gap on the verification cluster. Policy
attention spent there is not scenario-specific; it raises the ceiling
everywhere at once.

The third policy finding is where the taxonomy hands off to other
regimes. Against the state-level deployer and the high-capability
fine-tuner, the inference-time toolbox is structurally limited, and
these gaps cannot be closed by any inference-stage mechanism. They are
the province of hardware-level governance \citep{ansari2026hardware},
export controls on weights and accelerators, and the open-weight surface
that escapes platform intermediation entirely and converges toward
frontier capability
\citep{alexander2026efficiency, puri2026small, gomes2026open}.
Inference-time governance is a complement to these regimes, not a
substitute for them, and policy that treats it as a standalone solution
will leave exactly the highest-capability adversaries ungoverned.

\hypertarget{limitations}{%
\subsection{Limitations}\label{limitations}}

This work has several limitations, stated plainly so that the findings
are not read as stronger than they are.

The evidence base is four vendors (AWS Bedrock, Anthropic, OpenAI,
Google Vertex AI), with Azure OpenAI excluded as structurally
overlapping with the OpenAI pass (Appendix D). The four-vendor base is
representative of the frontier commercial supply side but does not
capture the open-weight and self-hosted surface, which is precisely the
surface the marketplace and self-regulation scenarios cannot reach. The
productisation findings are accurate for the commercial frontier and
silent beyond it. A further limitation concerns how that vendor evidence
was verified. The vendor content sweep was keyword-level, not
line-by-line: each documented capability was confirmed to appear in its
cited source, but the sources were not read in full. The caveat is
sharpest for AWS Bedrock, whose documentation was captured from a single
User Guide of roughly 4,800 pages \citep{aws_bedrock_ug_2026}; the AWS
claims rest on keyword-anchored extraction from that document rather
than an exhaustive reading of it.

The ratings are interpretive, and two interpretive choices are
reviewer-exposed. The readiness ratings for the enforcement cluster (E2,
E5, E6, E7) are stated in adversary-scoped form: currently deployable
against the surface they target, inadequate against the fine-tuner at
higher capability. A reviewer who prefers a single scalar rating would
downgrade these mechanisms; we hold that the scoped rating is the format
the two-dimensional adversary model requires, but the choice is a
judgement, not a fact. Separately, the scenario matrix rates three cells
as Blocks (V1, V2, V3 under industry self-regulation), distinguishing
structural incompatibility from mere irrelevance; the distinction is
defensible but a reviewer who rejects it can collapse Blocks into
Irrelevant without disturbing the substantive finding that the
verification cluster cannot be operated through voluntary governance
alone.

The ratings were assigned by a single author and re-rated by a second
rater on a randomly sampled subset of seven of the twenty mechanisms,
with exact agreement on five and one-step disagreements on the remaining
two (V4 and V7); quadratic-weighted Cohen's kappa was 0.74
(substantial), though the unweighted value is 0.53 (moderate) and the
estimate is fragile at this sample size, where a single rating change
moves the weighted statistic by roughly 0.13. The methodology and the
two disagreements are documented in Appendices B and C.
Single-primary-rater taxonomy construction is a known limitation that
the inter-rater check mitigates but does not eliminate; the subset is
seven mechanisms rather than the full twenty, the marginal distribution
is skewed toward the ``currently deployable'' rating, and the agreement
statistic should be read as corroborating rather than decisive. The
inter-rater check covered a subset of the readiness ratings only; the
adversary matrix (Section 4.3) and the scenario matrix (Section 6) were
single-author codings and were not independently re-rated. Notably,
both disagreements fell on the two mechanisms (V4 and V7) that the
author's sealed reasoning had flagged in advance as the most likely to
diverge, and that the second rater independently identified as the
hardest to rate; the convergence on where the uncertainty lies is itself
informative, and because both disagreements were one-step, they fall
below the pre-specified threshold (a gap greater than one rating) for a
formal rating revision, so the primary ratings stand and the
disagreements are documented rather than resolved by change.

Two mechanism-level distinctions are provisional. V3 (per-output
workload certificates) is operationally adjacent to V4 (tool-call
signing) and V1 (TEE attestation), and may collapse into V4 if vendors
productise a single primitive; the convergence argument does not depend
on V3 surviving as distinct. V5 carries a split rating (deployable as
descriptive evaluation reporting, research-stage as cryptographic
attestation) rather than a single rating, which is honest about the
vendor evidence but complicates the headline count. Finally, the
adversary and scenario matrices rate against the present productisation
state and a near-term productisation hypothesis for
partially-productised mechanisms; they are snapshots of a fast-moving
commercial surface, dated to the May 2026 vendor documentation, and will
require re-rating as the surface evolves.

\hypertarget{conclusion}{%
\subsection{Conclusion}\label{conclusion}}

Compute governance has been a governance of training because, for a
time, training compute was where capability was made. That assumption no
longer holds: capability migrates to the deployment stage through
inference scaling, agentic scaffolding, and consumer-hardware
proliferation, and the regulatory architecture has not followed. This
paper has provided the inference-time taxonomy that the literature had
the parts of but had not assembled: twenty mechanisms across monitoring,
verification, and enforcement, rated for readiness, tested against a
two-dimensional adversary model, mapped to four governance scenarios,
and connected to the hardware-level taxonomy through a substitution
analysis and a conditional substitution principle.

The central finding is a tension the paper has tried to state precisely
rather than resolve away. Commercial platforms already expose a broad set
of governance-relevant primitives, and their governance-grade coverage
falls as the demands of assurance, adversarial control, and local
deployment rise: the deployable substrate for inference-time governance
is more mature than the regulatory conversation assumes, and
its coverage of capable, non-cooperative adversaries is thinner than the
deployment count suggests. Both facts describe the same evidence. The
work that follows from them is concrete: productise the verification
cluster, which is the binding constraint in every governance scenario at
once; treat inference-time governance as one layer in a stack that
includes hardware-level governance and export controls rather than as a
substitute for them; and combine the governance scenarios so that each
covers what the others structurally cannot. The taxonomy is the start of
that work, and the most useful thing it offers a regulator is the
ability to tell which of its readiness claims applies to the threat
actually being governed.

\hypertarget{acknowledgements}{%
\section*{Acknowledgements}\label{acknowledgements}}
\addcontentsline{toc}{section}{Acknowledgements}

The author thanks Dr.~Muhammad Babar Imtiaz of the Software Research
Institute, Technological University of the Shannon (Athlone Campus,
Athlone, Co.~Westmeath, Ireland), whose research spans robotics, smart
manufacturing, artificial intelligence, and cyber security, for serving
as the independent second rater in the inter-rater reliability check
reported in Appendices B and C. Dr.~Imtiaz re-rated a randomly sampled
subset of the mechanism readiness ratings from a self-contained brief,
without prior involvement in the paper and without sight of the author's
own ratings until after his were submitted. That independence is what
allows the taxonomy's readiness ratings to be reported with a measured
agreement statistic rather than as a single author's judgement. Any
errors that remain are the author's own.

\appendix
The appendices follow: Appendix A (the mechanism-selection decision rule
and the rejected and merged candidates), Appendix B (the inter-rater
reliability methodology), Appendix C (the disagreement resolution),
Appendix D (the vendor documentation snapshots, accessed 22 May 2026,
with the Azure exclusion rationale), Appendix E (the full populated
adversary matrix, twenty mechanisms by twelve capability-by-role cells),
and Appendix F (the full populated scenario matrix, twenty mechanisms by
four scenarios).

\hypertarget{appendix-a-mechanism-selection-and-the-rejection-list}{%
\section{Appendix A: Mechanism Selection and the Rejection
List}\label{appendix-a-mechanism-selection-and-the-rejection-list}}

The twenty mechanisms in Section 3 were selected from a
twenty-seven-candidate list assembled during the corpus review, by
applying a decision rule fixed before the rule was applied to any
candidate. This appendix documents the rule, the candidates that did not
survive it, and the reconciliation of the final count.

\hypertarget{the-decision-rule}{%
\subsection{The decision rule}\label{the-decision-rule}}

A candidate was retained if and only if it satisfied a two-source rule
and passed a distinctness check.

The two-source rule required each retained mechanism to have at least
one academic source from the corpus that operationalises the mechanism,
and at least one vendor source: a documented vendor primitive, or a
documented absence with adversary-model implications. A candidate with
academic support but no vendor evidence is research-stage and is
recorded here in Appendix A rather than given a taxonomy slot; a
candidate with vendor productisation but no academic source is a vendor
finding without a mechanism slot.

The distinctness check required each candidate to be operationally
distinct from every other retained mechanism. Two candidates are not
distinct if the same regulatory action would implement both, if the same
vendor primitive instantiates both, or if the academic literature treats
them interchangeably. Where two candidates failed the distinctness
check, the one with stronger four-vendor evidence was retained and the
other was merged into its description.

The rule does not retain a candidate on grounds of being important or
interesting alone. This is the binding constraint: without it, the
selection reduces to author intuition.

\hypertarget{candidates-not-retained}{%
\subsection{Candidates not retained}\label{candidates-not-retained}}

Two candidates failed the two-source rule on the vendor side and are
dropped outright:

Speculative-decoding draft-model accounting. This transfers from the
companion hardware paper and from the test-time-scaling literature, but
no vendor productises draft-model speculative decoding as a billable or
monitorable primitive. OpenAI's Predicted Outputs is conceptually
adjacent but caller-supplied rather than draft-model-supplied, a
structurally different mechanism; it appears as a footnote in Section 4
rather than as a slot.

Intent-verified delegation chains. The academic support is a single
source proposing an intent-preserving delegation protocol, and no vendor
productises it. The underlying threat (multi-agent intent drift) is
real, but the proposed mechanism is research-stage, and the
authenticated-delegation primitives that do exist are operationally
distinct from it.

Three further candidates have evidence on both halves of the two-source
rule but fail the distinctness check and are merged rather than dropped:

Deployment-layer governance artefacts (the Policy Cards format) are
merged into E5. The artefact has no meaning without an interpreting
runtime, and that runtime is the policy-as-code enforcement engine
already captured as E5; no vendor productises the artefact as distinct
from the runtime.

The capability-update lifecycle (validate, sandbox, shadow, gate,
rollback) is merged into E1 and E7. The pattern composes existing
primitives (sandboxing is E6, gating is E5, rollback is E7) and is
methodologically interesting but is not a distinct mechanism.

Information-flow tracking with taint propagation is merged into E5 and
E6. It applies policy enforcement and tool authorisation across trust
boundaries with explicit taint semantics, an implementation pattern for
cross-boundary enforcement rather than a separate mechanism.

Two candidates are research-stage and are folded into a retained
mechanism's description rather than given their own slot:

White-box latent monitoring (residual-stream distance monitoring) is
folded into V7 as the white-box internal-state variant, contrasting with
V7's primary black-box output-trace framing. Readers interested in it
should treat V7 as the parent mechanism.

Longitudinal trust-factor scoring is folded into E4 and E5 as an
enforcement-modulator pattern. It adapts enforcement intensity on the
basis of longitudinal behaviour and is implementable as a configurable
parameter of rate limiting or policy enforcement rather than as a
distinct mechanism.

\hypertarget{reconciliation-of-the-count}{%
\subsection{Reconciliation of the
count}\label{reconciliation-of-the-count}}

The selection budget anticipated a symmetric structure mirroring the
companion hardware paper's seven monitoring, six verification, and seven
enforcement mechanisms. After the rule was applied, the monitoring
cluster yielded six mechanisms rather than seven, and the verification
cluster gained one mechanism (V7, chain-of-thought monitorability) that
has no companion-paper analogue. The final distribution is six
monitoring, seven verification, and seven enforcement mechanisms, for a
total of twenty.

We report twenty rather than forcing the count to a target. The
methodology already records V7 as a one-mechanism extension beyond the
companion taxonomy; the symmetric admission that the monitoring cluster
is one mechanism short of that taxonomy is the consistent move. The
count is driven by what the four-vendor evidence base supports, not by a
structural target.

\hypertarget{appendix-b-inter-rater-reliability-methodology}{%
\section{Appendix B: Inter-Rater Reliability
Methodology}\label{appendix-b-inter-rater-reliability-methodology}}

\hypertarget{b.1-design}{%
\subsection{Design}\label{b.1-design}}

To test whether the readiness ratings in Section 3 are reproducible
rather than idiosyncratic to a single rater, we conducted a second-rater
reliability check. Seven of the twenty mechanisms were drawn by random
selection from the final mechanism list, with the selection fixed before
either rater's ratings were examined. The seven mechanisms span all
three categories: from Verification, V1 (TEE-bounded inference
attestation), V2 (zero-knowledge proofs of inference), V4 (tool-call
cryptographic signing), and V7 (chain-of-thought monitorability); from
Monitoring, M3 (KV-cache and reasoning-token accounting); and from
Enforcement, E2 (output filtering and capability gating) and E4 (rate
limiting).

The second rater is an engineer with a reinforcement-learning and IoT
background and no prior involvement in the paper. The choice of an
out-of-field but technically expert rater is deliberate: the rating
scale is a technology-readiness judgement (``can this primitive be
deployed today, soon, only after research, or not at all''), not a
domain-specific governance judgement, and a rater without the author's
framing is a stronger test of whether the readiness signal is legible
from the mechanism descriptions alone.

\hypertarget{b.2-independence-controls}{%
\subsection{Independence
controls}\label{b.2-independence-controls}}

The author's ratings on the seven mechanisms were assigned and sealed on
20 June 2026, dated and stored separately, before the second rater's
ratings were received or examined. The second rater received a
self-contained brief (received 19 June, submitted 23 June 2026)
containing the four-point scale, the conservative-rating rule, and a
self-contained operational description of each of the seven mechanisms
with vendor and academic context, but not the author's ratings, not the
paper, and not the rating of the worked example beyond the single
illustrative case (M1, which is not among the seven). The author's
sealed sheet additionally recorded, for each mechanism, an anticipated
direction of divergence; these anticipations were sealed at the same
time as the ratings and are reported in Appendix C only after the fact.

\hypertarget{b.3-rating-scale-and-tie-breaking}{%
\subsection{Rating scale and
tie-breaking}\label{b.3-rating-scale-and-tie-breaking}}

Both raters used the identical four-point scale (1 = currently
deployable, 2 = near-term, 3 = requires research and development, 4 =
speculative) and the identical conservative-rating rule: when uncertain
between two adjacent ratings, choose the higher (more conservative)
number, and in particular rate against the relevant adversary rather
than only against a cooperative actor.

\hypertarget{b.4-agreement-statistic}{%
\subsection{Agreement statistic}\label{b.4-agreement-statistic}}

The two raters agreed exactly on five of the seven mechanisms (V1, V2,
M3, E2, E4) and differed by one point on two (V4 and V7), with the
second rater one step more conservative in each case. No disagreement
exceeded one point.

We report Cohen's kappa with quadratic weighting, as pre-specified in
the rater brief; quadratic weighting is appropriate because adjacent
disagreements on an ordered four-point scale are less serious than
far-apart ones, and the brief committed to this choice before any
ratings were compared. The quadratic-weighted kappa is 0.74, which falls
in the ``substantial'' band under the Landis and Koch (1977)
interpretation.

We disclose three caveats so that this figure is not over-read. First,
the estimate is sensitive to the weighting choice: the unweighted kappa
for the same data is 0.53 (the ``moderate'' band), and the
linear-weighted value is 0.63. The substantial-versus-moderate
characterisation therefore depends on a methodological decision made in
advance rather than on a robust property of the data. Second, the
estimate is fragile at a sample of seven: a single rating change moves
the weighted statistic by roughly 0.13 (had the raters also agreed on
V4, the weighted kappa would be 0.87). Third, the marginal distribution
is skewed, with the ``currently deployable'' rating accounting for five
of the author's seven ratings; kappa is known to behave unstably under
skewed marginals (the prevalence problem), which further counsels
against treating the point estimate as precise. We therefore read the
reliability check as corroborating the ratings rather than as a decisive
validation, and we report the raw agreement (five of seven exact, two of
seven one-step, zero of seven more than one step) alongside the kappa,
since the raw agreement is the more transparent summary at this sample
size.

\hypertarget{appendix-c-disagreement-resolution}{%
\section{Appendix C: Disagreement
Resolution}\label{appendix-c-disagreement-resolution}}

Two of the seven mechanisms produced a one-point disagreement. The brief
committed to a formal resolution procedure (re-reading source material
and adopting the more conservative rating) only for disagreements
greater than one point; both disagreements here are one-point, so the
primary ratings stand, and this appendix documents the disagreements
rather than revising the ratings. We discuss each, because each is
substantively informative about where the taxonomy's readiness
judgements are genuinely contestable.

\hypertarget{c.1-v4-tool-call-cryptographic-signing-author-1-second-rater-2}{%
\subsection{V4: tool-call cryptographic signing (author 1, second
rater
2)}\label{c.1-v4-tool-call-cryptographic-signing-author-1-second-rater-2}}

The author rated V4 currently deployable on the basis that two of four
vendors productise the signing primitive at production scale
(Anthropic's signed thinking-block field, compatible across Anthropic,
Bedrock, and Vertex surfaces; Vertex's \texttt{thought\_\allowbreak{}signature} on
Gemini 3, with stricter enforcement), and that a regulator mandating V4
could be complied with by the other vendors adopting an existing
primitive. The second rater rated V4 near-term, reasoning that the full
mechanism described (binding tool calls, arguments, and reasoning traces
into a single verifiable provenance chain) is not yet widely deployed,
and that the remaining gap, though largely engineering and
standardisation rather than research, places the complete workflow short
of ``currently deployable.''

The disagreement is a genuine and narrow one about the threshold for the
deployable rating: whether two-of-four-vendor productisation of the core
primitive suffices (the author's position), or whether the deployable
rating should require the complete end-to-end provenance chain to be in
production (the second rater's position). The author's sealed sheet
anticipated exactly this disagreement (``Babar may rate 2 if he weights
uniformity-across-vendors heavily''). The author's rating is retained
because the signing primitive itself is in production at frontier scale
at two vendors and is the component on which the provenance chain is
built; the second rater's more conservative reading is recorded as a
legitimate alternative threshold judgement, and the one-step gap is
below the revision threshold. The disagreement does not affect V4's role
in the analysis, which turns on the signing primitive's robustness to a
non-cooperative deployer (Section 4.3), a property both ratings grant.

\hypertarget{c.2-v7-chain-of-thought-monitorability-author-1-second-rater-2}{%
\subsection{V7: chain-of-thought monitorability (author 1, second
rater
2)}\label{c.2-v7-chain-of-thought-monitorability-author-1-second-rater-2}}

The author rated V7 currently deployable on the basis that three of four
vendors expose reasoning traces, summaries, or signed thinking at the
API or console layer, treating faithfulness (whether traces represent
the model's actual computation) as a question about the mechanism's
effectiveness rather than its deployability. The second rater rated V7
near-term, citing the difficulty of separating ``exposing reasoning
traces'' from ``trusting those traces,'' and noting that the
architectural divergence across vendors leaves uncertainty about whether
there is a single deployable primitive.

This is the more substantively interesting of the two disagreements, and
it maps onto a tension the paper addresses directly. The author's sealed
sheet anticipated both of the second rater's possible objections and
identified the faithfulness-based one as ``the more substantively
interesting disagreement''; the second rater independently flagged V7 as
the single hardest mechanism to rate and as the one most liable to
conflate two primitives (exposure versus faithful representation). The
paper's own treatment of V7 (Sections 3.2, 4.3, and 7.2) sides partly
with the second rater's instinct: it retains the deployable rating for
the exposure primitive but states plainly that V7 reaches adequate
adversarial coverage only against the low- and medium-capability user
and should not be relied on as a control against any more capable
adversary until faithfulness is resolved. In effect the disagreement is
reconciled not by changing the readiness rating but by the adversary
analysis, which records V7 as deployable-as-primitive and
weak-as-defence, a distinction the readiness scale alone cannot carry.
The one-step gap is below the revision threshold and the primary rating
stands.

\hypertarget{c.3-raters-free-form-observations}{%
\subsection{Rater's free-form
observations}\label{c.3-raters-free-form-observations}}

The second rater was asked two free-form questions about taxonomy
clarity. Both answers, returned independently of the author's sealed
anticipations, are recorded here because they corroborate where the
taxonomy is genuinely ambiguous. On clarity, the rater identified V7 as
the hardest to rate (the boundary between exposing reasoning traces and
trusting them) and V4 as requiring extra time (the mechanism combines
tool-call signing, trace provenance, and reasoning information). On
conflation, the rater suggested that V7 combines two related but
distinct problems (exposure versus faithful representation) and that V4
may combine tool-call signing with the broader notion of agent-trace
attestation, while judging neither issue serious enough to prevent
rating. These observations are consistent with the paper's own decision
to treat faithfulness as a separate axis from readiness for V7, and they
flag V4's internal composition as a candidate for finer decomposition in
future work, a point also noted in the Section 7.4 limitation on
provisional mechanism-level distinctions. The rater's attribution
preference was the default (anonymous in this appendix, named in the
acknowledgements).

\hypertarget{appendix-d-vendor-productisation-evidence}{%
\section{Appendix D: Vendor Productisation
Evidence}\label{appendix-d-vendor-productisation-evidence}}

The taxonomy's readiness ratings rest on a documented four-vendor
evidence base (AWS Bedrock, Anthropic, OpenAI, Google Vertex AI). Azure
OpenAI Service is excluded as structurally overlapping: it resells
OpenAI models under Microsoft's infrastructure and governance surface,
so its inference-time primitives are a subset of the union of the OpenAI
and cloud-provider rows already tabulated rather than an independent
fifth source. The tables below give the per-mechanism vendor features
for the three vendors curated to phase-2 standard (Anthropic, OpenAI,
Google Vertex AI), each with the documentation URL, the access date of
the captured snapshot, a match-confidence rating, and the adversary
scope against which the feature was assessed. AWS Bedrock evidence was
captured at survey tier (keyword inventory across the same documentation
corpus) and is summarised in the accompanying grep tables rather than
reproduced row-by-row here.

All snapshots were captured on the access date shown. Where a vendor
documents the absence of a primitive, that absence is recorded as
evidence with its adversary-model implication rather than omitted.

The tables cover the 16 mechanisms for which the phase-2 evidence base
contains at least one per-mechanism vendor row. Four mechanisms (M2
workload classification, M6 per-inference energy monitoring, V2
zero-knowledge proofs of inference, V3 hardware-backed workload
certificates) carry no per-mechanism row; their readiness ratings and
the cross-vendor or substrate-level evidence on which those ratings rest
are stated in their Section 3 entries. The phase-2 data also contains
rows for two candidates that the taxonomy merges rather than lists as
mechanisms (CUL, DLA; see Appendix A): their evidence is folded into E5
(deployment-layer governance artefacts) and into E1 and E7 (the
capability-update lifecycle) respectively, and is not rendered as a
separate table here.

\newgeometry{margin=1cm}
\begin{landscape}
\setlength{\emergencystretch}{3em}\spaceskip=3.5pt plus 2pt minus 1pt
\hypertarget{m1-per-query-flop-and-token-accounting}{%
\subsection{M1: Per-query usage and token accounting}\label{m1-per-query-flop-and-token-accounting}}

{\scriptsize\begin{longtable}[]{@{}
  >{\raggedright\arraybackslash}p{(\linewidth - 14\tabcolsep) * \real{0.07}}
  >{\raggedright\arraybackslash}p{(\linewidth - 14\tabcolsep) * \real{0.11}}
  >{\raggedright\arraybackslash}p{(\linewidth - 14\tabcolsep) * \real{0.30}}
  >{\raggedright\arraybackslash}p{(\linewidth - 14\tabcolsep) * \real{0.24}}
  >{\raggedright\arraybackslash}p{(\linewidth - 14\tabcolsep) * \real{0.10}}
  >{\raggedright\arraybackslash}p{(\linewidth - 14\tabcolsep) * \real{0.06}}
  >{\raggedright\arraybackslash}p{(\linewidth - 14\tabcolsep) * \real{0.05}}
  >{\raggedright\arraybackslash}p{(\linewidth - 14\tabcolsep) * \real{0.07}}
  @{}}
\toprule\noalign{}
\begin{minipage}[b]{\linewidth}\raggedright
Vendor
\end{minipage} & \begin{minipage}[b]{\linewidth}\raggedright
Feature
\end{minipage} & \begin{minipage}[b]{\linewidth}\raggedright
Documented capability
\end{minipage} & \begin{minipage}[b]{\linewidth}\raggedright
Documented limit
\end{minipage} & \begin{minipage}[b]{\linewidth}\raggedright
Adversary scope
\end{minipage} & \begin{minipage}[b]{\linewidth}\raggedright
Accessed
\end{minipage} & \begin{minipage}[b]{\linewidth}\raggedright
Conf.
\end{minipage} & \begin{minipage}[b]{\linewidth}\raggedright
Source
\end{minipage} \\
\midrule\noalign{}
\endhead
\bottomrule\noalign{}
\endlastfoot
Anthropic & Rate Limits / Usage tier accounting & Per-call token
accounting with cache-aware ITPM: API distinguishes input\_\allowbreak{}tokens
(post-cache), cache\_\allowbreak{}creation\_\allowbreak{}input\_\allowbreak{}tokens,
cache\_\allowbreak{}read\_\allowbreak{}input\_\allowbreak{}tokens; un\ldots{} & Tier 1 limits: 50 RPM; ITPM
30k-500k depending on model; OTPM 8k-80k. Tier 4: 4000 RPM, ITPM up to
10M, OTPM up to 800k\ldots. & Cooperative & 2026-05-22 & High &
\href{https://platform.claude.com/docs/en/support/rate-limits}{link} \\
Anthropic & Token counting API & Pre-flight token counting via separate
API endpoint returning the input-token count before submitting an
inference request. Allows clients to budget a\ldots{} & Token counts
apply to text inputs only; image and tool-result token counts
approximated. & Cooperative & 2026-05-22 & High &
\href{https://platform.claude.com/docs/en/build-with-claude/token-counting}{link} \\
Anthropic & Monitoring Usage and Cost API & Programmatic usage/\allowbreak{}cost
reporting endpoint; per-workspace and per-API-key breakdowns; per-model
and per-feature aggregation. & Aggregation latency not specified;
granularity below daily not documented. & Cooperative & 2026-05-22 &
High &
\href{https://platform.claude.com/docs/en/manage-claude/usage-and-cost-api}{link} \\
Anthropic & Extended thinking token usage breakdown &
usage.output\_\allowbreak{}tokens\_\allowbreak{}details.thinking\_\allowbreak{}tokens field reports the
internal-reasoning portion of billed output tokens. Distinct from text
output tokens. Wh\ldots{} & Streaming-side breakdown is end-of-message
only; no per-chunk thinking-token accounting. & Cooperative & 2026-05-22
& High &
\href{https://platform.claude.com/docs/en/build-with-claude/extended-thinking}{link} \\
Google Vertex AI & usageMetadata in response + per-metric quotas &
Per-call usageMetadata object returned in API responses
(promptTokenCount, totalTokenCount). Quotas measured per metric
(e.g.~embed\_\allowbreak{}content\_\allowbreak{}input\_\allowbreak{}toke\ldots{} & Specific token-accounting
fields tied to API version; legacy SDKs may surface different field
names. & Cooperative & 2026-05-22 & High &
\href{https://docs.cloud.google.com/gemini-enterprise-agent-platform/models/quotas}{link} \\
Google Vertex AI & Code execution token accounting & Code execution tool
returns separate accounting for code-execution inputs and outputs
distinct from text generation tokens. & Code-execution billing semantics
tied to sandbox tier and language runtime; specific costs in pricing
page. & Cooperative & 2026-05-22 & Medium &
\href{https://docs.cloud.google.com/gemini-enterprise-agent-platform/models/tools/code-execution}{link} \\
Google Vertex AI & Context cache token accounting & Context cache
primitive returns usage delta between cached and uncached prefixes;
per-cache token-count tracking. & Minimum cache size constraints
(model-specific); cached-token billing rate distinct from base input
rate. & Cooperative & 2026-05-22 & High &
\href{https://docs.cloud.google.com/gemini-enterprise-agent-platform/models/context-cache/context-cache-overview}{link} \\
OpenAI & Rate-limit usage tiers + RPM/\allowbreak{}RPD/\allowbreak{}TPM/\allowbreak{}TPD/\allowbreak{}IPM metrics +
per-o\ldots{} & Per-organization usage tiers (Free, Tier 1-5) with
documented qualification thresholds (\$5 / \$50 / \$100 / \$250 / \$1000
paid). Six rate-limit dimension\ldots{} & Usage limits ceiling at Tier 5
is \$200,000/\allowbreak{}month. Tier auto-graduation is opaque (only the threshold
is published; the e\ldots{} & Cooperative & 2026-05-22 & High &
\href{https://developers.openai.com/api/docs/guides/rate-limits}{link} \\
OpenAI & Counting tokens utility / tokenizer libraries & Pre-flight
token counting via tokenizer libraries (tiktoken) and token-count
utilities; lets clients budget against TPM rate limits. & Tokenizer is
BPE-specific per model family; image/\allowbreak{}audio token approximation
conventions documented but exact conversion\ldots{} & Cooperative &
2026-05-22 & High &
\href{https://developers.openai.com/api/docs/guides/counting-tokens}{link} \\
OpenAI & Reasoning model usage breakdown (reasoning\_\allowbreak{}tokens) &
usage.output\_\allowbreak{}tokens\_\allowbreak{}details.reasoning\_\allowbreak{}tokens field on reasoning-model
responses reports internal-reasoning portion of billed output tokens.
Distinct l\ldots{} & Reasoning summaries (not raw CoT) are exposed;
reasoning\_\allowbreak{}tokens count is provided but underlying tokens are not
visible\ldots{} & Cooperative + deployer-as-observer & 2026-05-22 & High
&
\href{https://developers.openai.com/api/docs/guides/reasoning}{link} \\
OpenAI & Conversation state / Responses API stored state & Server-side
state via Responses API (store parameter) provides per-conversation
persistence with associated billing. State tracking exposes per-turn
t\ldots{} & store=false disables persistence (no state); enabling state
has retention implications subject to data-retention control\ldots{} &
Cooperative & 2026-05-22 & High &
\href{https://developers.openai.com/api/docs/guides/conversation-state}{link} \\
OpenAI & Pricing (per-model per-token rates) & Per-model published rates
for input tokens, output tokens, reasoning tokens, cached tokens.
Distinct rates for batch, priority, flex tiers. & Per-model rates change
without API contract change; cited rates have implicit accessed-date
validity. & Cooperative & 2026-05-22 & High &
\href{https://developers.openai.com/api/docs/pricing}{link} \\
\end{longtable}}

\hypertarget{m3-kv-cache-and-reasoning-token-accounting}{%
\subsection{M3: KV-cache and reasoning-token
accounting}\label{m3-kv-cache-and-reasoning-token-accounting}}

{\scriptsize\begin{longtable}[]{@{}
  >{\raggedright\arraybackslash}p{(\linewidth - 14\tabcolsep) * \real{0.07}}
  >{\raggedright\arraybackslash}p{(\linewidth - 14\tabcolsep) * \real{0.11}}
  >{\raggedright\arraybackslash}p{(\linewidth - 14\tabcolsep) * \real{0.30}}
  >{\raggedright\arraybackslash}p{(\linewidth - 14\tabcolsep) * \real{0.24}}
  >{\raggedright\arraybackslash}p{(\linewidth - 14\tabcolsep) * \real{0.10}}
  >{\raggedright\arraybackslash}p{(\linewidth - 14\tabcolsep) * \real{0.06}}
  >{\raggedright\arraybackslash}p{(\linewidth - 14\tabcolsep) * \real{0.05}}
  >{\raggedright\arraybackslash}p{(\linewidth - 14\tabcolsep) * \real{0.07}}
  @{}}
\toprule\noalign{}
\begin{minipage}[b]{\linewidth}\raggedright
Vendor
\end{minipage} & \begin{minipage}[b]{\linewidth}\raggedright
Feature
\end{minipage} & \begin{minipage}[b]{\linewidth}\raggedright
Documented capability
\end{minipage} & \begin{minipage}[b]{\linewidth}\raggedright
Documented limit
\end{minipage} & \begin{minipage}[b]{\linewidth}\raggedright
Adversary scope
\end{minipage} & \begin{minipage}[b]{\linewidth}\raggedright
Accessed
\end{minipage} & \begin{minipage}[b]{\linewidth}\raggedright
Conf.
\end{minipage} & \begin{minipage}[b]{\linewidth}\raggedright
Source
\end{minipage} \\
\midrule\noalign{}
\endhead
\bottomrule\noalign{}
\endlastfoot
Anthropic & Prompt caching & Cache breakpoints in messages;
per-cache-breakpoint TTLs (5-minute default, longer TTLs available);
cache\_\allowbreak{}creation\_\allowbreak{}input\_\allowbreak{}tokens, cache\_\allowbreak{}read\_\allowbreak{}input\_\allowbreak{}toke\ldots{} &
5-minute default TTL; minimum cache breakpoint size (1024 tokens
typically); breakpoint count per message limited; cache\ldots{} &
Cooperative & 2026-05-22 & High &
\href{https://platform.claude.com/docs/en/build-with-claude/prompt-caching}{link} \\
Anthropic & Cache diagnostics (beta) & Diagnostic endpoint reporting
cache utilization, hit rate, breakpoint efficiency. Per-workspace and
per-API-key breakdowns. & Beta feature; metric semantics may change.
Aggregation latency not specified. & Cooperative & 2026-05-22 & Medium &
\href{https://platform.claude.com/docs/en/build-with-claude/cache-diagnostics}{link} \\
Anthropic & Extended thinking (thinking-token billing) &
thinking\_\allowbreak{}tokens field in usage object; separately billed from text
output tokens. Configurable budget via thinking.budget\_\allowbreak{}tokens (manual
mode) or thin\ldots{} & Manual budget mode deprecated on Opus 4.6/\allowbreak{}Sonnet
4.6; removed on Opus 4.7/4.8/\allowbreak{}Fable/\allowbreak{}Mythos (adaptive only). max\_\allowbreak{}tokens
d\ldots{} & Cooperative & 2026-05-22 & High &
\href{https://platform.claude.com/docs/en/build-with-claude/extended-thinking}{link} \\
Anthropic & Adaptive thinking + Effort parameter + Task budgets & effort
parameter controls thinking depth on adaptive-thinking models; task
budgets (beta) impose per-task token ceilings across turns. Server-side
enf\ldots{} & Task budgets are beta; concrete budget ceiling semantics
not fully documented in public docs. & Cooperative & 2026-05-22 & High &
\href{https://platform.claude.com/docs/en/build-with-claude/adaptive-thinking}{link} \\
Anthropic & Context window + Compaction + Context editing & Conversation
context management primitives: explicit context window per model;
server-side compaction; mid-conversation system messages and context
ed\ldots{} & Per-model context window caps (e.g., 200k-1M depending on
model class); compaction policy not user-controllable below hi\ldots{} &
Cooperative & 2026-05-22 & Medium &
\href{https://platform.claude.com/docs/en/build-with-claude/context-windows}{link} \\
Google Vertex AI & Context caching (Gemini-native) & Explicit
context-cache primitive (create, use, get info, update, delete). Caller
manages cache lifecycle. Supports caching across multi-turn
conversat\ldots{} & Per-model support varies (table in
security-controls); CMEK support per-model; cache lifetime configurable
but bounded. & Cooperative & 2026-05-22 & High &
\href{https://docs.cloud.google.com/gemini-enterprise-agent-platform/models/context-cache/context-cache-overview}{link} \\
Google Vertex AI & Thinking models with thinking\_\allowbreak{}level parameter &
Thinking enabled by default on supported models. Gemini 3 introduces
thinking\_\allowbreak{}level parameter (MINIMAL/\allowbreak{}LOW/\allowbreak{}MEDIUM/\allowbreak{}HIGH). Gemini 2.5 Flash
uses thinkin\ldots{} & Coarse-grained level enum on Gemini 3+ replaces
fine-grained budget. Pro models cannot disable thinking; Flash models
ca\ldots{} & Cooperative + deployer-as-observer & 2026-05-22 & High &
\href{https://docs.cloud.google.com/gemini-enterprise-agent-platform/models/thinking}{link} \\
Google Vertex AI & Prompt caching for Claude on Vertex AI & Claude
prompt-caching primitive accessible via Vertex AI route; per-cache
lifecycle similar to Anthropic API but exposed through Vertex AI
authenticat\ldots{} & Inherits Anthropic prompt-caching model semantics;
per-region availability follows Vertex AI region map for Claude. &
Cooperative & 2026-05-22 & Medium &
\href{https://docs.cloud.google.com/gemini-enterprise-agent-platform/models/partner-models/claude/prompt-caching}{link} \\
OpenAI & Prompt caching (in-memory + extended retention) & Automatic
caching for prompts $\geq$\$1024 tokens; cache routing by initial-prefix
hash. Two retention policies: in-memory (5-10 min idle, max 1 hour) and
ex\ldots{} & Cache routing is based on initial 256-token-hash typically;
\textgreater15 RPM same-prefix may overflow to additional machines
redu\ldots{} & Cooperative & 2026-05-22 & High &
\href{https://developers.openai.com/api/docs/guides/prompt-caching}{link} \\
OpenAI & Reasoning models (reasoning\_\allowbreak{}effort parameter,
reasoning\_\allowbreak{}toke\ldots{} & reasoning parameter on o-series and gpt-5.5
models with effort levels (low / medium / high). Reasoning tokens
consumed before final response; billed s\ldots{} & Effort is
coarse-grained (3 levels); no per-task token budget like Anthropic's
manual budget\_\allowbreak{}tokens. Raw CoT is NOT expo\ldots{} & Cooperative +
deployer-as-observer & 2026-05-22 & High &
\href{https://developers.openai.com/api/docs/guides/reasoning}{link} \\
OpenAI & Reasoning best practices (when-and-how-to-use) & Documented
guidance for selecting reasoning effort levels per task type.
Cost/\allowbreak{}latency trade-off explicit: reasoning models are slower and more
expensi\ldots{} & Guidance, not enforcement. Customer-side cost control
via reasoning\_\allowbreak{}effort selection. & Cooperative & 2026-05-22 & Medium &
\href{https://developers.openai.com/api/docs/guides/reasoning-best-practices}{link} \\
OpenAI & Compaction (Responses API) & Server-side compaction of
conversation history when approaching context limit; trades older
content for newer. & Compaction policy not user-tunable beyond
enable/\allowbreak{}disable; what is preserved vs compacted is opaque. & Cooperative
& 2026-05-22 & Medium &
\href{https://developers.openai.com/api/docs/guides/compaction}{link} \\
OpenAI & Predicted Outputs & Pre-supplied predicted output for
completion (speculative-style execution): caller provides expected
output text, model verifies/\allowbreak{}edits. Reduces latenc\ldots{} & Effective
only when prediction substantially matches output; otherwise billed but
no benefit. Not a general M4 (speculat\ldots{} & Cooperative &
2026-05-22 & Medium &
\href{https://developers.openai.com/api/docs/guides/predicted-outputs}{link} \\
\end{longtable}}

\hypertarget{m4-distributed-inference-monitoring-across-data-centres}{%
\subsection{M4: Distributed inference monitoring across data
centres}\label{m4-distributed-inference-monitoring-across-data-centres}}

{\scriptsize\begin{longtable}[]{@{}
  >{\raggedright\arraybackslash}p{(\linewidth - 14\tabcolsep) * \real{0.07}}
  >{\raggedright\arraybackslash}p{(\linewidth - 14\tabcolsep) * \real{0.11}}
  >{\raggedright\arraybackslash}p{(\linewidth - 14\tabcolsep) * \real{0.30}}
  >{\raggedright\arraybackslash}p{(\linewidth - 14\tabcolsep) * \real{0.24}}
  >{\raggedright\arraybackslash}p{(\linewidth - 14\tabcolsep) * \real{0.10}}
  >{\raggedright\arraybackslash}p{(\linewidth - 14\tabcolsep) * \real{0.06}}
  >{\raggedright\arraybackslash}p{(\linewidth - 14\tabcolsep) * \real{0.05}}
  >{\raggedright\arraybackslash}p{(\linewidth - 14\tabcolsep) * \real{0.07}}
  @{}}
\toprule\noalign{}
\begin{minipage}[b]{\linewidth}\raggedright
Vendor
\end{minipage} & \begin{minipage}[b]{\linewidth}\raggedright
Feature
\end{minipage} & \begin{minipage}[b]{\linewidth}\raggedright
Documented capability
\end{minipage} & \begin{minipage}[b]{\linewidth}\raggedright
Documented limit
\end{minipage} & \begin{minipage}[b]{\linewidth}\raggedright
Adversary scope
\end{minipage} & \begin{minipage}[b]{\linewidth}\raggedright
Accessed
\end{minipage} & \begin{minipage}[b]{\linewidth}\raggedright
Conf.
\end{minipage} & \begin{minipage}[b]{\linewidth}\raggedright
Source
\end{minipage} \\
\midrule\noalign{}
\endhead
\bottomrule\noalign{}
\endlastfoot
Anthropic & inference\_\allowbreak{}geo parameter / Claude on cloud platforms &
inference\_\allowbreak{}geo parameter (us, global) controls jurisdiction of inference
servers. Rate limits shared across geo values. Cloud-platform variants
(Bedroc\ldots{} & Rate limits NOT separated by inference\_\allowbreak{}geo (us and
global draw from same pool). Cloud-platform variants have different
r\ldots{} & Cooperative & 2026-05-22 & High &
\href{https://platform.claude.com/docs/en/build-with-claude/claude-on-vertex-ai}{link} \\
Google Vertex AI & Region-based deployment + Request-response logging &
Per-region quotas; explicit data-residency assertion (prompt logs stored
in same region or multi-region as customer-selected project). Per-region
or g\ldots{} & Global location used for some operations (e.g.~SDK with
GOOGLE\_\allowbreak{}GENAI\_\allowbreak{}USE\_\allowbreak{}ENTERPRISE) may not stay within a single
region\ldots{} & Cooperative + regulator-as-stakeholder & 2026-05-22 &
High &
\href{https://docs.cloud.google.com/gemini-enterprise-agent-platform/models/capabilities/request-response-logging}{link} \\
OpenAI & Data-residency controls (closest analogue; no native
multi-r\ldots{} & No native multi-region inference monitoring on the
standard OpenAI Platform. The closest analogue is project-level
data-residency controls: an eligibl\ldots{} & Data-residency sets the
processing region; it is not per-region inference telemetry or logging
of where individual infer\ldots{} & Cooperative & 2026-05-22 & Medium &
\href{https://developers.openai.com/api/docs/guides/your-data}{link} \\
\end{longtable}}

\hypertarget{m5-api-level-kyc-and-request-tracing}{%
\subsection{M5: Authenticated account identity and request tracing}\label{m5-api-level-kyc-and-request-tracing}}

{\scriptsize\begin{longtable}[]{@{}
  >{\raggedright\arraybackslash}p{(\linewidth - 14\tabcolsep) * \real{0.07}}
  >{\raggedright\arraybackslash}p{(\linewidth - 14\tabcolsep) * \real{0.11}}
  >{\raggedright\arraybackslash}p{(\linewidth - 14\tabcolsep) * \real{0.30}}
  >{\raggedright\arraybackslash}p{(\linewidth - 14\tabcolsep) * \real{0.24}}
  >{\raggedright\arraybackslash}p{(\linewidth - 14\tabcolsep) * \real{0.10}}
  >{\raggedright\arraybackslash}p{(\linewidth - 14\tabcolsep) * \real{0.06}}
  >{\raggedright\arraybackslash}p{(\linewidth - 14\tabcolsep) * \real{0.05}}
  >{\raggedright\arraybackslash}p{(\linewidth - 14\tabcolsep) * \real{0.07}}
  @{}}
\toprule\noalign{}
\begin{minipage}[b]{\linewidth}\raggedright
Vendor
\end{minipage} & \begin{minipage}[b]{\linewidth}\raggedright
Feature
\end{minipage} & \begin{minipage}[b]{\linewidth}\raggedright
Documented capability
\end{minipage} & \begin{minipage}[b]{\linewidth}\raggedright
Documented limit
\end{minipage} & \begin{minipage}[b]{\linewidth}\raggedright
Adversary scope
\end{minipage} & \begin{minipage}[b]{\linewidth}\raggedright
Accessed
\end{minipage} & \begin{minipage}[b]{\linewidth}\raggedright
Conf.
\end{minipage} & \begin{minipage}[b]{\linewidth}\raggedright
Source
\end{minipage} \\
\midrule\noalign{}
\endhead
\bottomrule\noalign{}
\endlastfoot
Anthropic & Trust Center / compliance certifications & SOC 2 Type II,
ISO 27001, ISO 42001, CSA STAR, HIPAA (Anthropic API, Claude Enterprise;
Partner-Managed on Bedrock/\allowbreak{}Vertex), NIST 800-171, FedRAMP High\ldots{} &
FedRAMP High only via specific cloud-platform deployments (AWS GovCloud,
Google Assured Workloads); not available on sta\ldots{} & Cooperative +
compliance-relevant adversary & 2026-05-22 & High &
\href{https://trust.anthropic.com/}{link} \\
Anthropic & Privacy Policy / data retention and disclosure &
Verification Data category (ID documents, biometric data, age
verification) collected in `certain circumstances.' Inputs/\allowbreak{}Outputs
retained subject to o\ldots{} & Retention duration not explicitly
bounded in policy; data-subject access rights jurisdiction-specific.
Inputs flagged fo\ldots{} & User-as-adversary, regulator-as-adversary &
2026-05-22 & Medium &
\href{https://anthropic.com/legal/privacy}{link} \\
Google Vertex AI & Cloud Audit Logs (Data Access audit logs) &
Per-project Cloud Audit Logs (Admin Activity always-on, Data Access
enabled separately). Audit log entries captured for AI Platform service
operations\ldots{} & Data Access audit logs not enabled by default;
opt-in per project. Retention default 30-400 days depending on log
bucket\ldots{} & Cooperative + regulator-as-stakeholder & 2026-05-22 &
High &
\href{https://docs.cloud.google.com/gemini-enterprise-agent-platform/models/enable-audit-logs}{link} \\
Google Vertex AI & Abuse monitoring with two-tier prompt logging &
Standard tier: prompt logs retained 90 days for AUP investigation when
classifier flags suspicious activity. Not used for training. Advanced AI
Safety\ldots{} & Opt-out via Google form for standard tier; Master
Agreement customers exempt by default. Advanced AI Safety logging
may\ldots{} & User-as-adversary, deployer-as-adversary,
regulator-as-stakeholder & 2026-05-22 & High &
\href{https://docs.cloud.google.com/gemini-enterprise-agent-platform/models/abuse-monitoring}{link} \\
Google Vertex AI & Compliance resource center (Google Cloud
certifications port\ldots{} & Google Cloud's compliance resource center
documents the provider's certifications, attestations, and independent
third-party audit reports, with offer\ldots{} & The captured page is a
navigation hub: it does not enumerate specific certifications (for
example SOC, ISO, HIPAA, FedRA\ldots{} & Regulator-as-stakeholder &
2026-05-22 & Medium &
\href{https://cloud.google.com/security/compliance}{link} \\
OpenAI & Abuse monitoring logs + retention controls & Default
abuse-monitoring logs for ALL API feature usage, retained 30 days unless
longer retention legally required. Logs contain prompts, responses,
c\ldots{} & Modified Abuse Monitoring and Zero Data Retention require
sales approval and additional contractual requirements. gpt-5\ldots. &
User-as-adversary, regulator-as-stakeholder & 2026-05-22 & High &
\href{https://developers.openai.com/api/docs/guides/your-data}{link} \\
OpenAI & Admin APIs (audit logs, invites, users, projects, API keys) &
Programmatic administration via separate Admin API key class (cannot
call non-admin endpoints). Endpoints: audit logs, invites, users,
projects, API k\ldots{} & Admin API key is itself a sensitive
credential. Audit log retention depth and granularity not documented in
public docs. & Cooperative + insider-as-adversary (limited) & 2026-05-22
& High &
\href{https://developers.openai.com/api/docs/guides/admin-apis}{link} \\
\end{longtable}}

\hypertarget{v1-tee-bounded-inference-attestation}{%
\subsection{V1: TEE-bounded inference
attestation}\label{v1-tee-bounded-inference-attestation}}

{\scriptsize\begin{longtable}[]{@{}
  >{\raggedright\arraybackslash}p{(\linewidth - 14\tabcolsep) * \real{0.07}}
  >{\raggedright\arraybackslash}p{(\linewidth - 14\tabcolsep) * \real{0.11}}
  >{\raggedright\arraybackslash}p{(\linewidth - 14\tabcolsep) * \real{0.30}}
  >{\raggedright\arraybackslash}p{(\linewidth - 14\tabcolsep) * \real{0.24}}
  >{\raggedright\arraybackslash}p{(\linewidth - 14\tabcolsep) * \real{0.10}}
  >{\raggedright\arraybackslash}p{(\linewidth - 14\tabcolsep) * \real{0.06}}
  >{\raggedright\arraybackslash}p{(\linewidth - 14\tabcolsep) * \real{0.05}}
  >{\raggedright\arraybackslash}p{(\linewidth - 14\tabcolsep) * \real{0.07}}
  @{}}
\toprule\noalign{}
\begin{minipage}[b]{\linewidth}\raggedright
Vendor
\end{minipage} & \begin{minipage}[b]{\linewidth}\raggedright
Feature
\end{minipage} & \begin{minipage}[b]{\linewidth}\raggedright
Documented capability
\end{minipage} & \begin{minipage}[b]{\linewidth}\raggedright
Documented limit
\end{minipage} & \begin{minipage}[b]{\linewidth}\raggedright
Adversary scope
\end{minipage} & \begin{minipage}[b]{\linewidth}\raggedright
Accessed
\end{minipage} & \begin{minipage}[b]{\linewidth}\raggedright
Conf.
\end{minipage} & \begin{minipage}[b]{\linewidth}\raggedright
Source
\end{minipage} \\
\midrule\noalign{}
\endhead
\bottomrule\noalign{}
\endlastfoot
Google Vertex AI & Confidential Computing platform (Confidential VM +
Confident\ldots{} & Four products: Confidential VM (encryption-in-use
via hardware TEE), Confidential Space (collaborative computation on
confidential data), Google Cloud\ldots{} & Confidential VM available in
Workbench BUT NOT documented as available for Gemini inference serving
at the user-facing l\ldots{} & Deployer-as-adversary,
regulator-as-stakeholder & 2026-05-22 & Medium &
\href{https://docs.cloud.google.com/confidential-computing/docs/confidential-computing-overview}{link} \\
\end{longtable}}

\hypertarget{v4-tool-call-cryptographic-signing-and-agent-trace-attestation}{%
\subsection{V4: Tool-call cryptographic signing and agent-trace
attestation}\label{v4-tool-call-cryptographic-signing-and-agent-trace-attestation}}

{\scriptsize\begin{longtable}[]{@{}
  >{\raggedright\arraybackslash}p{(\linewidth - 14\tabcolsep) * \real{0.07}}
  >{\raggedright\arraybackslash}p{(\linewidth - 14\tabcolsep) * \real{0.11}}
  >{\raggedright\arraybackslash}p{(\linewidth - 14\tabcolsep) * \real{0.30}}
  >{\raggedright\arraybackslash}p{(\linewidth - 14\tabcolsep) * \real{0.24}}
  >{\raggedright\arraybackslash}p{(\linewidth - 14\tabcolsep) * \real{0.10}}
  >{\raggedright\arraybackslash}p{(\linewidth - 14\tabcolsep) * \real{0.06}}
  >{\raggedright\arraybackslash}p{(\linewidth - 14\tabcolsep) * \real{0.05}}
  >{\raggedright\arraybackslash}p{(\linewidth - 14\tabcolsep) * \real{0.07}}
  @{}}
\toprule\noalign{}
\begin{minipage}[b]{\linewidth}\raggedright
Vendor
\end{minipage} & \begin{minipage}[b]{\linewidth}\raggedright
Feature
\end{minipage} & \begin{minipage}[b]{\linewidth}\raggedright
Documented capability
\end{minipage} & \begin{minipage}[b]{\linewidth}\raggedright
Documented limit
\end{minipage} & \begin{minipage}[b]{\linewidth}\raggedright
Adversary scope
\end{minipage} & \begin{minipage}[b]{\linewidth}\raggedright
Accessed
\end{minipage} & \begin{minipage}[b]{\linewidth}\raggedright
Conf.
\end{minipage} & \begin{minipage}[b]{\linewidth}\raggedright
Source
\end{minipage} \\
\midrule\noalign{}
\endhead
\bottomrule\noalign{}
\endlastfoot
Anthropic & Extended thinking signature field & Cryptographic signature
field on thinking blocks used to verify that thinking blocks were
generated by Claude when passed back to the API. signature\_\allowbreak{}d\ldots{} &
Signature is opaque (not human-interpretable); rotation policy and key
infrastructure not publicly documented. Filtering\ldots{} &
Deployer-as-adversary (limited) & 2026-05-22 & High &
\href{https://platform.claude.com/docs/en/build-with-claude/extended-thinking}{link} \\
Anthropic & Strict tool use / Programmatic tool calling / Fine-grained
t\ldots{} & Server-side schema enforcement on tool calls when
strict\_\allowbreak{}tool\_\allowbreak{}use is enabled. Programmatic tool calling supports
server-side validation of tool-call s\ldots{} & Schema enforcement is
structural (does Claude conform to the declared schema) not
cryptographic. No third-party-verifiab\ldots{} & Cooperative &
2026-05-22 & Medium &
\href{https://platform.claude.com/docs/en/agents-and-tools/tool-use/strict-tool-use}{link} \\
Google Vertex AI & thought\_\allowbreak{}signature on functionCall parts &
Cryptographic thought\_\allowbreak{}signature field attached to functionCall parts in
Gemini 3 model responses. Required to be returned by client in
subsequent requ\ldots{} & Signature is opaque to caller (encrypted
internal-state representation). skip\_\allowbreak{}thought\_\allowbreak{}signature\_\allowbreak{}validator
escape hatch e\ldots{} & Deployer-as-adversary (limited),
continuity-of-reasoning enforcement & 2026-05-22 & High &
\href{https://docs.cloud.google.com/gemini-enterprise-agent-platform/models/thought-signatures}{link} \\
\end{longtable}}

\hypertarget{v5-capability-evaluation-attestation}{%
\subsection{V5: Capability-evaluation reporting}\label{v5-capability-evaluation-attestation}}

{\scriptsize\begin{longtable}[]{@{}
  >{\raggedright\arraybackslash}p{(\linewidth - 14\tabcolsep) * \real{0.07}}
  >{\raggedright\arraybackslash}p{(\linewidth - 14\tabcolsep) * \real{0.11}}
  >{\raggedright\arraybackslash}p{(\linewidth - 14\tabcolsep) * \real{0.30}}
  >{\raggedright\arraybackslash}p{(\linewidth - 14\tabcolsep) * \real{0.24}}
  >{\raggedright\arraybackslash}p{(\linewidth - 14\tabcolsep) * \real{0.10}}
  >{\raggedright\arraybackslash}p{(\linewidth - 14\tabcolsep) * \real{0.06}}
  >{\raggedright\arraybackslash}p{(\linewidth - 14\tabcolsep) * \real{0.05}}
  >{\raggedright\arraybackslash}p{(\linewidth - 14\tabcolsep) * \real{0.07}}
  @{}}
\toprule\noalign{}
\begin{minipage}[b]{\linewidth}\raggedright
Vendor
\end{minipage} & \begin{minipage}[b]{\linewidth}\raggedright
Feature
\end{minipage} & \begin{minipage}[b]{\linewidth}\raggedright
Documented capability
\end{minipage} & \begin{minipage}[b]{\linewidth}\raggedright
Documented limit
\end{minipage} & \begin{minipage}[b]{\linewidth}\raggedright
Adversary scope
\end{minipage} & \begin{minipage}[b]{\linewidth}\raggedright
Accessed
\end{minipage} & \begin{minipage}[b]{\linewidth}\raggedright
Conf.
\end{minipage} & \begin{minipage}[b]{\linewidth}\raggedright
Source
\end{minipage} \\
\midrule\noalign{}
\endhead
\bottomrule\noalign{}
\endlastfoot
Google Vertex AI & Gen AI evaluation service (Adaptive Rubrics + Static
Rubrics\ldots{} & Enterprise-grade evaluation framework. Distinctive
primitive: adaptive rubrics (per-prompt tailored pass/\allowbreak{}fail tests).
Supports model migrations, head-\ldots{} & Evaluation results are
model-graded data, not cryptographically signed attestations. No
external-verifier-readable certi\ldots{} & Cooperative & 2026-05-22 &
Medium &
\href{https://docs.cloud.google.com/gemini-enterprise-agent-platform/models/evaluation-overview}{link} \\
OpenAI & Evals API + Graders + Working with Evals + Getting started &
Programmatic evaluation framework via Evals API; eval\_\allowbreak{}id resource with
reusable graders and data sources. Multiple grader types (string match,
partial\ldots{} & Evals platform deprecated 2026-06-03; becomes
read-only Oct 31 2026, shut down Nov 30 2026. Existing eval IDs not
portab\ldots{} & Cooperative & 2026-05-22 & Medium &
\href{https://developers.openai.com/api/docs/guides/evals}{link} \\
\end{longtable}}

\hypertarget{v6-retrieval-corpus-and-rag-attestation}{%
\subsection{V6: Retrieval provenance and citation logging}\label{v6-retrieval-corpus-and-rag-attestation}}

{\scriptsize\begin{longtable}[]{@{}
  >{\raggedright\arraybackslash}p{(\linewidth - 14\tabcolsep) * \real{0.07}}
  >{\raggedright\arraybackslash}p{(\linewidth - 14\tabcolsep) * \real{0.11}}
  >{\raggedright\arraybackslash}p{(\linewidth - 14\tabcolsep) * \real{0.30}}
  >{\raggedright\arraybackslash}p{(\linewidth - 14\tabcolsep) * \real{0.24}}
  >{\raggedright\arraybackslash}p{(\linewidth - 14\tabcolsep) * \real{0.10}}
  >{\raggedright\arraybackslash}p{(\linewidth - 14\tabcolsep) * \real{0.06}}
  >{\raggedright\arraybackslash}p{(\linewidth - 14\tabcolsep) * \real{0.05}}
  >{\raggedright\arraybackslash}p{(\linewidth - 14\tabcolsep) * \real{0.07}}
  @{}}
\toprule\noalign{}
\begin{minipage}[b]{\linewidth}\raggedright
Vendor
\end{minipage} & \begin{minipage}[b]{\linewidth}\raggedright
Feature
\end{minipage} & \begin{minipage}[b]{\linewidth}\raggedright
Documented capability
\end{minipage} & \begin{minipage}[b]{\linewidth}\raggedright
Documented limit
\end{minipage} & \begin{minipage}[b]{\linewidth}\raggedright
Adversary scope
\end{minipage} & \begin{minipage}[b]{\linewidth}\raggedright
Accessed
\end{minipage} & \begin{minipage}[b]{\linewidth}\raggedright
Conf.
\end{minipage} & \begin{minipage}[b]{\linewidth}\raggedright
Source
\end{minipage} \\
\midrule\noalign{}
\endhead
\bottomrule\noalign{}
\endlastfoot
Anthropic & Search results / Citations & Search results: when Claude
uses web search tool or search\_\allowbreak{}results content blocks, returned content
carries source URL and title. Citations: explicit\ldots{} & Citation
accuracy is a model behavior, not a cryptographic guarantee; citations
can be incorrect (hallucinated). No tamp\ldots{} & Cooperative &
2026-05-22 & Medium &
\href{https://platform.claude.com/docs/en/build-with-claude/search-results}{link} \\
Anthropic & Web search tool / Web fetch tool & First-party server tools
for retrieval; subprocessor (TurboPuffer) documented in Trust Center.
Tool responses carry source URLs. & Subprocessor change risk
(TurboPuffer was added in May 2026); no per-query content attestation;
no corpus-integrity hash\ldots{} & Cooperative & 2026-05-22 & Medium &
\href{https://platform.claude.com/docs/en/agents-and-tools/tool-use/web-search-tool}{link} \\
Google Vertex AI & RAG Engine (Grounding overview) & First-party RAG
primitive: vector store, retrieval, citation generation. Per-RAG-Engine
quotas explicit in quotas page. Citations returned with retrie\ldots{} &
Citation accuracy is a model behavior, not cryptographic provenance. No
per-document tamper-evidence on retrieval corpus\ldots{} & Cooperative &
2026-05-22 & High &
\href{https://docs.cloud.google.com/gemini-enterprise-agent-platform/models/grounding/ground-responses-using-rag}{link} \\
Google Vertex AI & Grounding with Google Search + Grounding with Google
Maps & First-party search-grounding tool returns results with explicit
source URLs and snippet attribution. Google Maps grounding provides
geographic-source\ldots{} & Search-grounding source list not
user-curatable; relies on Google Search index. Per-query freshness
inherits from Search\ldots{} & Cooperative & 2026-05-22 & High &
\href{https://docs.cloud.google.com/gemini-enterprise-agent-platform/models/grounding/grounding-with-google-search}{link} \\
Google Vertex AI & Security controls table per-model (CMEK / VPC-SC /
Data resi\ldots{} & Four security controls explicitly tabulated
per-model per-operation: data residency at rest, CMEK, VPC Service
Controls, Access Transparency. Per-mode\ldots{} & Support varies by
model and operation (some cells in table empty). RAG Engine and
Grounding only supported under securit\ldots{} & Deployer-as-adversary,
regulator-as-stakeholder & 2026-05-22 & High &
\href{https://docs.cloud.google.com/gemini-enterprise-agent-platform/models/security-controls}{link} \\
OpenAI & Web search tool (first-party) + Citation formatting &
First-party web-search tool returns results with URL, title, snippet.
Citation formatting guidance for citing sources in model output. &
Per-query retrieval has source URLs but no cryptographic provenance.
Subprocessor for web search not enumerated in the f\ldots{} &
Cooperative & 2026-05-22 & Medium &
\href{https://developers.openai.com/api/docs/guides/web-search}{link} \\
OpenAI & Vector stores + File search (Assistants tools) + MCP for
dee\ldots{} & Vector store ingest endpoint with per-store rate limit
(300 RPM). File search tool against vector stores. MCP-for-deep-research
extends retrieval via\ldots{} & Vector store ingestion rate limited per
store ID (300 RPM); no cryptographic corpus-integrity guarantees. MCP
retrieval\ldots{} & Cooperative & 2026-05-22 & Medium &
\href{https://developers.openai.com/api/docs/guides/mcp-deep-research}{link} \\
\end{longtable}}

\hypertarget{v7-chain-of-thought-monitorability}{%
\subsection{V7: Chain-of-thought
monitorability}\label{v7-chain-of-thought-monitorability}}

{\scriptsize\begin{longtable}[]{@{}
  >{\raggedright\arraybackslash}p{(\linewidth - 14\tabcolsep) * \real{0.07}}
  >{\raggedright\arraybackslash}p{(\linewidth - 14\tabcolsep) * \real{0.11}}
  >{\raggedright\arraybackslash}p{(\linewidth - 14\tabcolsep) * \real{0.30}}
  >{\raggedright\arraybackslash}p{(\linewidth - 14\tabcolsep) * \real{0.24}}
  >{\raggedright\arraybackslash}p{(\linewidth - 14\tabcolsep) * \real{0.10}}
  >{\raggedright\arraybackslash}p{(\linewidth - 14\tabcolsep) * \real{0.06}}
  >{\raggedright\arraybackslash}p{(\linewidth - 14\tabcolsep) * \real{0.05}}
  >{\raggedright\arraybackslash}p{(\linewidth - 14\tabcolsep) * \real{0.07}}
  @{}}
\toprule\noalign{}
\begin{minipage}[b]{\linewidth}\raggedright
Vendor
\end{minipage} & \begin{minipage}[b]{\linewidth}\raggedright
Feature
\end{minipage} & \begin{minipage}[b]{\linewidth}\raggedright
Documented capability
\end{minipage} & \begin{minipage}[b]{\linewidth}\raggedright
Documented limit
\end{minipage} & \begin{minipage}[b]{\linewidth}\raggedright
Adversary scope
\end{minipage} & \begin{minipage}[b]{\linewidth}\raggedright
Accessed
\end{minipage} & \begin{minipage}[b]{\linewidth}\raggedright
Conf.
\end{minipage} & \begin{minipage}[b]{\linewidth}\raggedright
Source
\end{minipage} \\
\midrule\noalign{}
\endhead
\bottomrule\noalign{}
\endlastfoot
Anthropic & Extended thinking (thinking blocks with display modes) &
Reasoning traces returned as first-class thinking content blocks. Three
display modes: `enabled' (full text), `summarized' (model-generated
summary),\ldots{} & Default display behavior varies by model
(Fable/\allowbreak{}Mythos/\allowbreak{}Mythos Preview default to `omitted'). Summarized mode
loses raw-t\ldots{} & Deployer / regulator-as-observer (partial) &
2026-05-22 & High &
\href{https://platform.claude.com/docs/en/build-with-claude/extended-thinking}{link} \\
Anthropic & Adaptive thinking / Effort parameter & Reasoning depth
dynamically controlled via effort parameter. Model decides when to think
and how much based on each request. Monitorability extends to\ldots{} &
effort parameter is integer-coarse; no fine-grained predictability of
when thinking will be invoked. & Cooperative & 2026-05-22 & Medium &
\href{https://platform.claude.com/docs/en/build-with-claude/adaptive-thinking}{link} \\
Google Vertex AI & Thinking mode + Agent Studio thinking visibility &
Thinking process generated and viewable in Agent Studio console
alongside response. Programmatic access via thought\_\allowbreak{}signature
(encrypted) for reasonin\ldots{} & Raw thinking content displayed in
Agent Studio console; programmatic raw-trace exposure via API less
explicit than Anthr\ldots{} & Deployer-as-observer,
regulator-as-observer & 2026-05-22 & Medium &
\href{https://docs.cloud.google.com/gemini-enterprise-agent-platform/models/thinking}{link} \\
OpenAI & Reasoning models: reasoning summaries (not raw CoT) & Reasoning
models expose REASONING SUMMARIES not raw chains of thought. The
reasoning\_\allowbreak{}tokens count is reported; the underlying tokens are not
visible t\ldots{} & Caller-side monitorability is structurally LIMITED:
only summaries, not raw traces. This is the design opposite of
Anthr\ldots{} & User-as-adversary, regulator-as-observer & 2026-05-22 &
High &
\href{https://developers.openai.com/api/docs/guides/reasoning}{link} \\
\end{longtable}}

\hypertarget{e1-per-deployment-licensing-and-access-control}{%
\subsection{E1: Per-deployment licensing and access
control}\label{e1-per-deployment-licensing-and-access-control}}

{\scriptsize\begin{longtable}[]{@{}
  >{\raggedright\arraybackslash}p{(\linewidth - 14\tabcolsep) * \real{0.07}}
  >{\raggedright\arraybackslash}p{(\linewidth - 14\tabcolsep) * \real{0.11}}
  >{\raggedright\arraybackslash}p{(\linewidth - 14\tabcolsep) * \real{0.30}}
  >{\raggedright\arraybackslash}p{(\linewidth - 14\tabcolsep) * \real{0.24}}
  >{\raggedright\arraybackslash}p{(\linewidth - 14\tabcolsep) * \real{0.10}}
  >{\raggedright\arraybackslash}p{(\linewidth - 14\tabcolsep) * \real{0.06}}
  >{\raggedright\arraybackslash}p{(\linewidth - 14\tabcolsep) * \real{0.05}}
  >{\raggedright\arraybackslash}p{(\linewidth - 14\tabcolsep) * \real{0.07}}
  @{}}
\toprule\noalign{}
\begin{minipage}[b]{\linewidth}\raggedright
Vendor
\end{minipage} & \begin{minipage}[b]{\linewidth}\raggedright
Feature
\end{minipage} & \begin{minipage}[b]{\linewidth}\raggedright
Documented capability
\end{minipage} & \begin{minipage}[b]{\linewidth}\raggedright
Documented limit
\end{minipage} & \begin{minipage}[b]{\linewidth}\raggedright
Adversary scope
\end{minipage} & \begin{minipage}[b]{\linewidth}\raggedright
Accessed
\end{minipage} & \begin{minipage}[b]{\linewidth}\raggedright
Conf.
\end{minipage} & \begin{minipage}[b]{\linewidth}\raggedright
Source
\end{minipage} \\
\midrule\noalign{}
\endhead
\bottomrule\noalign{}
\endlastfoot
Anthropic & Workspaces / Spend limits / Service tiers & Per-organization
usage tiers (Tier 1-4 + Monthly Invoicing). Tier advancement gated by
cumulative credit purchase. Customer-set spend limits per works\ldots{}
& Tier ceilings hard-capped (Tier 4 = \$200k/\allowbreak{}month standard; Monthly
Invoicing required for higher). Spend limit increases\ldots{} &
Cooperative & 2026-05-22 & High &
\href{https://platform.claude.com/docs/en/support/rate-limits}{link} \\
Google Vertex AI & Provisioned Throughput (fixed-cost, fixed-term
subscription) & Reserve throughput for supported generative AI models
via fixed monthly/\allowbreak{}weekly subscription. Specify model + locations.
Per-region availability; Singl\ldots{} & Pre-purchase commitment;
overage handling depends on consumption-option setting. Not all models
supported. & Cooperative & 2026-05-22 & High &
\href{https://docs.cloud.google.com/gemini-enterprise-agent-platform/models/provisioned-throughput}{link} \\
Google Vertex AI & Access control (IAM permissions for Vertex AI) &
Per-project / per-resource IAM permissions: aiplatform.admin,
aiplatform.user, aiplatform.viewer, fine-grained per-resource
permissions. aiplatform.co\ldots{} & Granular permissions for specific
resources; some Advanced AI features gated on per-project addendum
consent. & Cooperative + insider-as-adversary & 2026-05-22 & High &
\href{https://docs.cloud.google.com/gemini-enterprise-agent-platform/models/access-control}{link} \\
Google Vertex AI & Control access to Model Garden models &
Per-organization / per-project Model Garden access restrictions; gating
of partner models and open models. Advanced AI Safety Addendum consent
require\ldots{} & Per-model enable workflow; revoke via Model Garden
console. & Cooperative + insider-as-adversary & 2026-05-22 & High &
\href{https://docs.cloud.google.com/gemini-enterprise-agent-platform/models/control-model-access}{link} \\
OpenAI & Usage tiers (Free, Tier 1-5) + Organization/\allowbreak{}Project
structur\ldots{} & Six-tier model with auto-graduation: Free tier
(\$100/\allowbreak{}month), Tiers 1-5 (\$5/\$50/\$100/\$250/\$1000 paid).
Per-organization and per-project rate limits. Ti\ldots{} &
Auto-graduation triggers not transparent (tier transitions based on
opaque spending and usage history). Custom limits ab\ldots{} &
Cooperative & 2026-05-22 & High &
\href{https://developers.openai.com/api/docs/guides/rate-limits}{link} \\
OpenAI & RBAC permissions + IP allowlist + mTLS + OIDC & Role-based
access control with org-level and project-level roles. Custom-role
eligibility per permission area. IP allowlist, mTLS, OIDC
configuration\ldots{} & Some permissions (External Keys, IP allowlist,
mTLS, OIDC) restricted to Org Owner role only (not delegable to custom
ro\ldots{} & Cooperative + insider-as-adversary & 2026-05-22 & High &
\href{https://developers.openai.com/api/docs/guides/permissions}{link} \\
\end{longtable}}

\hypertarget{e2-output-filtering-and-capability-gating}{%
\subsection{E2: Output filtering and capability
gating}\label{e2-output-filtering-and-capability-gating}}

{\scriptsize\begin{longtable}[]{@{}
  >{\raggedright\arraybackslash}p{(\linewidth - 14\tabcolsep) * \real{0.07}}
  >{\raggedright\arraybackslash}p{(\linewidth - 14\tabcolsep) * \real{0.11}}
  >{\raggedright\arraybackslash}p{(\linewidth - 14\tabcolsep) * \real{0.30}}
  >{\raggedright\arraybackslash}p{(\linewidth - 14\tabcolsep) * \real{0.24}}
  >{\raggedright\arraybackslash}p{(\linewidth - 14\tabcolsep) * \real{0.10}}
  >{\raggedright\arraybackslash}p{(\linewidth - 14\tabcolsep) * \real{0.06}}
  >{\raggedright\arraybackslash}p{(\linewidth - 14\tabcolsep) * \real{0.05}}
  >{\raggedright\arraybackslash}p{(\linewidth - 14\tabcolsep) * \real{0.07}}
  @{}}
\toprule\noalign{}
\begin{minipage}[b]{\linewidth}\raggedright
Vendor
\end{minipage} & \begin{minipage}[b]{\linewidth}\raggedright
Feature
\end{minipage} & \begin{minipage}[b]{\linewidth}\raggedright
Documented capability
\end{minipage} & \begin{minipage}[b]{\linewidth}\raggedright
Documented limit
\end{minipage} & \begin{minipage}[b]{\linewidth}\raggedright
Adversary scope
\end{minipage} & \begin{minipage}[b]{\linewidth}\raggedright
Accessed
\end{minipage} & \begin{minipage}[b]{\linewidth}\raggedright
Conf.
\end{minipage} & \begin{minipage}[b]{\linewidth}\raggedright
Source
\end{minipage} \\
\midrule\noalign{}
\endhead
\bottomrule\noalign{}
\endlastfoot
Anthropic & Refusals and fallback / Streaming refusals & Server-side
safety classifiers may refuse to generate output. Refusal returned as a
structured stop\_\allowbreak{}reason or as a streaming refusal event. Fallback
c\ldots{} & Refusal taxonomy not exhaustively published; appeals process
not documented. & User-as-adversary & 2026-05-22 & High &
\href{https://platform.claude.com/docs/en/build-with-claude/refusals-and-fallback}{link} \\
Anthropic & Strengthen guardrails (mitigate prompt injections / reduce
h\ldots{} & Customer-side prompt-engineering guidance; structured-output
enforcement; input validation patterns; classifier-screening
recommendations. Anthropic S\ldots{} & Documentation primarily advisory
(best-practice guidance); concrete vendor-implemented filters per
category not disclose\ldots{} & User-as-adversary & 2026-05-22 & High &
\href{https://platform.claude.com/docs/en/test-and-evaluate/strengthen-guardrails/mitigate-jailbreaks-and-prompt-injections}{link} \\
Anthropic & Usage Policy / Acceptable Use Policy & Universal Usage
Standards covering 10+ harm categories (CSAM, weapons, critical
infrastructure, election integrity, etc.). High-Risk Use Case
Requirem\ldots{} & Policy carve-outs for certain governmental customers
(contractual use restrictions tailored to mission); the carve-out
t\ldots{} & User-as-adversary, deployer-as-adversary (limited) &
2026-05-22 & High & \href{https://anthropic.com/legal/aup}{link} \\
Anthropic & Computer use tool: prompt-injection classifier &
Anthropic-trained classifiers automatically run on computer-use prompts
to flag prompt injections in screenshots; steer model to ask for user
confirma\ldots{} & Classifier accuracy not publicly characterised.
Opt-out available, which means hostile-deployer can disable. Beta
featur\ldots{} & User-as-adversary, deployer-as-adversary & 2026-05-22 &
Medium &
\href{https://platform.claude.com/docs/en/agents-and-tools/tool-use/computer-use-tool}{link} \\
Google Vertex AI & Configurable safety filters (4 harm categories + 5
block thr\ldots{} & Four configurable HARM\_\allowbreak{}CATEGORY enums:
HATE\_\allowbreak{}SPEECH, DANGEROUS\_\allowbreak{}CONTENT, HARASSMENT, SEXUALLY\_\allowbreak{}EXPLICIT. Five
HARM\_\allowbreak{}BLOCK threshold values from BLOCK\_\allowbreak{}LOW\_\allowbreak{}\ldots{} & Probability vs
severity dimensions documented; Cloud console UI only uses probability
scores. BLOCK\_\allowbreak{}NONE option exists b\ldots{} & User-as-adversary &
2026-05-22 & High &
\href{https://docs.cloud.google.com/gemini-enterprise-agent-platform/models/capabilities/configure-safety-filters}{link} \\
Google Vertex AI & Gemini for filtering and moderation & Use Gemini as a
moderation classifier on caller-side text/\allowbreak{}image inputs. Distinct from
server-side filters; caller invokes for pre/\allowbreak{}post-screening. Alig\ldots{}
& Caller responsible for invocation; cost is per-Gemini-call rather than
free moderation endpoint. & User-as-adversary & 2026-05-22 & High &
\href{https://docs.cloud.google.com/gemini-enterprise-agent-platform/models/capabilities/gemini-for-filtering-and-moderation}{link} \\
Google Vertex AI & Process blocked responses (caller-side handling) &
Documented caller-side patterns for handling blocked responses:
finish\_\allowbreak{}reason inspection, blockReason enum handling, fallback-response
patterns. Strea\ldots{} & Caller responsibility to implement fallback;
no first-party fallback-text generation. & User-as-adversary,
deployer-as-defender & 2026-05-22 & High &
\href{https://docs.cloud.google.com/gemini-enterprise-agent-platform/models/capabilities/process-blocked-responses}{link} \\
Google Vertex AI & Suspected CSAM classifier on Claude-via-Vertex
(distinct fro\ldots{} & Vertex-platform-level CSAM classifier filters
image inputs to Anthropic models hosted on Vertex AI. SEPARATE from
Anthropic's own Trust \& Safety filte\ldots{} & Filters images only, not
text. Filters inputs only, not outputs. Vertex-platform-level controls
applied IN ADDITION TO p\ldots{} & User-as-adversary,
regulator-as-stakeholder & 2026-05-22 & High &
\href{https://docs.cloud.google.com/gemini-enterprise-agent-platform/models/partner-models/claude/safety}{link} \\
Google Vertex AI & Responsible AI / Safety overview policies & Google's
published AI Principles + Generative AI Prohibited Use Policy + Cloud
AUP. Multi-layer enforcement architecture: model-side training,
safety\ldots{} & Policy text high-level; specific classifier weights and
thresholds proprietary. & User-as-adversary, regulator-as-stakeholder &
2026-05-22 & Medium &
\href{https://docs.cloud.google.com/gemini-enterprise-agent-platform/models/responsible-ai}{link} \\
OpenAI & Moderation API (omni-moderation-latest) & Free dedicated
moderation endpoint accepting text and images (up to 20MB images).
Documented categories: harassment, hate, self-harm, sexual,
violence\ldots{} & Audio not classified by current moderation model.
Categories documented but classifier weights and thresholds
proprietar\ldots{} & User-as-adversary & 2026-05-22 & High &
\href{https://developers.openai.com/api/docs/guides/moderation}{link} \\
OpenAI & Guardrails framework (Agents SDK): input / output / tool
gua\ldots{} & Four control types: input guardrails (block disallowed
user requests), output guardrails (validate/\allowbreak{}redact final output), tool
guardrails (check argume\ldots{} & Guardrails are
application-implemented; Agents SDK provides scaffolding but the actual
rules are caller-defined. Anthrop\ldots{} & User-as-adversary,
deployer-as-defender & 2026-05-22 & High &
\href{https://developers.openai.com/api/docs/guides/guardrails}{link} \\
OpenAI & Safety best practices guidance (KYC, HITL, input/\allowbreak{}output
cons\ldots{} & Documented best practices: use Moderation API,
adversarial testing/\allowbreak{}red-teaming, human-in-the-loop, prompt engineering,
KYC (account registration / cre\ldots{} & Advisory; not enforced by
platform. KYC implementation entirely customer-side. &
User-as-adversary, regulator-as-stakeholder & 2026-05-22 & Medium &
\href{https://developers.openai.com/api/docs/guides/safety-best-practices}{link} \\
OpenAI & Under 18 API Guidance (COPPA-aligned) & Organizations serving
minors must comply with COPPA and similar laws. Cannot process personal
data of children under 13 (or applicable age of digital\ldots{} &
Compliance is organization responsibility; OpenAI does not directly
enforce age verification. Sole-responsibility-on-cus\ldots{} &
User-as-adversary (minor protection) & 2026-05-22 & Medium &
\href{https://developers.openai.com/api/docs/guides/under-18-guidance}{link} \\
\end{longtable}}

\hypertarget{e3-jurisdiction-bounded-inference-geofencing}{%
\subsection{E3: Jurisdiction-bounded inference
(geofencing)}\label{e3-jurisdiction-bounded-inference-geofencing}}

{\scriptsize\begin{longtable}[]{@{}
  >{\raggedright\arraybackslash}p{(\linewidth - 14\tabcolsep) * \real{0.07}}
  >{\raggedright\arraybackslash}p{(\linewidth - 14\tabcolsep) * \real{0.11}}
  >{\raggedright\arraybackslash}p{(\linewidth - 14\tabcolsep) * \real{0.30}}
  >{\raggedright\arraybackslash}p{(\linewidth - 14\tabcolsep) * \real{0.24}}
  >{\raggedright\arraybackslash}p{(\linewidth - 14\tabcolsep) * \real{0.10}}
  >{\raggedright\arraybackslash}p{(\linewidth - 14\tabcolsep) * \real{0.06}}
  >{\raggedright\arraybackslash}p{(\linewidth - 14\tabcolsep) * \real{0.05}}
  >{\raggedright\arraybackslash}p{(\linewidth - 14\tabcolsep) * \real{0.07}}
  @{}}
\toprule\noalign{}
\begin{minipage}[b]{\linewidth}\raggedright
Vendor
\end{minipage} & \begin{minipage}[b]{\linewidth}\raggedright
Feature
\end{minipage} & \begin{minipage}[b]{\linewidth}\raggedright
Documented capability
\end{minipage} & \begin{minipage}[b]{\linewidth}\raggedright
Documented limit
\end{minipage} & \begin{minipage}[b]{\linewidth}\raggedright
Adversary scope
\end{minipage} & \begin{minipage}[b]{\linewidth}\raggedright
Accessed
\end{minipage} & \begin{minipage}[b]{\linewidth}\raggedright
Conf.
\end{minipage} & \begin{minipage}[b]{\linewidth}\raggedright
Source
\end{minipage} \\
\midrule\noalign{}
\endhead
\bottomrule\noalign{}
\endlastfoot
Anthropic & inference\_\allowbreak{}geo parameter / cloud-platform regional offerings
& inference\_\allowbreak{}geo parameter (currently `us' or `global'). Cloud-platform
deployments via AWS Bedrock (multiple regions), Google Vertex AI
(multiple region\ldots{} & inference\_\allowbreak{}geo options limited at this time
(`us'/`global'); rate limits NOT separated by geo.
Cloud-platform-specific re\ldots{} & Cooperative +
regulatory-as-stakeholder & 2026-05-22 & High &
\href{https://platform.claude.com/docs/en/build-with-claude/claude-in-amazon-bedrock}{link} \\
Google Vertex AI & Data residency + VPC Service Controls + Per-region
quotas & Per-project data-residency: prompt logs stored in
customer-selected region or multi-region. VPC Service Controls perimeter
enforcement around inferenc\ldots{} & Some Gemini 3 services available
only in global location; not all models region-bounded. VPC-SC support
per-model per-op\ldots{} & Regulator-as-stakeholder, sovereignty-aware
deployer & 2026-05-22 & High &
\href{https://docs.cloud.google.com/gemini-enterprise-agent-platform/models/security-controls}{link} \\
OpenAI & Allowed-geography list + Bedrock/\allowbreak{}Foundry regional deployment &
Tier-1 access conditional on a supported-countries (allowed-geography)
list. OpenAI models in Amazon Bedrock provide AWS-managed regional
deployment w\ldots{} & Allowed-geography list not enumerated in this
file (separate page). Bedrock/\allowbreak{}Foundry deployment is OpenAI's explicit
mult\ldots{} & Cooperative + regulator-as-stakeholder & 2026-05-22 &
Medium &
\href{https://developers.openai.com/api/docs/guides/amazon-bedrock}{link} \\
OpenAI & Standard-platform data-residency controls & Project-level
data-residency controls let an eligible account set a processing region
for new projects on the standard OpenAI Platform. & Eligibility-gated;
an uplift applies to models released on or after March 2026; sets the
processing region rather than a\ldots{} & Cooperative +
regulator-as-stakeholder & 2026-05-22 & High &
\href{https://developers.openai.com/api/docs/guides/your-data}{link} \\
\end{longtable}}

\hypertarget{e4-rate-limiting-as-governance-primitive}{%
\subsection{E4: Rate limiting as governance
primitive}\label{e4-rate-limiting-as-governance-primitive}}

{\scriptsize\begin{longtable}[]{@{}
  >{\raggedright\arraybackslash}p{(\linewidth - 14\tabcolsep) * \real{0.07}}
  >{\raggedright\arraybackslash}p{(\linewidth - 14\tabcolsep) * \real{0.11}}
  >{\raggedright\arraybackslash}p{(\linewidth - 14\tabcolsep) * \real{0.30}}
  >{\raggedright\arraybackslash}p{(\linewidth - 14\tabcolsep) * \real{0.24}}
  >{\raggedright\arraybackslash}p{(\linewidth - 14\tabcolsep) * \real{0.10}}
  >{\raggedright\arraybackslash}p{(\linewidth - 14\tabcolsep) * \real{0.06}}
  >{\raggedright\arraybackslash}p{(\linewidth - 14\tabcolsep) * \real{0.05}}
  >{\raggedright\arraybackslash}p{(\linewidth - 14\tabcolsep) * \real{0.07}}
  @{}}
\toprule\noalign{}
\begin{minipage}[b]{\linewidth}\raggedright
Vendor
\end{minipage} & \begin{minipage}[b]{\linewidth}\raggedright
Feature
\end{minipage} & \begin{minipage}[b]{\linewidth}\raggedright
Documented capability
\end{minipage} & \begin{minipage}[b]{\linewidth}\raggedright
Documented limit
\end{minipage} & \begin{minipage}[b]{\linewidth}\raggedright
Adversary scope
\end{minipage} & \begin{minipage}[b]{\linewidth}\raggedright
Accessed
\end{minipage} & \begin{minipage}[b]{\linewidth}\raggedright
Conf.
\end{minipage} & \begin{minipage}[b]{\linewidth}\raggedright
Source
\end{minipage} \\
\midrule\noalign{}
\endhead
\bottomrule\noalign{}
\endlastfoot
Anthropic & Rate limits (RPM, ITPM, OTPM, per model class) &
Per-organization rate limits in three dimensions (RPM, ITPM, OTPM) per
model class. Token-bucket algorithm with continuous replenishment.
Tier-based e\ldots{} & Tier-based limits hard-capped. Acceleration
limits separately enforced (sharp usage increases trigger 429 even below
nom\ldots{} & Cooperative, deployer-budget-constraint & 2026-05-22 &
High &
\href{https://platform.claude.com/docs/en/support/rate-limits}{link} \\
Anthropic & Message Batches API limits & Separate RPM, processing-queue,
and per-batch-request limits for batch API. Tier-scaled (Tier 1 = 50
RPM/100k queue/100k per batch; Tier 4 = 4000/500k\ldots{} & Per-batch
maximum of 100k requests across all tiers. Processing-queue depth scales
with tier. & Cooperative & 2026-05-22 & High &
\href{https://platform.claude.com/docs/en/build-with-claude/batch-processing}{link} \\
Anthropic & Managed Agents endpoints (separate limits) &
Per-organization separate limits for Managed Agents: 300 RPM create, 600
RPM read operations. & Distinct from Messages API rate limits; Managed
Agents documented elsewhere. & Cooperative & 2026-05-22 & Medium &
\href{https://platform.claude.com/docs/en/support/rate-limits}{link} \\
Anthropic & Fast mode rate limits (research preview) & Dedicated rate
limits for fast mode (speed:`fast' on Opus 4.6/4.7/4.8); separate from
standard Opus limits. anthropic-fast-* response headers expose f\ldots{}
& Research-preview feature; rate-limit caps not exhaustively published.
& Cooperative & 2026-05-22 & Medium &
\href{https://platform.claude.com/docs/en/build-with-claude/fast-mode}{link} \\
Google Vertex AI & Per-metric quotas (token-based, request-based,
resource-base\ldots{} & Distinctive primitive: per-metric quota system
with named metrics
(e.g.~embed\_\allowbreak{}content\_\allowbreak{}input\_\allowbreak{}tokens\_\allowbreak{}per\_\allowbreak{}minute\_\allowbreak{}per\_\allowbreak{}base\_\allowbreak{}model,
online\_\allowbreak{}prediction\_\allowbreak{}reques\ldots{} & Quota grants subject to approval
process; some quotas globally bounded (not adjustable). Tuned-model
inference shares ba\ldots{} & Cooperative & 2026-05-22 & High &
\href{https://docs.cloud.google.com/gemini-enterprise-agent-platform/models/quotas}{link} \\
Google Vertex AI & Standard / Priority / Flex PayGo tiers + Batch
inference & Three on-demand tiers (Standard PayGo, Priority PayGo, Flex
PayGo) plus reserved Provisioned Throughput. Batch inference for Gemini
models uses dynami\ldots{} & Tier choice per-request via service-tier
configuration. Batch queueing during high-demand windows. & Cooperative
& 2026-05-22 & High &
\href{https://docs.cloud.google.com/gemini-enterprise-agent-platform/models/capabilities/batch-inference}{link} \\
OpenAI & Six-dimensional rate limits (RPM/\allowbreak{}RPD/\allowbreak{}TPM/\allowbreak{}TPD/\allowbreak{}IPM/\allowbreak{}audio-min) &
Six rate-limit dimensions per organization per model: requests per
minute (RPM), requests per day (RPD), tokens per minute (TPM), tokens
per day (TPD)\ldots{} & Hit-on-any-dimension: first dimension exceeded
triggers 429. Vector-store ingestion limits separate (300 RPM per
vector\ldots{} & Cooperative, deployer-budget-constraint & 2026-05-22 &
High &
\href{https://developers.openai.com/api/docs/guides/rate-limits}{link} \\
OpenAI & Priority processing (service\_\allowbreak{}tier=priority) & Per-request
opt-in to priority service tier in exchange for higher pricing.
Project-level default tier configurable. Ramp-rate limit: requests may
be\ldots{} & Downgrade-to-standard is silent (response shows
service\_\allowbreak{}tier=`default'). Ramp-rate limit threshold (1M TPM and 50\%
incre\ldots{} & Cooperative & 2026-05-22 & High &
\href{https://developers.openai.com/api/docs/guides/priority-processing}{link} \\
OpenAI & Flex processing (service\_\allowbreak{}tier=flex) & Reduced-cost
asynchronous tier for non-urgent workloads. May return
resource-unavailable error during contention; client retries by setting
service\_\allowbreak{}ti\ldots{} & Resource-unavailable is the user-visible failure
mode; no SLA at flex tier. & Cooperative & 2026-05-22 & High &
\href{https://developers.openai.com/api/docs/guides/flex-processing}{link} \\
OpenAI & Batch API (separate queue limits) & Asynchronous batch endpoint
with separate queue-limit dimension (queued input tokens per model).
Tokens from pending batch jobs count against batch qu\ldots{} & 24-hour
completion window; not for latency-sensitive workloads. Per-model queue
limits dashboard-only. & Cooperative & 2026-05-22 & High &
\href{https://developers.openai.com/api/docs/guides/batch}{link} \\
\end{longtable}}

\hypertarget{e5-agentic-action-permissions-and-policy-as-code-gates}{%
\subsection{E5: Agentic action permissions and policy-as-code
gates}\label{e5-agentic-action-permissions-and-policy-as-code-gates}}

{\scriptsize\begin{longtable}[]{@{}
  >{\raggedright\arraybackslash}p{(\linewidth - 14\tabcolsep) * \real{0.07}}
  >{\raggedright\arraybackslash}p{(\linewidth - 14\tabcolsep) * \real{0.11}}
  >{\raggedright\arraybackslash}p{(\linewidth - 14\tabcolsep) * \real{0.30}}
  >{\raggedright\arraybackslash}p{(\linewidth - 14\tabcolsep) * \real{0.24}}
  >{\raggedright\arraybackslash}p{(\linewidth - 14\tabcolsep) * \real{0.10}}
  >{\raggedright\arraybackslash}p{(\linewidth - 14\tabcolsep) * \real{0.06}}
  >{\raggedright\arraybackslash}p{(\linewidth - 14\tabcolsep) * \real{0.05}}
  >{\raggedright\arraybackslash}p{(\linewidth - 14\tabcolsep) * \real{0.07}}
  @{}}
\toprule\noalign{}
\begin{minipage}[b]{\linewidth}\raggedright
Vendor
\end{minipage} & \begin{minipage}[b]{\linewidth}\raggedright
Feature
\end{minipage} & \begin{minipage}[b]{\linewidth}\raggedright
Documented capability
\end{minipage} & \begin{minipage}[b]{\linewidth}\raggedright
Documented limit
\end{minipage} & \begin{minipage}[b]{\linewidth}\raggedright
Adversary scope
\end{minipage} & \begin{minipage}[b]{\linewidth}\raggedright
Accessed
\end{minipage} & \begin{minipage}[b]{\linewidth}\raggedright
Conf.
\end{minipage} & \begin{minipage}[b]{\linewidth}\raggedright
Source
\end{minipage} \\
\midrule\noalign{}
\endhead
\bottomrule\noalign{}
\endlastfoot
Anthropic & Tool use / Define tools / Handle tool calls & Structured
tools parameter defines available tool surface (name, description, JSON
schema). Model emits tool\_\allowbreak{}use blocks; client executes and returns
t\ldots{} & Tool authorisation is application-layer (Anthropic does not
gate per-tool access at the API). No tool registry shared ac\ldots{} &
Cooperative & 2026-05-22 & High &
\href{https://platform.claude.com/docs/en/agents-and-tools/tool-use/overview}{link} \\
Anthropic & Structured outputs & Schema-enforced JSON output via
structured-outputs API: model output validated against JSON schema
server-side before return. & Schema enforcement is structural, not
semantic; policy-level constraints must be encoded in schema (e.g., enum
types for\ldots{} & Cooperative & 2026-05-22 & High &
\href{https://platform.claude.com/docs/en/build-with-claude/structured-outputs}{link} \\
Anthropic & Computer use tool: confirmation gates & Default behavior:
classifier-detected prompt injections in screenshots steer the model to
ask for user confirmation before proceeding. Opt-out availab\ldots{} &
Confirmation behavior tied to injection-classifier verdict; not
generally policy-driven. Opt-out blanket removes the gat\ldots{} &
User-as-adversary & 2026-05-22 & Medium &
\href{https://platform.claude.com/docs/en/agents-and-tools/tool-use/computer-use-tool}{link} \\
Google Vertex AI & Function calling (JSON-schema tool definitions) &
Structured function calling via JSON-schema tool definitions. Multi-step
flow: model emits functionCall -\textgreater{} caller executes
-\textgreater{} caller returns functionR\ldots{} & Schema is JSON Schema
subset; not all JSON Schema features supported. Strict-mode enforcement
varies by model version. & Cooperative & 2026-05-22 & High &
\href{https://docs.cloud.google.com/gemini-enterprise-agent-platform/models/tools/function-calling}{link} \\
Google Vertex AI & Controlled generation (structured outputs) &
Server-side schema-enforced JSON output. Per-call schema declaration;
model output guaranteed to conform to declared schema. & Schema
enforcement is structural; semantic constraints (e.g., valid enum
values) must be encoded in schema. & Cooperative & 2026-05-22 & High &
\href{https://docs.cloud.google.com/gemini-enterprise-agent-platform/models/capabilities/control-generated-output}{link} \\
Google Vertex AI & Computer use tool & First-party computer-use tool for
desktop/\allowbreak{}browser automation. Built-in computer tool returns structured UI
actions; caller executes in isolated enviro\ldots{} & Caller responsible
for sandbox; tool returns actions, not enforcement. Specific safety
guidance about untrusted-input ha\ldots{} & User-as-adversary,
third-party-as-adversary & 2026-05-22 & High &
\href{https://docs.cloud.google.com/gemini-enterprise-agent-platform/models/computer-use}{link} \\
Google Vertex AI & Code execution tool & First-party code-execution tool
runs caller-requested Python in sandboxed environment. Per-region quota:
1000 RPM sandbox\_\allowbreak{}environment\_\allowbreak{}execute\_\allowbreak{}requests\ldots{} & Sandbox
runtime details not exhaustively published. Python-only at time of
writing. & User-as-adversary & 2026-05-22 & High &
\href{https://docs.cloud.google.com/gemini-enterprise-agent-platform/models/tools/code-execution}{link} \\
OpenAI & Function calling + custom tools + Responses API & Function
calling via JSON-schema tool definitions in tools parameter.
Custom-tools for free-form text inputs/\allowbreak{}outputs. Multi-step flow: model
emits too\ldots{} & Tool selection is application-layer; OpenAI does not
gate per-tool access at API. Caller responsible for actually
execut\ldots{} & Cooperative & 2026-05-22 & High &
\href{https://developers.openai.com/api/docs/guides/function-calling}{link} \\
OpenAI & Structured outputs (JSON schema enforcement) & Server-side
schema-enforced JSON output guaranteeing the model adheres to
caller-supplied JSON schema. Explicit refusals are programmatically
detectab\ldots{} & Schema enforcement is structural; semantic
constraints must be encoded via enums/\allowbreak{}required fields. Refusals are
model-sid\ldots{} & Cooperative & 2026-05-22 & High &
\href{https://developers.openai.com/api/docs/guides/structured-outputs}{link} \\
OpenAI & Orchestration (handoffs + agents-as-tools) & Multi-agent
orchestration patterns: handoffs (specialist takes over conversation),
agents-as-tools (manager calls specialists as bounded
capabilities)\ldots{} & Handoff API in SDK only; routing logic is
application-side. No platform-level enforcement of agent-graph
constraints. & Cooperative & 2026-05-22 & Medium &
\href{https://developers.openai.com/api/docs/guides/orchestration}{link} \\
OpenAI & Safety in building agents (no untrusted developer
messages,\ldots{} & Documented threat model: prompt injection,
private-data leakage. Mitigation guidance: do not put untrusted
variables in developer messages, use struct\ldots{} & Mitigations
advisory not enforced. Agent Builder being deprecated Nov 30 2026;
guidance applies to Agents SDK going forw\ldots{} & User-as-adversary,
third-party-as-adversary & 2026-05-22 & Medium &
\href{https://developers.openai.com/api/docs/guides/safety-in-building-agents}{link} \\
\end{longtable}}

\hypertarget{e6-tool-use-authorization-and-sandboxing}{%
\subsection{E6: Tool-use authorization and
sandboxing}\label{e6-tool-use-authorization-and-sandboxing}}

{\scriptsize\begin{longtable}[]{@{}
  >{\raggedright\arraybackslash}p{(\linewidth - 14\tabcolsep) * \real{0.07}}
  >{\raggedright\arraybackslash}p{(\linewidth - 14\tabcolsep) * \real{0.11}}
  >{\raggedright\arraybackslash}p{(\linewidth - 14\tabcolsep) * \real{0.30}}
  >{\raggedright\arraybackslash}p{(\linewidth - 14\tabcolsep) * \real{0.24}}
  >{\raggedright\arraybackslash}p{(\linewidth - 14\tabcolsep) * \real{0.10}}
  >{\raggedright\arraybackslash}p{(\linewidth - 14\tabcolsep) * \real{0.06}}
  >{\raggedright\arraybackslash}p{(\linewidth - 14\tabcolsep) * \real{0.05}}
  >{\raggedright\arraybackslash}p{(\linewidth - 14\tabcolsep) * \real{0.07}}
  @{}}
\toprule\noalign{}
\begin{minipage}[b]{\linewidth}\raggedright
Vendor
\end{minipage} & \begin{minipage}[b]{\linewidth}\raggedright
Feature
\end{minipage} & \begin{minipage}[b]{\linewidth}\raggedright
Documented capability
\end{minipage} & \begin{minipage}[b]{\linewidth}\raggedright
Documented limit
\end{minipage} & \begin{minipage}[b]{\linewidth}\raggedright
Adversary scope
\end{minipage} & \begin{minipage}[b]{\linewidth}\raggedright
Accessed
\end{minipage} & \begin{minipage}[b]{\linewidth}\raggedright
Conf.
\end{minipage} & \begin{minipage}[b]{\linewidth}\raggedright
Source
\end{minipage} \\
\midrule\noalign{}
\endhead
\bottomrule\noalign{}
\endlastfoot
Anthropic & Computer use tool sandboxing guidance & Computer use
requires a sandboxed computing environment; Anthropic's reference
implementation provides Docker container, agent loop, web interface.
To\ldots{} & Sandboxing is customer responsibility; reference
implementation is example, not enforced primitive. Per-action
permissio\ldots{} & User-as-adversary & 2026-05-22 & Medium &
\href{https://platform.claude.com/docs/en/agents-and-tools/tool-use/computer-use-tool}{link} \\
Anthropic & Bash tool / Text editor tool & Anthropic-defined tool
surfaces with structured schemas for command execution (bash) and file
editing (text editor). Application implements actual exe\ldots{} &
Anthropic provides tool schema only; sandboxing entirely customer-side.
& User-as-adversary & 2026-05-22 & Medium &
\href{https://platform.claude.com/docs/en/agents-and-tools/tool-use/bash-tool}{link} \\
Anthropic & MCP connector / Remote MCP servers & Model Context Protocol
integration; MCP connector + Remote MCP servers expose tools via
standardised protocol. Per-server authentication and scoping h\ldots{} &
MCP security model documented separately; scoping granularity
per-server, not per-tool-invocation within a server. & Cooperative &
2026-05-22 & Medium &
\href{https://platform.claude.com/docs/en/agents-and-tools/mcp-connector}{link} \\
Google Vertex AI & VPC Service Controls + private endpoints & VPC-SC
perimeter around Vertex AI inference endpoints. Private endpoints via
Private Service Connect. Per-model VPC-SC support documented in
security-\ldots{} & Some operations (e.g.~global-location) not
VPC-SC-bounded. Per-model variance in support. & Deployer-as-defender &
2026-05-22 & High &
\href{https://docs.cloud.google.com/gemini-enterprise-agent-platform/models/networking}{link} \\
Google Vertex AI & Sandbox environment quotas (Code Execution) &
Per-region sandbox\_\allowbreak{}environment quotas: 1000 RPM execute, 1000 entities,
500 RPM write. Sandbox tied to Agent Runtime quotas (10 RPM
create/\allowbreak{}delete, 100\ldots{} & Quotas adjustable via standard quota
request flow but bounded. & User-as-adversary & 2026-05-22 & High &
\href{https://docs.cloud.google.com/gemini-enterprise-agent-platform/models/quotas}{link} \\
Google Vertex AI & Request-response logging (configurable) &
Per-endpoint request-response logging to BigQuery / Cloud Logging;
opt-in. Logs include prompts, responses, latency, status. Useful for
post-hoc audit\ldots{} & Logging is opt-in; caller responsibility.
Storage costs separate. & Deployer-as-defender & 2026-05-22 & High &
\href{https://docs.cloud.google.com/gemini-enterprise-agent-platform/models/capabilities/request-response-logging}{link} \\
OpenAI & Sandbox agents (TypeScript and Python Agents SDKs, beta) &
First-party sandbox primitive in Agents SDK. SandboxAgent provides
isolated Unix-like execution environment with filesystem, shell,
installed packages\ldots{} & Beta product. API details and supported
capabilities may change. Provider-specific isolation (specific sandbox
backend n\ldots{} & User-as-adversary, third-party-as-adversary &
2026-05-22 & High &
\href{https://developers.openai.com/api/docs/guides/sandbox-agents}{link} \\
OpenAI & Computer use tool (built-in loop, custom harness,
code-execu\ldots{} & Three integration paths for desktop/\allowbreak{}browser
automation: built-in Responses API computer tool (model returns
structured UI actions); custom Playwright/\ldots{} & gpt-5.4+ trained
for the code-execution path; behavior may diverge across model versions.
Isolation is caller responsibi\ldots{} & User-as-adversary,
third-party-as-adversary & 2026-05-22 & High &
\href{https://developers.openai.com/api/docs/guides/computer-use}{link} \\
OpenAI & Local shell + Shell + Apply Patch + Code interpreter &
First-party hosted tools for shell access and code execution. Apply
Patch as a structured-edit tool. Code interpreter as a hosted Python
execution san\ldots{} & Hosted tools run in OpenAI infrastructure;
caller does not control sandbox details. Resource limits not in public
docs a\ldots{} & User-as-adversary & 2026-05-22 & Medium &
\href{https://developers.openai.com/api/docs/guides/tools-code-interpreter}{link} \\
OpenAI & MCP integration (hosted MCP + SDK-managed MCP + Secure MCP
T\ldots{} & Three MCP integration paths: hosted MCP (model calls remote
server through OpenAI's surface), SDK-managed MCP (caller's runtime owns
connection, appro\ldots{} & Hosted MCP server reputation is third-party;
OpenAI provides routing not server attestation. Approvals are SDK-level
not\ldots{} & Third-party-as-adversary & 2026-05-22 & Medium &
\href{https://developers.openai.com/api/docs/guides/integrations-and-observability}{link} \\
\end{longtable}}

\hypertarget{e7-inference-time-off-switch-and-kill-chain}{%
\subsection{E7: Inference-time off-switch and
kill-chain}\label{e7-inference-time-off-switch-and-kill-chain}}

{\scriptsize\begin{longtable}[]{@{}
  >{\raggedright\arraybackslash}p{(\linewidth - 14\tabcolsep) * \real{0.07}}
  >{\raggedright\arraybackslash}p{(\linewidth - 14\tabcolsep) * \real{0.11}}
  >{\raggedright\arraybackslash}p{(\linewidth - 14\tabcolsep) * \real{0.30}}
  >{\raggedright\arraybackslash}p{(\linewidth - 14\tabcolsep) * \real{0.24}}
  >{\raggedright\arraybackslash}p{(\linewidth - 14\tabcolsep) * \real{0.10}}
  >{\raggedright\arraybackslash}p{(\linewidth - 14\tabcolsep) * \real{0.06}}
  >{\raggedright\arraybackslash}p{(\linewidth - 14\tabcolsep) * \real{0.05}}
  >{\raggedright\arraybackslash}p{(\linewidth - 14\tabcolsep) * \real{0.07}}
  @{}}
\toprule\noalign{}
\begin{minipage}[b]{\linewidth}\raggedright
Vendor
\end{minipage} & \begin{minipage}[b]{\linewidth}\raggedright
Feature
\end{minipage} & \begin{minipage}[b]{\linewidth}\raggedright
Documented capability
\end{minipage} & \begin{minipage}[b]{\linewidth}\raggedright
Documented limit
\end{minipage} & \begin{minipage}[b]{\linewidth}\raggedright
Adversary scope
\end{minipage} & \begin{minipage}[b]{\linewidth}\raggedright
Accessed
\end{minipage} & \begin{minipage}[b]{\linewidth}\raggedright
Conf.
\end{minipage} & \begin{minipage}[b]{\linewidth}\raggedright
Source
\end{minipage} \\
\midrule\noalign{}
\endhead
\bottomrule\noalign{}
\endlastfoot
Anthropic & Usage Policy enforcement: throttle/\allowbreak{}suspend/\allowbreak{}terminate &
Per-organization Safeguards Team enforcement: throttle, suspend, or
terminate access on Usage Policy violation. Outputs may be blocked or
modified. & Enforcement criteria not exhaustively published. Appeal
process not documented. Throttle/\allowbreak{}suspend/\allowbreak{}terminate decisions not\ldots{}
& User-as-adversary, deployer-as-adversary & 2026-05-22 & Medium &
\href{https://anthropic.com/legal/aup}{link} \\
OpenAI & Model deprecation notice + transition windows & Advance notice
(email + page entry) before model retirement. Documented transition
windows: e.g., Agent Builder + Evals deprecation announced June
202\ldots{} & Notice period not fixed at single duration; varies per
deprecation. Customer-driven migration; no automatic re-routing
t\ldots{} & Cooperative + deployer-as-stakeholder & 2026-05-22 & High &
\href{https://developers.openai.com/api/docs/deprecations}{link} \\
\end{longtable}}

\end{landscape}
\restoregeometry

\hypertarget{appendix-e-adversary-coverage-matrix}{%
\section{Appendix E: Adversary-Coverage
Matrix}\label{appendix-e-adversary-coverage-matrix}}

Each cell records how a mechanism rates against one adversary cell,
defined by capability tier (C1 low, C2 medium, C3 high) crossed with
adversary role (R1 developer, R2 deployer, R3 user, R4 fine-tuner).
Ratings are A (adequate), P (partial), I (inadequate), N (not
applicable). The derivation and the per-mechanism reasoning are in
Section 4.3.

{\footnotesize\begin{longtable}[]{@{}
  >{\raggedright\arraybackslash}p{(\columnwidth - 24\tabcolsep) * \real{0.0769}}
  >{\raggedright\arraybackslash}p{(\columnwidth - 24\tabcolsep) * \real{0.0769}}
  >{\raggedright\arraybackslash}p{(\columnwidth - 24\tabcolsep) * \real{0.0769}}
  >{\raggedright\arraybackslash}p{(\columnwidth - 24\tabcolsep) * \real{0.0769}}
  >{\raggedright\arraybackslash}p{(\columnwidth - 24\tabcolsep) * \real{0.0769}}
  >{\raggedright\arraybackslash}p{(\columnwidth - 24\tabcolsep) * \real{0.0769}}
  >{\raggedright\arraybackslash}p{(\columnwidth - 24\tabcolsep) * \real{0.0769}}
  >{\raggedright\arraybackslash}p{(\columnwidth - 24\tabcolsep) * \real{0.0769}}
  >{\raggedright\arraybackslash}p{(\columnwidth - 24\tabcolsep) * \real{0.0769}}
  >{\raggedright\arraybackslash}p{(\columnwidth - 24\tabcolsep) * \real{0.0769}}
  >{\raggedright\arraybackslash}p{(\columnwidth - 24\tabcolsep) * \real{0.0769}}
  >{\raggedright\arraybackslash}p{(\columnwidth - 24\tabcolsep) * \real{0.0769}}
  >{\raggedright\arraybackslash}p{(\columnwidth - 24\tabcolsep) * \real{0.0769}}@{}}
\toprule\noalign{}
\begin{minipage}[b]{\linewidth}\raggedright
Mechanism
\end{minipage} & \begin{minipage}[b]{\linewidth}\raggedright
C1.R1
\end{minipage} & \begin{minipage}[b]{\linewidth}\raggedright
C1.R2
\end{minipage} & \begin{minipage}[b]{\linewidth}\raggedright
C1.R3
\end{minipage} & \begin{minipage}[b]{\linewidth}\raggedright
C1.R4
\end{minipage} & \begin{minipage}[b]{\linewidth}\raggedright
C2.R1
\end{minipage} & \begin{minipage}[b]{\linewidth}\raggedright
C2.R2
\end{minipage} & \begin{minipage}[b]{\linewidth}\raggedright
C2.R3
\end{minipage} & \begin{minipage}[b]{\linewidth}\raggedright
C2.R4
\end{minipage} & \begin{minipage}[b]{\linewidth}\raggedright
C3.R1
\end{minipage} & \begin{minipage}[b]{\linewidth}\raggedright
C3.R2
\end{minipage} & \begin{minipage}[b]{\linewidth}\raggedright
C3.R3
\end{minipage} & \begin{minipage}[b]{\linewidth}\raggedright
C3.R4
\end{minipage} \\
\midrule\noalign{}
\endhead
\bottomrule\noalign{}
\endlastfoot
M1 & N & I & A & N & N & P & A & N & N & I & A & N \\
M2 & N & I & P & N & N & I & P & P & N & I & P & I \\
M3 & N & P & A & N & N & P & A & P & N & I & A & P \\
M4 & N & P & A & N & N & I & A & N & N & I & P & N \\
M5 & N & A & A & N & N & P & P & N & N & I & I & N \\
M6 & N & I & I & N & N & I & I & N & N & I & I & N \\
V1 & P & P & N & P & P & A & N & A & I & P & N & P \\
V2 & N & I & N & N & N & I & N & N & N & I & N & N \\
V3 & P & P & N & P & P & P & N & P & I & I & N & I \\
V4 & N & A & A & P & N & A & A & P & N & P & A & P \\
V5 & A & P & N & A & P & P & N & A & I & I & N & A \\
V6 & N & P & P & N & N & I & P & N & N & I & P & N \\
V7 & P & P & A & P & P & I & A & I & I & I & P & I \\
E1 & N & A & A & N & N & P & A & N & N & I & P & N \\
E2 & P & A & A & P & P & I & P & I & I & I & P & I \\
E3 & N & A & A & N & N & P & P & N & N & I & I & N \\
E4 & N & A & A & N & N & P & P & N & N & P & P & N \\
E5 & N & A & A & P & N & I & P & P & N & I & P & I \\
E6 & N & A & A & P & N & P & P & P & N & I & P & I \\
E7 & N & A & A & P & N & P & P & P & N & I & P & I \\
\end{longtable}}

\hypertarget{appendix-f-mechanism-scenario-mapping-matrix}{%
\section{Appendix F: Mechanism-Scenario Mapping
Matrix}\label{appendix-f-mechanism-scenario-mapping-matrix}}

Each cell records a mechanism's role in one governance scenario: L
(primary lever), S (supporting), I (irrelevant), B (structurally
blocked). Scenarios are S1 domestic regulation, S2 bilateral or
multilateral coordination, S4 industry self-regulation, S5
compute-marketplace governance. The derivation and per-scenario
reasoning are in Section 6.

{\footnotesize\begin{longtable}[]{@{}
  >{\raggedright\arraybackslash}p{(\columnwidth - 8\tabcolsep) * \real{0.2000}}
  >{\raggedright\arraybackslash}p{(\columnwidth - 8\tabcolsep) * \real{0.2000}}
  >{\raggedright\arraybackslash}p{(\columnwidth - 8\tabcolsep) * \real{0.2000}}
  >{\raggedright\arraybackslash}p{(\columnwidth - 8\tabcolsep) * \real{0.2000}}
  >{\raggedright\arraybackslash}p{(\columnwidth - 8\tabcolsep) * \real{0.2000}}@{}}
\toprule\noalign{}
\begin{minipage}[b]{\linewidth}\raggedright
Mechanism
\end{minipage} & \begin{minipage}[b]{\linewidth}\raggedright
S1 Domestic
\end{minipage} & \begin{minipage}[b]{\linewidth}\raggedright
S2 Bi/Multilateral
\end{minipage} & \begin{minipage}[b]{\linewidth}\raggedright
S4 Self-regulation
\end{minipage} & \begin{minipage}[b]{\linewidth}\raggedright
S5 Marketplace
\end{minipage} \\
\midrule\noalign{}
\endhead
\bottomrule\noalign{}
\endlastfoot
M1 & L & S & S & L \\
M2 & I & I & I & I \\
M3 & S & S & S & S \\
M4 & S & L & I & S \\
M5 & L & S & S & L \\
M6 & I & I & I & I \\
V1 & S & L & B & S \\
V2 & I & S & B & I \\
V3 & I & S & B & I \\
V4 & S & L & S & L \\
V5 & S & L & S & S \\
V6 & S & S & S & S \\
V7 & S & S & L & S \\
E1 & L & S & S & L \\
E2 & L & S & S & L \\
E3 & L & L & I & L \\
E4 & L & S & S & L \\
E5 & S & I & S & L \\
E6 & S & I & S & L \\
E7 & L & S & S & L \\
\end{longtable}}

\bibliography{vendor_and_resolved_5_Aug_2026,paper_1_inference_governance_reused_5_Aug_2026,third_search_curated_5_Aug_2026,fourth_search_curated_5_Aug_2026,sixth_search_5_Aug_2026,additional_papers_5_Aug_2026}
\end{document}